\documentclass[reprint,superscriptaddress,groupdaddress,prx,floatfix,twocolumn,]{revtex4-2}
\usepackage[utf8]{inputenc}
\usepackage[english]{babel}
\usepackage{amsmath}
\usepackage{amsfonts}
\usepackage{amssymb}
\usepackage{natbib}
\usepackage{textcomp}
\usepackage{textgreek}
\usepackage{gensymb}
\usepackage{dsfont}
\usepackage[section]{placeins}
\usepackage{tabularx}
\usepackage{physics}
\usepackage{pgfplots}
\usetikzlibrary{intersections, pgfplots.fillbetween}
\usepackage{graphicx}
\usepackage{bm}
\usepackage{amssymb}
\usepackage{calc}
\usepackage{printlen}
\usepackage{hyperref}
\hypersetup{
    colorlinks=true,
    linkcolor=blue,
    filecolor=magenta,
    urlcolor=magenta,
    citecolor=violet}
\setcitestyle{numbers,square,comma}
\usepackage[capitalise]{cleveref}

\begin{document}

\newcommand{\T}[1]{\mathrm{#1}}
\newcommand{\tL}{\T{L}}
\newcommand{\tR}{\T{R}}
\newcommand{\LT}{\mathcal{L}_{\T{T}}}
\newcommand{\LS}{\mathcal{L}_{\T{S}}}
\newcommand{\LB}{\mathcal{L}_{\T{B}}}
\newcommand{\LQD}{\mathcal{L}_{\T{QD}}}
\newcommand{\LQP}{\mathcal{L}_{\T{QP}}}
\newcommand{\LCP}{\mathcal{L}_{\T{CP}}}
\newcommand{\Ltot}{\mathcal{L}_{\T{tot}}}
\newcommand{\Km}{\mathcal{K}}
\newcommand{\Lm}{\mathcal{L}}
\newcommand{\Tm}{\mathcal{T}}
\newcommand{\Pm}{\mathcal{P}}
\newcommand{\Qm}{\mathcal{Q}}
\newcommand{\rhoB}{\hat{\rho}_{\T{B}}}
\newcommand{\rhot}{\hat{\rho}_{\T{tot}}}
\newcommand{\HCP}{\hat{H}_{\T{CP},l}}
\newcommand{\HQP}{\hat{H}_{\T{QP},l}}
\newcommand{\HQD}{\hat{H}_{\T{QD}}}
\newcommand{\HT}{\hat{H}_{\T{T}}}
\newcommand{\cor}[1]{\textcolor{red}{#1}}
\newcommand{\kron}[2]{\delta_{#1 , #2}}
\newcommand{\mbf}[1]{\mathbf{#1}}
\newcommand{\up}{\uparrow}
\newcommand{\down}{\downarrow}
\newcommand{\opc}{\hat{c}}
\newcommand{\opd}{\hat{d}}
\newcommand{\oprho}{\hat{\rho}}
\newcommand{\nl}{\nonumber\\ }
\preprint{APS/123-QED}
\title{A particle conserving approach to AC-DC driven interacting quantum dots with superconducting leads}
\author{Julian Siegl}
\email{Julian.Siegl@physik.uni-regensburg.de}
\affiliation{Theoretische\; Physik,\,	Universit\"{a}t\, Regensburg,\, 93053\, Regensburg,\, Germany}
\author{Jordi Pic\'{o}-Cort\'{e}s}
\email{Jordi.Pico-Cortes@physik.uni-regensburg.de}
\affiliation{Theoretische\; Physik,\,	Universit\"{a}t\, Regensburg,\, 93053\, Regensburg,\, Germany}
\affiliation{Instituto de Ciencia de Materiales de Madrid (CSIC) E-28049, Spain}
\author{Milena Grifoni}
\affiliation{Theoretische\; Physik,\,	Universit\"{a}t\, Regensburg,\, 93053\, Regensburg,\, Germany}
\date{\today}

\begin{abstract}
  The combined action of a DC bias and a microwave drive on the transport characteristic of a superconductor-quantum dot-superconductor junction is investigated. To cope with time dependent non-equilibrium effects and interactions in the quantum dot, we develop a general formalism for the dynamics of the density operator based on a particle conserving approach to superconductivity. Without invoking a broken $U(1)$ symmetry, we identify a dynamical phase connected to the coherent transfer of Cooper pairs across the junction. In the weak coupling limit, we show that besides quasiparticle transport, proximity induced superconducting correlations manifest in anomalous pair tunneling involving the transfer of a Cooper pair. The resulting generalized master equation in presence of the microwave drive showcases the characteristic bichromatic response due to the combination of the AC Josephson effect and an AC voltage. Analytical expressions for all harmonics in the driving frequency of both the current and the reduced dot operator are given for arbitrary driving strength. For the net DC current the resulting photon assisted processes give rise to rich current-voltage characteristics. In addition to photon assisted subgap transport we find regions of total current inversion in the stability diagram. There, the junction acts as a pump with the net DC current flowing against the applied DC bias. The first harmonic of the current, being closely related to the nonlinear dynamic susceptibility of the junction, is discussed at finite applied DC bias.
\end{abstract}

\maketitle

\section{Introduction\label{sec: Introduction}}
Superconducting circuits based on Josephson junctions are currently one of the leading platforms for quantum information technology~\cite{Makhlin2010,Ladd2010,Wendin2017}. Charge transport in such junctions exhibits a rich phenomenology due to the presence of both Cooper pairs and gapped Bogoliubov quasiparticles. The latter are a fundamental tool in quantum technologies employing hybrid nanostructures, enabling high resolution transport spectroscopy~\cite{Giaever1960,Grove2009,Dirks2009}. However, quasiparticle poisoning~\cite{Bespalov2016,Frombach2020} can hinder certain applications. Signatures of quasiparticles extend, e.g. by thermal excitation, also into the gap, where they contribute distinctive features~\cite{Whan1996,Pfaller2013,Ratz2014,Gaass2014,Gramich2016}, in addition to the ones produced by Andreev processes~\cite{Andreev1964,Flensberg1988}.

Adding an AC drive to nanostructures results in a plethora of additional effects and rich transport characteristics~\cite{Grifoni1998,Platero2004}. The presence of a drive further opens up the possibility to manipulate the system properties using Floquet engineering~\cite{Shirley1965,Oka2019}. For a nanostructure connected to superconducting leads, quasiparticle transport is modified by the possibility of photon assisted tunneling processes~\cite{Tien1963,Kouwenhoven1994,Kouwenhoven1994b,Whan1996}. The resulting transport signatures have been measured e.g. in scanning tunneling microscope experiments with both a superconducting tip and a superconducting substrate~\cite{Roychowdhury2015,Kot2020,Peters2020}, where it is shown that they allow discerning the nature of the charge carriers employing the separation between the AC-induced sidebands. For a Josephson junction the combination of a time-dependent bias voltage and a supercurrent further results in the appearance of Shapiro steps~\cite{Shapiro1963}, a bichromatic effect arising from the interplay between the AC bias frequency and the intrinsic Josephson frequency of the junction. The microscopic origin of these steps has been investigated in several systems, most notably in quantum point contacts, where multiple Andreev reflections lead to subharmonic Shapiro steps~\cite{Dubos2001,Cuevas2002}. Recently, microwave driven Josephson junctions have attracted interest as a way of detecting $4\pi-$periodic supercurrents, one of the key signatures of topological superconductors~\cite{Rokhinson2012,Wiedenmann2016,Laroche2019,Fischer2022}.

\begin{figure}
\centering
\includegraphics[scale=1]{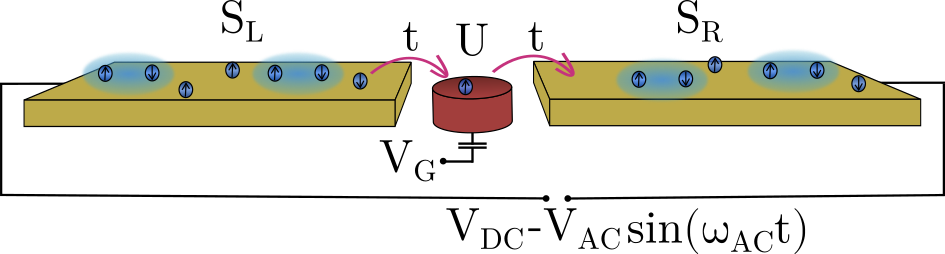}
\caption{Schematic setup of a gated quantum dot (QD) coupled to two superconducting leads (labeled L and R). A bias voltage with DC and AC components is applied between the left and right superconductor. These are characterized by superconducting gaps $\Delta_{\T{L}}$ and $\Delta_{\T{R}}$, respectively. A gate voltage is applied capacitively to the quantum dot via an electric gate.}
\label{fig: schematic}
\end{figure}

In this work we investigate the transport properties of a junction formed by an interacting quantum dot (QD) connected to two superconducting leads (an S-QD-S junction) in the presence of an AC drive. Quantum dots offer an ideal realization of the weak link required for a Josephson junction, providing an excellent platform to probe the relationship between superconducting correlations and interactions~\cite{Clerk2000,Jrgensen2007,MartinRodero2011}. One major characteristic of weakly coupled, small sized quantum dots is the energy cost to add extra charge to them. These charging effects due to the Coulomb repulsion are antagonistic to Cooper pairing. Thus, transfer of Cooper pairs is disfavored compared to quasiparticle transport for strong interaction. Conversely, coherent processes involving the transport of Cooper pairs are dominant below the gap, a situation that has been studied profusely in the infinite gap limit~\cite{Pala2007,Governale2008,Hiltscher2012}, and other approximation schemes ~\cite{Vecino2003,MartinRodero2011}. Several implementations of QD-based Josephson junctions have been devised, including semiconductor nanowires~\cite{vanDam2006} and carbon-based weak links such as nanotubes~\cite{Buitelaar2003,Vecino2004,Cleuziou2006,Eichler2007,Dirks2009, Grove2009,Pillet2010,Gaass2014}, fullerenes~\cite{Winkelmann2009}, and gated monolayers of graphene~\cite{Dirks2011}. QD-based Josephson junctions have moreover been proposed as a platform for quantum computing, employing Andreev spin qubits~\cite{Padurariu2010,Park2017,Pavesic2022,2205.03843}.

From a theoretical point of view, non-linear transport properties of interacting S-QD-S weak links subject to both DC and AC biases have been poorly investigated so far. This is partly due to the many-body character of the problem, together with the necessity to keep track of the number of Cooper pairs and quasiparticles transferred from one lead to another. Density operator-based methods are commonly used to investigate strongly interacting problems since they intrinsically enable an exact description of the many body nature of the interaction \cite{Bloch1957,Redfield1965,Konig1997,Konig1998,Pedersen2005,Timm2008,Leijnse2008,Koller2010,Karlstrom2013}. In our work we extend previous density operator-based treatments of transport through interacting S-QD-S junctions~\cite{Hiltscher2012,Pfaller2013,Governale2008,Pala2007,Ratz2014} to include the effects of a time-dependent AC bias. Superconducting correlations in the leads are treated within a particle conserving approach to transport in Josephson junctions~\cite{Josephson1962}. It allows one to systematically account for the non-equilibrium phenomena due to finite voltage bias. In particular, current conservation is ensured without the need of choosing particular symmetric configurations of the S-QD-S junction or self consistent procedures~\cite{MartinRodero2011}. The proximity effect results in the formation of coherences in the charge sector of the dot. The dynamics of such coherences and their role in transport has attracted significant interest lately \cite{Governale2008,Kamp2019,Kamp2021,Heckschen2022}. The particle conserving approach enables a rigorous discussion of these coherences for finite bias where the need to distinguish the individual condensates becomes apparent. The formalism is valid for generic interaction strength and amplitudes of the gap.

With this aim, we employ the Nakajima-Zwanzig projector operator formalism~\cite{Nakajima1958,Zwanzig1960} to obtain an exact generalized master equation (GME) for the reduced density operator of the interacting quantum dot as well as an integral equation for the current. Moreover, we derive formally exact expressions for both the current and the reduced density operator in the steady state which display the expected periodicity in both the intrinsic Josephson frequency and the frequency of the AC bias. With focus on the weak-tunneling limit, we retain only the lowest order terms in the coupling to the electrodes. We investigate the strength of the proximity induced coherences on the dot. For strong interaction, we describe their contribution to transport as a renormalization of the tunneling rates. There, they contribute an anomalous pair tunneling rate already at lowest order in the tunnel coupling. Our theory accounts for driven quasiparticle transport and recovers from a microscopic derivation similar results for the net DC current in presence of the drive as found in phenomenological approaches~\cite{Whan1996,Tien1963}. The full treatment of the photon assisted tunneling processes, as considered in this work, allow us to identify regions of the stability diagram where the strongly peaked density of states of the superconductors results, together with the AC bias, in dominant backward tunneling rates across the junction. For these regions our theory predicts total current inversion, in which the current flows against the applied DC voltage bias. 

Beyond the average current, we discuss the first harmonic of the current. This observable, describing the dynamic response of the current through the junction to the applied drive, is related to the conductivity in the limit of weak AC-drive amplitudes and DC bias. We investigate it as a function of applied gate and DC-bias voltages. In addition to replicas of the Coulomb resonance, we find intricate behavior in the gap region around the charge degeneracy points. There, for high enough driving frequencies, photon assisted sequential quasiparticle tunneling yields the dominant contribution to the AC response down to zero bias.

The paper is structured as follows: In \cref{sec: Model} we introduce the model of the junction in a formalism preserving particle conservation in the superconducting leads. We turn to a discussion of the transport theory for general AC-DC driven junctions in \cref{sec: Transport for an AC driven superconducting junction} and give equations for the steady state properties of the junction in \cref{sec: Steady state}. After truncating the resulting equations to second order in the tunnel coupling, we consider the case of a quantum dot junction in \cref{sec: Second order tunneling and current Kernels for a QD Josephson junction} and study the transport signatures for both the DC and AC-DC driven situations in \cref{sec: DC case} and \cref{sec: I-V characteristics} respectively. Finally, in \cref{sec: Conclusion} conclusions are drawn and further avenues of research are discussed.

\section{Model\label{sec: Model}}
We consider an S-QD-S junction, as exemplified by the setup of \cref{fig: schematic}, consisting of a gated quantum dot which is weakly coupled to two superconducting leads labeled by $l=\tL,\tR$. The total Hamiltonian is of the form
\begin{equation}
    \hat{H}_{\T{tot}}(t)=\HQD+\sum_{l}\hat{H}_{l}(t)+\HT,
    \label{eq: 95}
\end{equation}
where $\HQD$ and $\hat{H}_{l}$ are the Hamiltonians of, respectively, the dot and lead $l$. The tunneling Hamiltonian $\HT$ accounts for the tunnel coupling between the leads and the dot. The latter is modeled by the single impurity Anderson model (SIAM)~\cite{Anderson1961}, with
\begin{equation}
	\HQD=\sum_{\sigma}(\epsilon_{\sigma}+a_{\T{G}}eV_{\T{G}})\opd^{\dagger}_{\sigma}\opd_{\sigma}+U\opd^{\dagger}_{\up}\opd_{\up}\opd^{\dagger}_{\down}\opd_{\down},
	\label{eq: 1}
\end{equation}
where $\sigma\in \lbrace \up,\down\rbrace$, $\epsilon_{\sigma}$ denotes the spin dependent single particle energy, $eV_{\T{G}}$ is the energy shift due to the gate voltage and $U$ is a Hubbard like interaction~\cite{Hubbard1963}. The lever arm $a_{\T{G}}$ accounts for the imperfect coupling of the gate electrode to the dot. In the following we will consider, without loss of generality, $a_{\T{G}}=1$. The Fock space of the dot is spanned by the set $\{\ket{\chi}=\ket{0},\ket{\up},\ket{\down},\ket{2}\}$, comprising the empty, singly occupied with spin $\sigma$, and doubly occupied states, respectively.

For the leads, we start by considering a Hamiltonian  of the form
\begin{align}
    \hat{H}_{l}(t) = & \sum_{\sigma,\bm{k}} ( \xi_{l,\bm{k}} + \mu_{l}(t) ) \opc^{\dagger}_{l,\bm{k},\sigma} \opc_{l,\bm{k},\sigma} \nonumber \\
    + & \sum_{\substack{\sigma,\sigma',\bm{q}\\\bm{k},\bm{k}'}}
    V_{l}(\bm{q}) \opc^{\dagger}_{l,\bm{k}+\bm{q},\sigma} \opc^{\dagger}_{l,\bm{k}'-\bm{q},\sigma'} \opc_{l,\bm{k}',\sigma'} \opc_{l,\bm{k},\sigma}.
    \label{eq: 62}
\end{align}
Here, $\opc_{l,\bm{k},\sigma}$ is the annihilation operator for an electron from lead $l$ with momentum $\bm{k}$ and spin $\sigma$, with an associated energy $\xi_{l,\bm{k}}$ with respect to the chemical potential $\mu_{l}(t)$ and electron-electron interaction $V_{l}(\bm{q})$. The chemical potentials read
\begin{equation}
    \mu_{l}(t)= a_{l} [eV_{\T{DC}} - eV_{\T{AC}}\sin(\omega_{\T{AC}}t)],
    \label{eq: chem-pot}
\end{equation}
where $a_\T{L}-a_\T{R}=1$ and $0\leq a_\T{L}\leq 1$. These factors account for an asymmetric bias drop across the junction.

In the DC driven case, the transfer of a Cooper pair from the left to the right lead has an energy cost of $2(\mu_\T{L} - \mu_\T{R})=2eV_\T{DC}$.  In presence of an AC-drive, the time-dependency of the bias is reflected in any transfer of Cooper pairs between the leads. As such, keeping track of the number of Cooper pairs is fundamental for a full description of non-equilibrium transport.  Therefore, from here onward we will consider a particle-conserving approach to superconductivity, as introduced by Josephson and independently by Bardeen~\cite{Josephson1962,Bardeen1962}. From a fundamental point of view, it is clear that the electrodes' Hamiltonian \cref{eq: 62} conserves the particle number and hence is invariant under a $U(1)$-gauge transformation. One of the benefits of the theory to be discusses below is that such fundamental symmetry is not violated.

\subsection{Particle conserving formulation \label{subsec:Particle conserving formulation}}

An attractive interaction in \cref{eq: 62} results in the formation of a Cooper pair condensate in which electrons are bound in time-reversed pairs~\cite{Cooper1956}. In the particle-conserving approach, the ground state is completely described by the number of Cooper pairs in the condensate $M_l$, so that we can denote it simply by $\ket{M_l}$~\cite{Leggett2008}. Cooper pairs can be broken (e.g. due to thermal effects), and the minimal energy necessary to do so is twice the superconducting gap. As a result, the excitation spectrum of the superconductors is characteristically gapped. In order to describe these excited states, we rely on a mean-field description of the interaction. We introduce the superconducting gap for lead $l$ as the anomalous average $\Delta_{l,\bm{k}} = \sum_{\bm{k}'} V_{l}(\bm{k}-\bm{k}')\langle \hat{S}^{\dagger}_{l} \opc_{l,\bm{k}^\prime,\up} \opc_{l,\bar{\bm{k}}^\prime,\down} \rangle$. Here we have employed the Cooper pair creation and annihilation operators
\begin{align}
     \hat{S}_{l}^{\dagger}\ket{M_l}=\ket{M_l+1},\,\,\,\hat{S}_{l}\ket{M_l}=\ket{M_l-1},
     \label{eq: S operators}
\end{align}
which fix particle conservation in the above average. They satisfy $\hat{S}_l^{\dagger}\hat{S}_l=1-\hat{P}_{l,0}$~\cite{Pfaller2013}, where $\hat{P}_{l,0}$ projects to the state without Cooper pairs in lead $l$. In the following, we shall consider macroscopic leads, such that the action of $\hat{P}_{l,0}$ can be neglected. The normal averages in the mean field approximation result in Hartree- and Fock terms, which are diagonal in both spin and momentum and can thus be absorbed into the energies $\xi_{l,\bm{k}}$~\cite{Leggett2008}. We focus in the following on the case of isotropic dispersion such that $\xi_{l,\bm{k}}=\xi_{l,k}$. Thus, we are left with the mean field Hamiltonian
\begin{align}
    \hat{H}^{\text{MF}}_{l}(t) = & \sum_{\bm{k},\sigma} (\xi_{l,k}+\mu_{l}(t))\opc^{\dagger}_{l,\bm{k},\sigma}\opc_{l,\bm{k},\sigma} \nonumber \\
    - & \sum_{\bm{k}} \left( \Delta_{l,\bm{k}} \hat{S}_{l} \opc_{l,\bm{k},\up}^\dagger \opc_{l,\bar{\bm{k}},\down}^\dagger
    + \text{H.c.}\right).
    \label{eq: HMF}
\end{align}
For a constant interaction, which we assume in the following, $\Delta_{l,\bm{k}} = \Delta_l$ and the gap is independent of $\bm{k}$, yielding s-type superconductivity.

We next investigate the properties of this Hamiltonian under the $U(1)$-gauge transformation $\hat{c}^p_{l,\bm{k},\sigma} \rightarrow \hat{c}^p_{l,\bm{k},\sigma} e^{ip\varphi}$. The definition of $\hat{S}_l$ in \cref{eq: S operators} is implicit. Nonetheless, it maps a state with a given number of Cooper pairs onto a state with one Cooper pair less. Thus, even if its exact representation in terms of the $\hat{c}_{l,\bm{k},\sigma}$ operators remains unknown, it will be a linear combination of terms each containing two more annihilation than creation operators. From this, one concludes the transformation of $\hat{S}_l$ to be $\hat{S}^p_l \rightarrow  \hat{S}^p_l e^{i2p\varphi}$. Inserting this transformation behavior endows both the gap and the mean field Hamiltonian with the full $U(1)$ symmetry of the original Hamiltonian.

The Hamiltonian in \cref{eq: HMF} can be diagonalized employing the (particle-conserving) Bogoliubov-Valatin transformations~\cite{Josephson1962,Bardeen1962}
\begin{equation}
    \hat{c}_{l,\bm{k},\sigma}^{\dagger} = u_{l,k} \hat{\gamma}_{l,\bm{k},\sigma}^{\dagger} + \text{sgn}\left(\sigma\right) v_{l,k}^{*} \hat{S}_{l}^{\dagger} \hat{\gamma}_{l,\bar{\bm{k}},\bar{\sigma}} + \mathcal{O}\bigl(\hat{P}_{l,0}\bigr),
    \label{eq: bog-val}
\end{equation}
where
\begin{align}
    u_{l,k} & =\sqrt{(1/2)\left(1+\xi_{l,k}/E_{l,k}\right)},\\
    v_{l,k} & =e^{i\arg({\Delta_l})}\sqrt{\left(1/2\right)\left(1-\xi_{l,k}/E_{l,k}\right)}.
    \label{eq: e-h-populations}
\end{align}
Here, we have introduced the quasiparticle excitation energy $E_{l,k} = \sqrt{\xi_{l,k}^2+|\Delta_l|^2}$. \cref{eq: bog-val} defines the Bogoliubov quasiparticle operators $\hat{\gamma}_{l,\bm{k},\sigma},\hat{\gamma}^{\dagger}_{l,\bm{k},\sigma}$, which describe the fermionic excitations of the system.
These excitations are fermionic for macroscopic leads, since
\begin{align}
    \{\hat{\gamma}_{l,\bm{k},\sigma},\hat{\gamma}^\dagger_{l',\bm{k}',\sigma'}\}
    =\delta_{l,l'}[\delta_{\bm{k},\bm{k}'}\delta_{\sigma,\sigma'}+\mathcal{O}(\hat{P}_{l,0})].
    \label{eq: comm-rel}
\end{align}
The ground state is the vacuum for the quasiparticles, since $\hat{\gamma}_{l,\bm{k},\sigma}\ket{M_l}=0,\,\forall l,\bm{k},\sigma$. In turn, for a given particle number $N_l$, the excitation spectrum can be obtained by applying the Cooper pair and quasiparticle operators (which commute, up to factors $\mathcal{O}\bigl(\hat{P}_{l,0}\bigr)$) to the ground state. For arbitrary $N_l$, the Fock space is spanned by states of the form $\ket{M_l,\left\{\nu_{l,\bm{k},\sigma}\right\}}$, where $\nu_{l,\bm{k},\sigma}$ is the occupation of a given quasiparticle mode. Employing these properties, it can further be shown that $[\hat{N}_l,\hat{S}^\dagger_l]=2\hat{S}^\dagger_l$, with $\hat{N}_l$ the fermion number operator~\cite{Josephson1962}.

\subsection{Particle conserving mean field Hamiltonian of an S-QD-S junction \label{subsec: Particle conserving mean field Hamiltonian of an S-QD-S junction}}

Once the Hamiltonian \cref{eq: 62} is diagonalized, we may split it into quasiparticle and Cooper pair parts $\hat{H}_{l}=\hat{H}_{\T{QP},l}+\hat{H}_{\T{CP},l}$. Here
\begin{equation}
	\HQP(t)=\sum_{\bm{k},\sigma} (E_{l,k} + \mu_{l}(t)) \hat{\gamma}^{\dagger}_{l,\bm{k},\sigma} \hat{\gamma}_{l,\bm{k},\sigma},
	\label{eq: 2}
\end{equation}
describes the quasiparticle excitations, and
\begin{equation}
	\HCP(t) = \mu_{l}(t) \sum_{\bm{k},\sigma} \big( \opc^{\dagger}_{l,\bm{k},\sigma} \opc_{l,\bm{k},\sigma} -\hat{\gamma}^{\dagger}_{l,\bm{k},\sigma} \hat{\gamma}_{l,\bm{k},\sigma} \big),
	\label{eq: 3}
\end{equation}
accounts for the Cooper pairs in lead $l$.

The coupling between the dot and the leads is mediated by the tunneling Hamiltonian
\begin{align}
    \hat{H}_{\T{T}}= &\sum_{l,\bm{k},\sigma} \left( t_l \hat{c}^\dagger_{l,\bm{k},\sigma} \hat{d}_\sigma
    + \text{H.c.} \right)\nonumber \\
    = & \sum_{l,\bm{k},\sigma} \left[t_l\left(
    u_{l,k} \hat{\gamma}^\dagger_{l,\bm{k},\sigma}
    + \text{sgn}(\sigma) v^*_{l,k} \hat{S}^\dagger_l \hat{\gamma}_{l,\bar{\bm{k}},\bar{\sigma}}
    \right) \hat{d}_\sigma + \text{H.c.}\right].
    \label{eq: H Tunnel}
\end{align}
Introducing a Fock index $p=\pm$, such that $\hat{f}^{+}:=\hat{f}^\dagger$ and $\hat{f}^- := \hat{f}$ (as well as $h^{+}:=h^*,h^-:=h$ for complex numbers), the tunneling Hamiltonian can be written as
\begin{align}
	\hat{H}_{\T{T}}= & \sum_{\substack{l,\bm{k},\sigma,p}} pt_{l}^{\bar{p}} \opc^{p}_{l,\bm{k},\sigma} \opd^{\bar{p}}_{\sigma} \nonumber\\
	= & \sum_{l,\bm{k},\sigma,p} pt_{l}^{\bar{p}} \left((u^{\bar{p}}_{l,k} \hat{\gamma}^{p}_{l,\bm{k},\sigma}
	+\T{sgn}(\sigma) v^{p}_{l,k} \hat{S}^{p}_{l} \hat{\gamma}^{\bar{ p}}_{l,\bar{\bm{k}},\bar{\sigma}} \right) \opd^{\bar{p}}_{\sigma}.
	\label{eq: 4}
\end{align}
The  first and second terms of \cref{eq: 4} provide two transport channels in the quasiparticle representation. The second term, in particular, involves simultaneously quasiparticles and Cooper pairs in the process.

\section{Transport theory for an AC driven superconducting junction\label{sec: Transport for an AC driven superconducting junction}}

In the following a transport theory is presented to study the superconducting junction introduced in the previous section. We include both DC and AC biases as well as anomalous and normal contributions arising at finite  $|\Delta_{l}|$. The formalism extends previous works \cite{Governale2008,Hiltscher2012,Pfaller2013,Ratz2014,Gaass2014} while recovering the results therein in the appropriate parameter regimes.

\subsection{Current\label{subsec: current}}

We will be mainly concerned with studying the current through the junction. As a convention, we take the current to be the flow of charge from the quantum dot into the left lead \footnote{Due to particle conservation in this formalism, any displacement currents vanish in the time average. Therefore, alternative conventions differ in at most a phase(sign) for the AC(DC) part of the current.}. The current operator is then given by
\begin{equation}
	\hat{I}_{\tL} = -e\dot{\hat{N}}_{\T{L}} =\frac{-ie}{\hbar} [\hat{H}(t) ,\hat{N}_{\T{L}}] =\frac{ie}{\hbar} \sum_{\substack{\bm{k},\sigma,p}} t_{\tL}^{\bar{p}} \opc^{p}_{\tL,\bm{k},\sigma} \opd^{\bar{p}}_{\sigma},
    \label{eq: 37}
\end{equation}
where $e=|e|$ the elementary charge. The expectation value of the current can be obtained as
\begin{equation}
    I_{\tL}(t)=\T{Tr}\lbrace\hat{I}_{\tL}\rhot(t)\rbrace,
    \label{eq: exp val of I}
\end{equation}
where $\rhot(t)$ is the density operator for the junction. It satisfies the Liouville-von-Neumann equation
\begin{equation}
    \frac{d}{dt}\oprho_{\T{tot}}(t) = \frac{1}{i\hbar} [\hat{H}_{\T{tot}}(t), \rhot(t)] = \Ltot(t)\rhot(t),
    \label{eq: Liouvillian}
\end{equation}
where $i\hbar\Ltot(t)\hat{O}= [\hat{H}(t) ,\hat{O}]$ defines the Liouvillian superoperator. Similarly, we introduce $\LQD, \LCP(t), \LQP(t)$ and $\LT$ by restricting $\hat{H}_{\T{tot}}$ to the respective part of the Hamiltonian in \cref{eq: Liouvillian}.

Josephson junctions are characterized by the DC and AC Josephson effects~\cite{Josephson1962,Josephson1974}, the latter of which entails the dynamics of the condensates caused by a finite DC bias. In order to treat the periodicity coming from the AC Josephson effect and the one coming from the AC voltage bias on the same formal footing, we perform the following unitary transformation on the density operator
\begin{align}
	\oprho_{\T{tot}}'(t)=\mathcal{U}(t)\oprho_{\T{tot}}(t)=\exp(-\int_{t_0}^{t}dt'\LCP(t'))\oprho_{\T{tot}}(t).
	\label{eq: 9}
\end{align}
We denote the transformed operators with a prime from here onward. The Liouville-von Neumann equation for the transformed density operator reads
\begin{equation}
    \frac{d}{dt}\oprho'_{\T{tot}}(t)=[\LQD+\LQP(t)+\LT'(t)]\rhot'(t).
    \label{eq: Liouvillian-trans}
\end{equation}
The transformation removes the Cooper pair Liouvillian at the cost of turning the tunneling Liouvillian time-dependent. Meanwhile, $\LQD$ and $\LQP(t)$ are unchanged, since they commute with the Cooper pair Liouvillian. On the other hand, employing that $\LCP(t) \hat{S}^p_l =2p\mu_l(t)\hat{S}^p_l$, the tunneling Liouvillian becomes
\begin{align}
	\LT^\prime(t)= &\sum_{l,\bm{k},\sigma,p,\alpha}  \frac{pt^{\bar{p}}_{l}}{i\hbar}
	\bigg[  u^{\bar{p}}_{l,k} \hat{\gamma}^{p,\alpha}_{l,\bm{k},\sigma} \nonumber \\
	+ & \T{sgn}(\sigma) v^{p}_{l,k} \hat{\gamma}^{\bar{p},\alpha}_{l,\bar{\bm{k}},\bar{\sigma}} e^{\frac{2ip}{\hbar}\int_{t_0}^{t}dt'\mu_l\left(t'\right)} \hat{S}_{l}^{p,\alpha} \bigg] \opd^{\bar{p},\alpha}_{\sigma},
	\label{eq: 12}
\end{align}
where we introduced the superoperator index $\alpha$, defined by
\begin{equation}
    \hat{X}^{\alpha}\hat{Y}=
    \begin{cases}
        \hat{X}\hat{Y}, & \T{if }\, \alpha=+,\\
        \hat{Y}\hat{X}, &\T{if }\, \alpha=-.
    \end{cases}
\end{equation}
These enable us to write the action of the Liouvillian in a compact fashion
\begin{align}
    i\hbar\Lm_{q}(t)\hat{O}=\sum_{\alpha}\alpha\hat{H}_q^{\alpha}(t)\hat{O},
\end{align}
for $q=$ tot, QD, QP, CP, and T.

Notice that \cref{eq: 12} contains a time integration over the lead chemical potentials. At a constant voltage bias $V_\T{DC}$, the AC Josephson effect predicts a steady state current through the junction which will be, in general, periodic in time with the Josephson frequency $\omega_\T{J} = 2eV_\T{DC}/\hbar$. Due to the two individual bias drops in our setup, it is here more convenient to define the vector
\begin{equation}
    \bm{\omega}_\T{DC}=\frac{2eV_\T{DC}}{\hbar}(a_\tL,a_\tR).
\end{equation}
The tunneling Liouvillian thus becomes a sum of functions of the time $t$ being periodic in the linear combinations of  $\bm{\omega}_\T{DC}$ and $\omega_\T{AC}$. This property can be made explicit by employing the Jacobi-Anger expansion~\cite{Abramowitz1965} to find
\begin{align}
    e^{\frac{i2p}{\hbar}\int_{t_0}^{t}dt'\mu_{l}\left(t'\right)}
    =e^{ip\bm{u}_{l}\cdot\bm{\omega}_{\text{DC}}(t-t_{0})} \sum_{n} i^{n} J_{n}\left( a_l\epsilon_\text{AC} \right) e^{in\omega_{\text{AC}}t},
    \label{eq: integral-mu}
\end{align}
where $\epsilon_\text{AC} = 2eV_{\T{AC}}/(\hbar\omega_{\T{AC}})$ is a parameter quantifying the strength of the drive, $J_n(z)$ is the $n$th order Bessel function of the first kind, $\bm{u}_l=(\delta_{l,\tL},\delta_{l,\tR})$, and we chose $t_{0}=\pi/2\omega_{\text{AC}}$ for convenience \footnote{Any other choice is identical up to an additional phase factor absorbed into $v_{l,k}$}.

\cref{eq: Liouvillian-trans} is the starting point to derive a generalized master equation (GME) for the reduced density operator which we will employ to describe the transport properties of the system.

\subsection{Generalized master equation\label{subsec: Identifying the proper reference state of the bath}}

The density operator describes the full coherent dynamics of the dot and the leads, with the latter including the infinite degrees of freedom of the quasiparticles and the Cooper pairs. In turn, the current formula \cref{eq: exp val of I} involves the total trace of the product of the current and density operator. We conveniently write it as
\begin{equation}
    \T{Tr}\lbrace\cdots\rbrace=\T{Tr}_\T{sys}\lbrace\T{Tr}_\T{QP}\lbrace\cdots\rbrace\rbrace,
\end{equation}
where $\T{Tr}_\T{QP}\lbrace\cdots\rbrace$ is the trace over the quasiparticle degrees of freedom, while $\T{Tr}_\T{sys} \lbrace \cdots \rbrace$ is the trace over the system (i.e. quantum dot and Cooper pairs). This opens up the possibility to drastically reduce the number of degrees of freedom that have to be considered by introducing the reduced density operator
\begin{equation}
    \oprho'(t)= \T{Tr}_{\T{QP}} \lbrace \rhot'(t) \rbrace.
    \label{eq: def rho}
\end{equation}
\cref{eq: Liouvillian-trans} leads to the generalized master equation for the reduced density operator given by 
\begin{equation}
    \frac{d}{dt}\oprho'(t) = \Lm_{\T{QD}} \oprho'(t)
    + \int_{0}^{t}ds \Km'_{\T{T}}(t,s) \oprho'(s),
    \label{eq: 43}
\end{equation}
where $\Km'_{\T{T}}(t,s)$ denotes the tunneling kernel superoperator and we assumed a factorized initial state of system and quasiparticles. The assumption is natural in that it amounts to starting from a quenched system where we turn on the coupling between dot and lead at time $t=0$. For such initial conditions, the correlations between dot and lead degrees of freedom are vanishing at initial times. The tunneling kernel superoperator includes the effect of the quasiparticles and introduces irreversibility in the time evolution. The current can be obtained from $\oprho'$ by 
\begin{equation}
	I_{\tL}(t)=\int_{0}^{t}ds\T{Tr}_\T{sys}\lbrace\Km^\prime_{\T{I},\text{L}}(t,s)\oprho'(s)\rbrace,
	\label{eq: 49}
\end{equation}
where we have introduced the current kernel $\Km'_{\T{I},\text{L}}$. Both \cref{eq: 43,eq: 49} and the general form of $\Km'_{j}(t,s)$, where $j=\T{T},(\T{I,L})$ stands respectively for the tunneling and current kernels, are derived in \cref{App: Derivatio of the kernels}.

\subsection{Weak coupling limit \label{subsec: Second order tunneling and current kernels}}

The kernel superoperators can be expanded as a power series in the tunneling amplitudes. In the weak tunneling limit, only the lowest order contribution in the tunneling coupling is accounted for. Due to particle conservation in the leads, this corresponds to the sequential tunneling limit described by the terms $\propto\mathcal{L}_\T{T}^{\prime2}(t)$ and $\propto\hat{I}_\tL'\mathcal{L}_\T{T}^{\prime}(t)$, for $\T{T},(\T{I,L})$ respectively. The tunneling kernel to sequential tunneling order is then given by
\begin{align}
    \Km^{\prime(2)}_{\T{T}}(t,s) \oprho'(s) = \T{Tr}_{\T{QP}}\lbrace \LT'(t) \mathcal{G}'_0(t,s) \LT'(s)
    \oprho'(s) \otimes \oprho_{\T{QP}} \rbrace,
    \label{eq: Tunneling Kernel}
\end{align}
where
\begin{align}
    \mathcal{G}'_0(t,s) = \exp(\int_s^t ds' \LQD + \LQP(s')),
    \label{eq: freeprop}
\end{align}
is the free propagator and $\oprho_{\T{QP}}$ is the grand canonical density operator of the quasiparticles at thermal equilibrium (see \cref{App: Derivatio of the kernels}). The time-dependence of the chemical potential translates here into a time-dependence of $\LQP(t)$ (c.f. \cref{eq: 2}). This, together with the time-dependence of $\LT'(t)$ (i.e. \cref{eq: 12}), breaks time translation invariance and results in a kernel which is a function of two time variables. While the presence of a dissipative bath ensures the presence of a steady state~\cite{Grifoni1998}, it is this property that enables such a state to be not necessarily static even in the long time limit. Similarly, the current kernel to sequential tunneling order is given by
\begin{align}
   \Km^{\prime(2)}_{\T{I},\T{L}}(t,s)\oprho'(s)=\T{Tr}_\T{QP}\lbrace \hat{I}'_{\T{L}}(t)\mathcal{G}'_0(t,s)\LT'(s)\oprho'(s)\otimes\oprho_{\T{QP}}\rbrace.
   \label{eq: Currentkernel2}
\end{align}
Note that the current operator is also transformed by \cref{eq: 9}. The similarity between the current and tunneling kernels will allow us to treat them in the same manner in the following.

\subsection{Dynamics in the Cooper pair number representation \label{subsec: Eliminating the Cooper pairs}}

After tracing out the quasiparticle degrees of freedom, the density matrix still includes the Cooper pair numbers, which makes the density operator infinitely-sized. In particular, the density matrix can be written as
\begin{equation}
    \hat{\rho}^{\prime} \left(t\right) = \sum_{\substack{\bm{M},\Delta\bm{M}\\\chi,\chi'}} \varrho^{\prime}_{\chi,\chi'}(\Delta\bm{M},\bm{M};t) \ketbra{\chi,\bm{M}+\Delta\bm{M}}{\chi',\bm{M}},
    \label{eq: structure of rho}
\end{equation}
where the vector $\bm{M}=(M_{\T{L}},M_{\T{R}})$ labels the Cooper pair numbers of the two condensates, while $\Delta\bm{M}=(\Delta M_{\T{L}}, \Delta M_{\T{R}})$ measures the difference of the Cooper pair content of the condensates in the bra and ket parts of the density operator. This is the Cooper pair number representation.

The superselection rules discussed in \cref{App: Selection rules} force all components $\varrho^{\prime}_{\chi,\chi'} (\Delta\bm{M},\bm{M};t)$ to vanish, except where the ket and bra parts have the same total number of particles. Note that for an off diagonal state in the Cooper pair Liouville space, this implies the possibility of having coherences of the types $\ketbra{0}{2},\ketbra{2}{0}$ in the Liouville space of the dot. The dynamics of such coherences has been studied in earlier works~\cite{Pala2007,Kamp2019,Kamp2021} for zero bias. It highlights the necessity to account for the respective state of the condensate contributing to a coherence. E.g. the coherence $\ketbra{0}{2}$ on the dot is associated to a state of $\Delta\bm{M} \in \lbrace \mathbb{Z}(1,-1) + \bm{u}_{l}\rbrace_{l=\T{L,R}}$ in the condensate.

In order to isolate the action of the kernels on the Liouville space of the Cooper pairs, we write them as
\begin{align}
    \mathcal{K}'_{j}\left(t,s\right) \hat{\mathcal{O}} & = \sum_{\bm{N}^{+},\bm{N}^{-}} \kappa'_{j}\left(\bm{N}^{+},\bm{N}^{-};t,s\right)
    \bm{\hat{S}}^{\bm{N}^{+}} \hat{\mathcal{O}} \bm{\hat{S}}^{\bm{N}^{-}},
    \label{eq: action of K}
\end{align}
where we use the shorthand notation $\bm{\hat{S}}^{\bm{N}} = \hat{S}_\T{L}^{N_\T{L}} \hat{S}_\T{R}^{N_\T{R}}$,  $\hat{S}_l^{N} = (\hat{S}_l^{\T{sign}(N)})^{|N|}$.

These expressions can be simplified considerably, since the current and most physical observables of interest are independent of the total Cooper pair number and only take into account \textit{differences} of Cooper pairs. In particular, inserting \cref{eq: structure of rho,eq: action of K} into \cref{eq: 49} yields the following expression for the current
\begin{align}
&I_{\text{L}}\left(t\right)\\ \nonumber =&\int_{0}^{t}ds\sum_{\bm{N}^{+}\bm{N}^{-}}\text{Tr}_{\T{sys}}\{\kappa_{j}^{\prime}\left(\bm{N}^{+},\bm{N}^{-};t,s\right)\hat{\bm{S}}^{\bm{N}^{+}+\bm{N}^{-}}\hat{\rho}^{\prime}\left(s\right)\}
    \label{eq: Current in Cooper pair rep}
\end{align}
where we used the cyclic property of the trace to move the Cooper pair operators acting on the right to the left. Due to this, the total Cooper pair number $\bm{M}$ is not changed (c.f. \cref{eq: structure of rho}). Now, applying the trace over the Cooper pair sector yields the condition
\begin{align}
     \bm{N}^+ + \bm{N}^- = - \Delta\bm{M}.
\end{align}
which we employ to fix e.g. $\bm{N}^-$. This motivates the following definitions
\begin{align}
    \kappa'_{j}\left(-\bm{\Delta M};t,s\right) & = \sum_{\bm{N}^{+}} \kappa'_{j}\left(\bm{N}^{+},-\bm{\Delta M}-\bm{N}^{+};t,s\right),
    \label{eq: simplified KI}\\
    \hat{\varrho}'(\Delta \bm{M};t) & = \sum_{\bm{M},\chi,\chi'} \varrho'_{\chi,\chi'}(\Delta \bm{M},\bm{M};t) \ketbra{\chi}{\chi'}.
    \label{eq: simplified rho}
\end{align}
The $\hat{\varrho}'(\Delta \bm{M};t)$ defined here are operators acting on the dot sector only. They are a generalization of the partial trace over the Cooper pair degrees of freedom. In particular, $\hat{\varrho}'(0;t)$ is the partial trace proper and thus the reduced density operator of the dot with respect to the full leads (i.e. Cooper pairs and quasiparticles).
The rest of the $\hat{\varrho}'(\Delta \bm{M};t)$ are sums over the subdiagonals or supradiagonals of the density matrix. Note that the dot operators $\hat{\varrho}'(\Delta \bm{M};t)$ are in general not reduced density operators. In fact, their traces are arbitrary except for the case of $\Delta\bm{M}=0$ where it is one.

With these definitions, the current can be written as
\begin{align}
    I_{\text{L}}(t)=& \int_0^tds \sum_{\Delta\bm{M}} \T{Tr}_\T{QD}\bigg\lbrace \kappa_{\T{I,L}}^\prime(-\Delta\bm{M};t,s) \hat{\varrho}^\prime(\Delta\bm{M};s) \bigg\rbrace.
    \label{eq: Current simplified}
\end{align}
\cref{eq: Current simplified} is extremely convenient. It reduces the transport problem to the evaluation of operators which act only in the dot space, with the entirety of the Liouville space of the Cooper pairs reduced to an additional parameter $\Delta\bm{M}$. The presence of $\Delta\bm{M}\neq0$ entries in \cref{eq: Current simplified} indicates the key role that coherent effects between the Cooper pair condensates have in transport.

The GME for $\hat{\varrho}(\Delta\bm{M};t)$ is obtained by taking the matrix element $\bra{\Delta\bm{M}+\bm{M}}\hat{O}\ket{\bm{M}}$ on either side of \cref{eq: 43} and summing over $\bm{M}$ in much the same way as for the current. We find
\begin{align}
    &\frac{d}{dt} \hat{\varrho}^\prime(\Delta\bm{ M};t) = \LQD \hat{\varrho}^\prime(\Delta\bm{ M};t) \nonumber\\
    + & \sum_{\Delta \bm{M}'} \int_0^t ds \kappa_{\T{T}}^\prime(\Delta\bm{ M}-\Delta\bm{ M}';t,s) \hat{\varrho}^\prime(\Delta\bm{M}';s), \label{eq: EOM simplified}
\end{align}
where the time evolution of $\hat{\varrho}(\Delta\bm{M};t)$ for different $\Delta\bm{M},\Delta\bm{M}'$ is coupled via the action of the kernel $\kappa_{\T{T}}^\prime(\Delta\bm{ M}-\Delta\bm{ M}';t,s)$.

\subsection{Dynamics in the phase representation \label{subsec: CPpart2}}

\cref{eq: EOM simplified} is of Toeplitz form, which suggests turning to a representation in the conjugate variable of $\Delta\bm{M}$. We denote the adjoint variable to the Cooper pair imbalance $\Delta\bm{M}$ by the phase vector $\bm{\varphi}=(\varphi_{\T{L}},\varphi_{\T{R}})$. The transforms of kernels and the reduced dot operators in the phase representation are given by
\begin{align}
    \kappa_{j}^{\circ}(\bm{\varphi};t,s) & = \sum_{\Delta\bm{M}} e^{i\Delta\bm{M}\cdot\bm{\varphi}} \kappa_{j}^{\prime}(\Delta\bm{M};t,s), \label{eq:kernel-ph}\\
    \hat{\varrho}^{\circ}(\bm{\varphi};t) & = \sum_{\Delta\bm{M}} e^{i\Delta\bm{M}\cdot\bm{\varphi}} \hat{\varrho}^{\prime}(\Delta\bm{M};t). \label{eq:dm-ph}
\end{align}
We use the symbol $\circ$ to denote quantities in the phase representation.
Multiplying the GME in \cref{eq: EOM simplified} with $\exp(i\Delta\bm{M}\cdot\bm{\varphi})$ and summing over $\Delta\bm{M}$ yields
\begin{equation}
    \frac{d}{dt} \hat{\varrho}^{\circ}(\bm{\varphi};t) = \mathcal{L}_{\text{QD}} {\hat{\varrho}}^{\circ}(\bm{\varphi};t)
    + \int_{0}^{t}ds \kappa_{\text{T}}^{\circ}(\bm{\varphi};t,s) \hat{\varrho}^{\circ}(\bm{\varphi};s).
    \label{eq: GME in phase final}
\end{equation}
With this, we have managed to turn an infinite  set of coupled equations into a continuous family of uncoupled GMEs. This simplified form is possible because of the unitary transformation performed in \cref{subsec: current}. Due to it, the kernel does only act on the Cooper pair sector via $\hat{S}_l$ operators, which are diagonal in the phase representation.

In order to evaluate the current, we also bring \cref{eq: Current simplified} into the phase representation
\begin{align}
    I_{\text{L}}(t) = & \int_{0}^{t}ds \int_{\square}d\bm{\varphi} \text{Tr}_{\text{QD}}\bigg\lbrace \kappa_{\text{I},\text{L}}^{\circ}(\bm{\varphi};t,s) \hat{\varrho}^{\circ}(\bm{\varphi};s) \bigg\rbrace,
    \label{eq: Current in phase rep}
\end{align}
where we write $\int_{\square}d\bm{\varphi}=\frac{1}{4\pi^2}\int_{0}^{2\pi}d\varphi_{\tL}\int_{0}^{2\pi}d\varphi_{\tR}$ as a shorthand. Hence, the trace over the Cooper pair imbalance becomes an integral over the phase variables.

The phase introduced in this section has multiple interesting properties, which we now briefly comment on. From an interpretative point of view, the origin of the phase vector as the conjugate of a quantity measuring the coherences in the leads gives a direct interpretation to its role in transport. While the existence of a physical phase governing the dynamics of Josephson junctions has been well established experimentally since the 1960s~\cite{Shapiro1963}, the interpretation of such a phase as a difference of two well defined phases of the wave functions of the condensates composing the junction is inherently incompatible with the $U(1)$ symmetry of the gap~\cite{Bardeen1957,Peierls1991}. The above derivation shows that one can naturally define a phase entering the expression for the current without invoking the concept of a spontaneous symmetry breaking at the level of either the gap, nor the mean field Hamiltonian we started from.

\section{The steady state \label{sec: Steady state}}

In the last two sections we significantly simplified the problem by removing the effect of the Liouville space of the Cooper pairs up to a parametric dependence on a phase $\bm{\varphi}$.  In this section we focus on the steady state where the integro-differential equations for the dot operators acquire a simpler algebraic form.

The kernels in \cref{eq: Current in phase rep,eq: GME in phase final} depend on two time variables, which account for both the effect of the time-periodicity of the bias and the memory time of the quasiparticle baths. It is convenient to write the kernels in a form which showcases these two time-dependencies distinctively.
By employing the fact that in \cref{eq: 12} each $\hat{S}$ operator is associated to an exponential of the form given in \cref{eq: integral-mu}, we can elucidate this representation. After performing the integral in \cref{eq: freeprop}, we expand the constituents of the kernels in their periodicities as
\begin{align}
    \LT'(t) = &\sum_{n,\bm{m}} \ell_{\T{T};n,\bm{m}} e^{i(n\omega_\T{AC}+\bm{m}\cdot\bm{\omega}_\T{DC})t} \hat{S}^{\bm{m}},
    \label{eq: expansions1}\\
    \hat{I}_{\text{L}}^{\prime}\left(t\right) = &\sum_{n,\bm{m}} \hat{j}_{\text{L};n,\bm{m}} e^{i\left(n\omega_{\text{AC}}+\bm{m}\cdot\bm{\omega}_{\text{DC}}\right)t} \hat{S}^{\bm{m}},
    \label{eq: expansions2}\\
   \mathcal{G}'_0(t,s) = & \sum_n g'_{0,n}(t-s) e^{in\omega_\T{AC}s},
    \label{eq: expansions3}
\end{align}
where in \cref{eq: expansions1,eq: expansions2} $\bm{m}$ is restricted to $0,\pm\bm{u}_l$. The kernels to second order only contain contributions of the kind of \cref{eq: expansions1,eq: expansions2,eq: expansions3} and thus one can bring their phase transform into the form
\begin{align}
    \kappa^\circ_{j}\left(\bm{\varphi};t,s\right) = & \sum_{n,\bm{m}} k_{j,n,\bm{m}}\left(t-s\right) e^{i[\bm{m}\cdot(\bm{\omega}_{\T{DC}}s+\bm{\varphi})+n\omega_{\T{AC}}s]},
    \label{eq: Fourierdecomposed K}
\end{align}
which manifestly separates into time-translational invariant operators $k_{\T{j};n,\bm{m}}(t-s)$ and periodic factors~\footnote{This property also holds to all orders as higher order kernels at most contain the convolutions of terms of the form \cref{eq: expansions1,eq: expansions2,eq: expansions3} taken at different times.}. The time-translational invariant parts decay exponentially at long times, ensuring the existence of a steady state for the reduced dot operator.

In \cref{eq: Fourierdecomposed K} $\bm{m}$ has turned into a Fourier index in both the phase and the Josephson frequency, showing explicitly the AC Josephson effect. The bichromatic nature of the junctions dynamics is reflected in the presence of both $\omega_\T{AC}$ and $\bm{\omega}_\T{DC}$. Due to the dependence of the kernel components $k_{\T{T};n,\bm{m}}(t-s)$ on a single time variable with finite memory time, \cref{eq: GME in phase final} can be turned into an algebraic equation through a Laplace transform $\tilde{f}(\lambda)=\int_0^\infty dt e^{-\lambda t} f(t)$. We denote with a tilde the terms in Laplace space from here onward. Applying this transformation to \cref{eq: GME in phase final} yields
\begin{align}
     0  =& (\LQD - \lambda) \hat{\tilde{\rho}}^\circ(\bm{\varphi};\lambda)
    + \hat{\rho}^\circ(\bm{\varphi};0)
    \label{eq: GME-Laplace}   \\
      + &\sum_{n,\bm{m}} e^{i\bm{m}\cdot\bm{\varphi}} \tilde{k}_{\T{T};n,\bm{m}}\left(\lambda\right) \hat{\tilde{\rho}}^\circ(\bm{\varphi};\lambda-in\omega_{\text{AC}}-i\bm{m}\cdot\bm{\omega}_{\text{DC}}).\nonumber
\end{align}
Note that the second term in the right side is the reduced dot operator at initial time. Hence, it is a real time quantity (i.e. not a Laplace space quantity).

Knowledge of $\hat{\tilde{\rho}}^\circ(\bm{\varphi};\lambda)$ allows one to find the steady state solution through the application of the final value theorem generalized to periodic functions~\cite{Grifoni1998}. In particular, the reduced dot operator has the following asymptotic form
\begin{equation}
    \hat{\rho}^{\circ\infty}\left(\bm{\varphi};t\right) = \sum_{n,\bm{m}} \hat{\varrho}^\circ_{n,\bm{m}}(\bm{\varphi}) e^{i(n\omega_{\text{AC}}+\bm{m}\cdot\bm{\omega}_{\text{DC}})t},
    \label{eq: time-periodic-DO}
\end{equation}
where we have defined the operatorial Fourier coefficients of the quasiperiodic reduced dot operator
\begin{align}
    \hat{\varrho}^\circ_{n,\bm{m}}(\bm{\varphi}) = & \lim_{ \lambda \to in\omega_{\text{AC}} + i\bm{m}\cdot\bm{\omega}_{\text{DC}}}
    \nonumber \\  &
    (\lambda - in\omega_{\text{AC}} - i\bm{m}\cdot\bm{\omega}_{\text{DC}}) \hat{\tilde{\rho}}^\circ\left(\bm{\varphi};\lambda\right).
    \label{eq: Fourier-comp-DO}
\end{align}
In general, such a decomposition for bichromatic driving is complicated by the corresponding poles in Laplace space lying dense for incommensurate frequencies. This problem is avoided here as $\bm{m}$ is bounded by the large, but finite, number of Cooper pairs in the leads. Furthermore, the fact that large $\bm{m}$ are associated to highly correlated states without an energy gap protecting them ensures that any additional decoherence mechanism would limit their contribution to the steady state.

In order to obtain the $(n,\bm{m})$ Fourier component of the steady state reduced dot operator as in \cref{eq: time-periodic-DO}, the residua of $\hat{\tilde{\rho}}^\circ(\bm{\varphi};\lambda)$ at its poles have to be extracted by taking a limit in \cref{eq: GME-Laplace}. That is, one multiplies both sides of \cref{eq: GME-Laplace} by $\lambda - in\omega_{\text{AC}} - i\bm{m}\cdot\bm{\omega}_{\text{DC}}$ and then takes the limit $\lambda \to in\omega_{\text{AC}} + i\bm{m}\cdot\bm{\omega}_{\text{DC}}$. The GME then reduces to an algebraic equation for the Fourier components of the steady state reduced dot operator as
\begin{align}
    & 0 = (\LQD -in\omega_{\text{AC}}-i\bm{m}\cdot\bm{\omega}_{\text{DC}}) \hat{\varrho}^\circ_{n,\bm{m}}(\bm{\varphi})+ \sum_{n',\bm{m}'} e^{i\bm{m}'\cdot\bm{\varphi}} \nl
    &  \times \tilde{k}_{\T{T};n',\bm{m}'}(in\omega_{\text{AC}}+i\bm{m}\cdot\bm{\omega}_{\text{DC}}) \hat{\varrho}^\circ_{n-n',\bm{m}-\bm{m}'}(\bm{\varphi}).
    \label{eq: GME-intermediate}
\end{align}
Solving the phase dependent part of \cref{eq: GME-intermediate} results in
\begin{align}
    \hat{\varrho}^\circ_{n,\bm{m}}(\bm{\varphi}) = e^{i\bm{m}\cdot\bm{\varphi}} F(\bm{\varphi}) \hat{\varrho}_{n,\bm{m}}.
    \label{eq: Fourier-comp-DO-phase}
\end{align}
Here, the $F(\bm{\varphi})$ factor cannot be determined directly from \cref{eq: GME-intermediate} as its exact form depends on the initial preparation. By assuming the dot and the superconducting leads to be uncoupled at initial time $t=0$ and with no additional phase dependent contributions to the Hamiltonian, this envelope function is fixed to $F(\bm{\varphi})=1$. The origin of this envelope and its form is discussed in more detail in \cref{App: initial conditions}.

Once the phase dependent part of the reduced dot operator is determined, we are left with solving for the $\hat{\varrho}_{n,\bm{m}}$. Substituting \cref{eq: Fourier-comp-DO-phase} in \cref{eq: GME-intermediate} results in
\begin{widetext}
\begin{align}
    0 & =(\LQD -in\omega_{\text{AC}} - i\bm{m}\cdot\bm{\omega}_{\text{DC}}) \hat{\varrho}_{n,\bm{m}}
    + \sum_{n',\bm{m}'} \tilde{k}_{\T{T};n',\bm{m}'}\left(in\omega_{\text{AC}}+i\bm{m}\cdot\bm{\omega}_{\text{DC}}\right) \hat{\varrho}_{n-n',\bm{m}-\bm{m}'}.
    \label{eq: GME-final}
\end{align}
\end{widetext}
This expression describes exactly the steady state limit of the GME, with the intricate integro-differential form of \cref{eq: 43}, reduced to a set of coupled algebraic equations for the Fourier components $\hat{\varrho}_{n,\bm{m}}$.

The same arguments leading to \cref{eq: GME-final} can be employed to simplify the integral form of the current given in \cref{eq: 49}. In general, the current in the steady state $I_{\T{L},\infty}$ will also be quasiperiodic, with the form
\begin{equation}
    I_{\tL,\infty}(t) = \sum_{n,\bm{m}} j_{\tL,n,\bm{m}} e^{i(n\omega_{\T{AC}}+\bm{m}\cdot\bm{\omega}_{\T{DC}})t}.
    \label{eq: cur-qp}
\end{equation}
The Fourier components of the current are defined, following the same arguments outlined below \cref{eq: Fourier-comp-DO}, as
\begin{equation}
    j_{\T{L};n,\bm{m}} = \lim_{\lambda \to in\omega_{\text{AC}} + i\bm{m}\cdot\bm{\omega}_{\text{DC}}} (\lambda -in\omega_{\text{AC}} -i\bm{m}\cdot\bm{\omega}_{\text{DC}}) \tilde{I}_\T{L}\left(\lambda\right).
    \label{eq: Fourier-comp-current}
\end{equation}
They can then be related, employing the current kernel, to the Fourier components of the steady state reduced dot operator via
\begin{align}
	&j_{\tL,n,\bm{m}} = \sum_{n',\bm{m}'} \int_{\square}d\bm{\varphi} F\left(\bm{\varphi}\right) e^{i\bm{m}\cdot\bm{\varphi}} \label{eq: 22}\\
	& \times \T{Tr}_\T{QD}\bigg\lbrace  \tilde{k}_{\T{I,L},n',m'}(in\omega_{\T{AC}}+i\bm{m}\cdot\bm{\omega}_\T{DC}) \hat{\varrho}_{n-n',\bm{m}-\bm{m}'} \bigg\rbrace.\nonumber
\end{align}

The current will be, in general, quasiperiodic in both the Josephson frequency $\omega_\T{J}=(1,-1)\cdot\bm{\omega}_\T{DC}$ and $\omega_\T{AC}$. This is a result of the superselection rules discussed in \cref{App: Selection rules}. Under certain circumstances the motion will be strictly periodic (i.e. not quasiperiodic). This will be clearly the case whenever $\omega_\T{J} = \omega_\T{AC}$. In the weak coupling limit that we will be considering next, the current is also periodic in $\omega_\text{AC}$, although for a different reason. Namely, that the Fourier components in $\bm{\omega}_\text{DC}$ other than $\bm{m}=0$ will start appearing in the next perturbative order (barring resonances). Regardless, we will focus on the DC component of the current, corresponding to $j_{\tL,0,0}$ and the nonlinear susceptibility $j_{\tL,1,0}$. They are given by
\begin{equation}
    j_{\text{L},n,0}=\text{Tr}_{\text{QD}}\bigg\lbrace\sum_{n',\bm{m}'}\tilde{k}_{\text{I,L};n',\bm{m}'}\left(in\omega_{\T{AC}}\right)\hat{\varrho}_{n-n',-\bm{m}'}\bigg\rbrace.
\end{equation}
In the following we will give these quantities in the weak coupling limit.

\section{Sequential tunneling kernels for a QD Josephson junction}\label{sec: Second order tunneling and current Kernels for a QD Josephson junction}

After discussing the formalism in its general form, in this section we will restrict ourselves to the weak coupling limit. We start by giving an analytical form of the kernels to the lowest non vanishing order in the tunneling amplitudes $|t_l|$. First we provide the normal and anomalous kernels in the time domain, then we obtain their Fourier components needed to compute the DC-current according to \cref{eq: 22}. Finally, analytical and numerical results for the steady state current are discussed in the following sections \cref{sec: DC case,sec: I-V characteristics}.

\subsection{Tunneling kernel in the time domain \label{subsec: Tunneling kernel in the time domain}}

\begin{figure}
	\centering
    \includegraphics{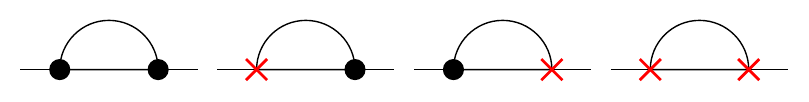}
	\caption{Diagrams contributing to the second order kernel. Besides quasiparticle vertices (dots), also combined vertices (crosses) containing both Cooper pair and quasiparticle creation or annihilation operators are possible.}
	\label{fig: diag: kernel 2 diagrams}
\end{figure}

As noted above, the tunneling kernel and the current kernel differ only in the last operator acting on the argument. This enables us to deduce the expression for the current kernel from the one of the tunneling kernel using diagrammatic rules. We therefore consider \cref{eq: Tunneling Kernel}, which after inserting the explicit form of the tunneling Liouvillian given in \cref{eq: 12} and some minor manipulations reads
\begin{widetext}
\begin{align}
    \Km'&^{(2)}_{\T{T}}(t,s) \oprho(s) =
    \sum_{\substack{l,l',\bm{k},\bm{k}',\sigma,\sigma'\\p,p',\alpha,\alpha'}} \frac{-\alpha\alpha'pp't^{\bar{p}}_{l}t^{\bar{p}'}_{l'}}{(i\hbar)^2} \opd^{\bar{p},\alpha}_{\sigma} \mathcal{G}'_{\T{QD}}(t,s) \opd^{\bar{p}',\alpha'}_{\sigma'} \oprho(s) \T{Tr}_{\T{QP}}\bigg\lbrace \bigg[ u^{\bar{p}}_{l,k} e^{\int_s^t ds' \frac{ip}{\hbar}(E_{l,k}+\mu_l(s'))} \hat{\gamma}^{p,\alpha}_{l,\bm{k},\sigma} + v^{p}_{l,k} e^{\frac{2ip}{\hbar}\int_{t_0}^{t}dt'\mu_l\left(t'\right)} \nonumber\\
    \times & \T{sgn}(\sigma) e^{\int_s^t ds' \frac{i\bar{p}}{\hbar}(E_{l,k}+\mu_l(s'))} \hat{S}_{l}^{p,\alpha} \hat{\gamma}^{\bar{p},\alpha}_{l,\bar{\bm{k}},\bar{\sigma}} \bigg] \bigg[ u^{\bar{p}'}_{l',k'} \hat{\gamma}^{p',\alpha'}_{l',\bm{k}',\sigma'} + \T{sgn}(\sigma') v^{p'}_{l',k'} e^{\frac{2ip'}{\hbar}\int_{t_0}^{s}dt'\mu_{l'}\left(t'\right)} \hat{S}_{l'}^{p',\alpha'} \hat{\gamma}^{\bar{p}',\alpha'}_{l',\bar{\bm{k}}',\bar{\sigma}'} \bigg] \oprho_{\T{QP}} \bigg\rbrace,
    \label{eq:kernel-too-big}
\end{align}
\end{widetext}
where we introduced ${\mathcal{G}'_{\T{QD}}(t,s)={\exp(\LQD(t-s))}}$ and used
\begin{align}
    &\hat{\gamma}^{p,\alpha}_{l,\bm{k},\sigma} \mathcal{G}'_0(t,s) \nonumber \\  = &\mathcal{G}'_0(t,s) \exp(\int_s^t ds' \frac{ip}{\hbar}(E_{l,k}+\mu_l(s'))) \hat{\gamma}^{p,\alpha}_{l,\bm{k},\sigma}, \\
    &\T{Tr}_{\T{QP}}\lbrace \mathcal{G}'_{0}(t,s)\cdots\rbrace = \mathcal{G}'_{\T{QD}}(t,s)\T{Tr}_{\T{QP}}\lbrace\cdots \rbrace,
    \label{eq: G0 comm}
\end{align}
to bring all the lead operators to the right of the reduced density matrix. Performing now the trace over the quasiparticles directly yields
\begin{align}
    \text{Tr}_{\text{QP}} \left\{ \hat{\gamma}_{l,\boldsymbol{k},\sigma}^{p,\alpha} \hat{\gamma}_{l',\boldsymbol{k}',\sigma'}^{p',\alpha'} \hat{\rho}_{\text{QP}}\right\}  & = \delta_{p\bar{p}'} \delta_{ll'} \delta_{\sigma\sigma'} \delta_{\boldsymbol{k},\boldsymbol{k}'} f^{p\alpha'}\left(E_{l,k}\right).
    \label{eq:trace-2nd-order}
\end{align}
From \cref{eq:kernel-too-big}, the terms after performing the trace can have either no $\hat{S}_l$ operator, one $\hat{S}_l$ operator or two $\hat{S}_l$ operators from the same lead and opposite hermiticity. Such two operators would compensate each other, but here they do not necessarily share the same superoperator index. The possible contributions are visualized in a diagrammatic form in \cref{fig: diag: kernel 2 diagrams}.
Identifying the terms of the kernel according to \cref{eq: action of K,eq: simplified KI}, the situation simplifies considerably, as the superoperator index of the Cooper pair operators is effectively rendered irrelevant. Thus, we are able to sum up the outer two and the central two diagrams in \cref{fig: diag: kernel 2 diagrams}, leaving us with only two types of non-vanishing contributions to the tunneling kernel. These two contributions are the normal and anomalous tunneling kernels, with the former representing the usual quasiparticle tunneling, while the latter involve one Cooper pair in one of the vertices. Hence, the anomalous kernel necessarily implies a coherence in the quantum dot space. After converting the sums over momenta $\bm{k}$ to integrals over energies $E$, these kernels are given by
\begin{widetext}
\begin{align}
    \kappa'&^{(2)}_{\T{T}}\left(0;t,s\right)=\sum_{l,\sigma,p,\alpha,\alpha'}\frac{\alpha\alpha'|t_{l}|^2}{(i\hbar)^2}\opd^{\bar{p},\alpha}_{\sigma}\mathcal{G}_\T{QD}(t,s)\opd^{p,\alpha'}_{\sigma}\int_{-\infty}^\infty dE D_l(E) e^{\frac{i}{\hbar}\int_s^t ds' (E+p\mu_l(s'))}f^{\alpha'}(E),\label{eq: time domain kappa normal}
\end{align}
\begin{align}
    \kappa'^{(2)}_{\T{T}}\left(p\bm{u}_l;t,s\right) = & \sum_{\sigma,\alpha,\alpha'} \frac{p\alpha\alpha'|t_{l}|^2}{(i\hbar)^2} \T{sgn}(\sigma) e^{-ip\phi_l} \opd^{\bar{p},\alpha}_{\sigma} \mathcal{G}_\T{QD}(t,s) \opd^{\bar{p},\alpha'}_{\bar{\sigma}} e^{\frac{ip}{\hbar} \int_{0}^{s}dt'2\mu_l\left(t'\right)} \int_{-\infty}^\infty dE A_l(E) e^{\frac{i}{\hbar}\int_s^t ds' (E+p\mu_l(s'))} f^{\alpha'}(E),
    \label{eq: time domain kappa anom}
\end{align}
\end{widetext}
where we introduced ${f^q(E)={1/(1+e^{q\beta E})}}$, ${\phi_l={\arg(\Delta_l (t_l^{*})^2)}+{\frac{2}{\hbar}\int_0^{t_0}ds\mu_l(s)}}$, with $D_{l}(E)$ and $A_{l}(E)$ the normal and anomalous densities of states (DOS), respectively. In the wide band limit, the latter are given by
\begin{align}
	D_{l}(E)= & D^{0} \T{Re}\bigg\lbrace \sqrt{\frac{\left(E-i\gamma\right)^{2}} {\left(E-i\gamma\right)^{2}-|\Delta_{l}|^{2}}} \bigg\rbrace,
	\label{eq:DOS-n}\\
	A_{l}(E)= & D^{0} \T{Re}\bigg\lbrace \sqrt{\frac{|\Delta_{l}|^{2}} {\left(E-i\gamma\right)^{2}-|\Delta_{l}|^{2}}} \bigg\rbrace \T{sgn}(E).
	\label{eq:DOS-sc}
\end{align}
where $D^0$ is the DOS in the normal state at the Fermi level, and $\gamma$ is a Dynes parameter~\cite{Yeyati1997,Dynes1978} which accounts for a finite broadening of the peaks of the superconducting DOS at $E=\pm|\Delta_l|$. Note that the broadening is introduced here as a phenomenological parameter. See, e.g. Ref.~\cite{Yeyati1997}, where this is discussed in the case of a non-interacting dot.

The differences between \cref{eq: time domain kappa normal} and \cref{eq: time domain kappa anom} reflect the physical origin of these terms. The normal kernel corresponds to quasiparticle transport through the junction. In that sense, it is equivalent to a non-superconducting lead with a particular DOS, given by \cref{eq:DOS-n}. The anomalous kernel describes the tunneling of quasiparticles together with Cooper pairs. It is the source of the proximity effect. That is, it results in the appearance of superconducting correlations in the dot  $\propto{ \hat{S}_l^\dagger \hat{d}_\down \hat{d}_\up},{\hat{S}_l \hat{d}^\dagger_\up \hat{d}^\dagger_\down}$.

Upon performing the integrals over time in \cref{eq: time domain kappa normal,eq: time domain kappa anom}, we can use the Jacobi-Anger expansion introduced in \cref{eq: integral-mu}, to find
\begin{widetext}
\begin{align}
    \kappa'_{\T{T}}\left(0;t,s\right) & = \sum_{\substack{l,\sigma,p,\alpha,\alpha',n}} \frac{\alpha\alpha'|t_{l}|^2p^n}{i\hbar} \opd^{\bar{p},\alpha}_{\sigma} Y^{\alpha'}_{l,n}(p\mu_l(0)-i\hbar\LQD,t-s) \opd^{p,\alpha'}_{\sigma} e^{in\omega_{\T{AC}}s},
    \label{eq: second order kernel time normal}
    \\
    \kappa'_{\T{T}}\left(p\bm{u}_l;t,s\right) & = \sum_{\substack{\sigma,\alpha,\alpha',n}} \frac{\alpha\alpha'|t_{l}|^2p^{n+1}}{i\hbar} \T{sgn}(\sigma) e^{-ip\phi'_l} \opd^{\bar{p},\alpha}_{\sigma} Z^{\alpha'}_{l,n}(p\mu_l(0)-i\hbar\LQD,t-s) \opd^{\bar{p},\alpha'}_{\bar{\sigma}} e^{in\omega_{\T{AC}}s} e^{ip\omega_{\T{DC},l}s},
    \label{eq: second order kernel time anom}
\end{align}
with the associated integrals
\begin{align}
    Y^q_{l,n}(\nu,\tau) & = \frac{(-1)^n}{i\hbar} J_n\bigg[ a_l\epsilon_{\T{AC}} \sin\bigg(\frac{\omega_{\T{AC}}}{2}\tau\bigg) \bigg] \int_{-\infty}^\infty dE D_l(E) f^{q}(E) e^{i\frac{E+\nu+(n\hbar\omega_{\T{AC}}/2)}{\hbar}\tau},
    \label{lap1}
    \\
    Z^q_{l,n}(\nu,\tau) & = \frac{(i)^n}{i\hbar} J_n\bigg[ a_l\epsilon_{\T{AC}} \cos\bigg(\frac{\omega_{\T{AC}}}{2}\tau\bigg)\bigg] \int_{-\infty}^\infty dE A_l(E) f^{q}(E) e^{i\frac{E+\nu+(n\hbar\omega_{\T{AC}}/2)}{\hbar}\tau}.
    \label{lap2}
\end{align}
\end{widetext}
In phase space the tunneling kernel takes the form
\begin{align}
    \kappa^{\circ(2)}(\bm{\varphi};t,s) & = \kappa'^{(2)}_{\T{T}}\left(0;t,s\right) + \sum_{p,l} e^{ip\varphi_l} \kappa'^{(2)}_{\T{T}}\left(p\bm{u}_l;t,s\right),
    \label{eq: seq tun kernel in phase}
\end{align}
which is an admixture of the normal and anomalous contributions.

\subsection{Fourier decomposition of the tunneling kernel\label{subsec: Fourier decomposition of the tunneling kernel}}

From the form of the kernels given in \cref{eq: second order kernel time normal,eq: second order kernel time anom} one can directly identify the expansion coefficients $k_{\T{T},n,\bm{m}}(t-s)$ defined in \cref{eq: Fourierdecomposed K}. Instead of giving them explicitly, we first note that all dependence on the time difference $t-s$ is contained in the functions in \cref{lap1,lap2}. Therefore, we conveniently perform the Laplace transformation on them as
\begin{align}
    \int_0^\infty d\tau e^{-\lambda\tau} Y^q_{l,n}(\nu,\tau) = \tilde{Y}^q_{l,n}(\nu+i\hbar\lambda)
    \label{eq: normal integral Laplace},
\end{align}
where we introduced ${\tilde{Y}^{q}_{l,n}(\nu)=\int_0^\infty d\tau Y^{q}_{l,n}(\nu,\tau)}$. Analogously, we call $\widetilde{Z}$ the Laplace transform of the function $Z$. The normal and anomalous integrals as defined in \cref{eq: normal integral Laplace} can be expressed for $V_\T{AC}=0$ in terms of a sum over the Matsubara frequencies and are given in \cref{App: sequential tunneling integrals}. In the resulting final form of the components of the kernel we can again identify the contribution from the normal ($\bm{m}=0$) sequential tunneling kernel
\begin{widetext}
\begin{align}
    \widetilde{k}_{\T{T},n,0}(\lambda) = & \sum_{l,\sigma,p,\alpha,\alpha'} \frac{p^{n}}{i\hbar} |t_{l}|^{2} \alpha \alpha' \opd^{\bar{p},\alpha}_{\sigma} \widetilde{Y}^{\alpha'}_{l,n}(i\hbar(\lambda-\LQD)+p\mu_{l}(0)) \opd^{p,\alpha'}_{\sigma},
    \label{eq: 74}
\end{align}
and the one from the anomalous ($\bm{m} = \pm \bm{u}_l$) sequential tunneling kernel given by
\begin{align}
    \widetilde{k}_{\T{T},n,p\bm{u}_l}(\lambda) = & \sum_{\sigma,\alpha,\alpha'} \frac{p^{n+1}}{i\hbar} |t_{l}|^{2} \T{sgn}(\sigma) \alpha \alpha' e^{-ip\phi_{l}} e^{-ipa_l\frac{\epsilon_{\T{AC}}}{2}} \opd^{\bar{p},\alpha}_{\sigma} \widetilde{Z}^{\alpha'}_{l,n}(i\hbar(\lambda-\LQD)+p\mu_{l}(0)) \opd^{\bar{p},\alpha'}_{\bar{\sigma}}.
    \label{eq: 75}
\end{align}
\end{widetext}

The strength of the contribution from the kernels can be estimated by introducing the rate of tunneling of normal electrons in and out of lead $l$
\begin{equation}
    \Gamma_l=\frac{2\pi}{\hbar}|t_l|^2 D^0.
\end{equation}
Near the coherence peaks the density of states increases by a factor $|\Delta_l|/2\gamma$. The weak coupling limit is justified provided that $\hbar\Gamma_l|\Delta_l|/2\gamma$ is the smallest energy scale in the system (for $l=\T{L},\T{R}$). Comparing \cref{eq: 4,eq: 37} we find that, due to their definitions in \cref{eq: Tunneling Kernel,eq: Currentkernel2}, the expressions for the current kernels follow from the tunneling kernels in \cref{eq: 74,eq: 75} by adding a term $ep$, fixing $\alpha=+$ and $l=\T{L}$.

\section{Transport characteristics of a DC-biased Josephson quantum dot \label{sec: DC case}}

We start by discussing the DC-biased case,  in the weak coupling limit putting emphasis on the role of the degrees of freedom of the Cooper pairs. This case has been investigated in previous works in the limit $|\Delta_l|\rightarrow\infty$~\cite{Governale2008,Hiltscher2012}, in which the quasiparticle degrees of freedom can be disregarded or without the anomalous contributions~\cite{Pfaller2013,Gaass2014}. The general situation of an AC-DC driven junction is discussed in \cref{sec: I-V characteristics}. As we will see, outside of certain parameter regions where the Cooper pairs induce resonant transitions between the $\ket{0}$ and $\ket{2}$ states of the quantum dot~\cite{Kamp2021}, the main effect of the superconducting correlations appearing due to the anomalous kernel (i.e. the proximity effect) is to renormalize the tunneling rates.

\subsection{Populations and coherences}

First, let us consider which terms in the quasiperiodic expansion of the reduced dot operator, \cref{eq: time-periodic-DO}, are relevant. Clearly, in the DC-biased case only the terms with $n=0$ have to be considered.

Let us consider first the DC component $\hat{\varrho}_{0,0} = 1$. Its diagonal elements are the populations of the quantum dot, $P_0,P_\up,P_\down,P_2$ (i.e. the occupation probabilities of each state). Due to conservation of probability, $\text{Tr}_\text{QD} \lbrace \hat{\varrho}_{0,0} \rbrace = 1$ so that the terms in $\hat{\varrho}_{0,0}$ are of order $1$ in an expansion in $\Gamma_l$.

\begin{figure}
    \includegraphics[width=1\columnwidth]{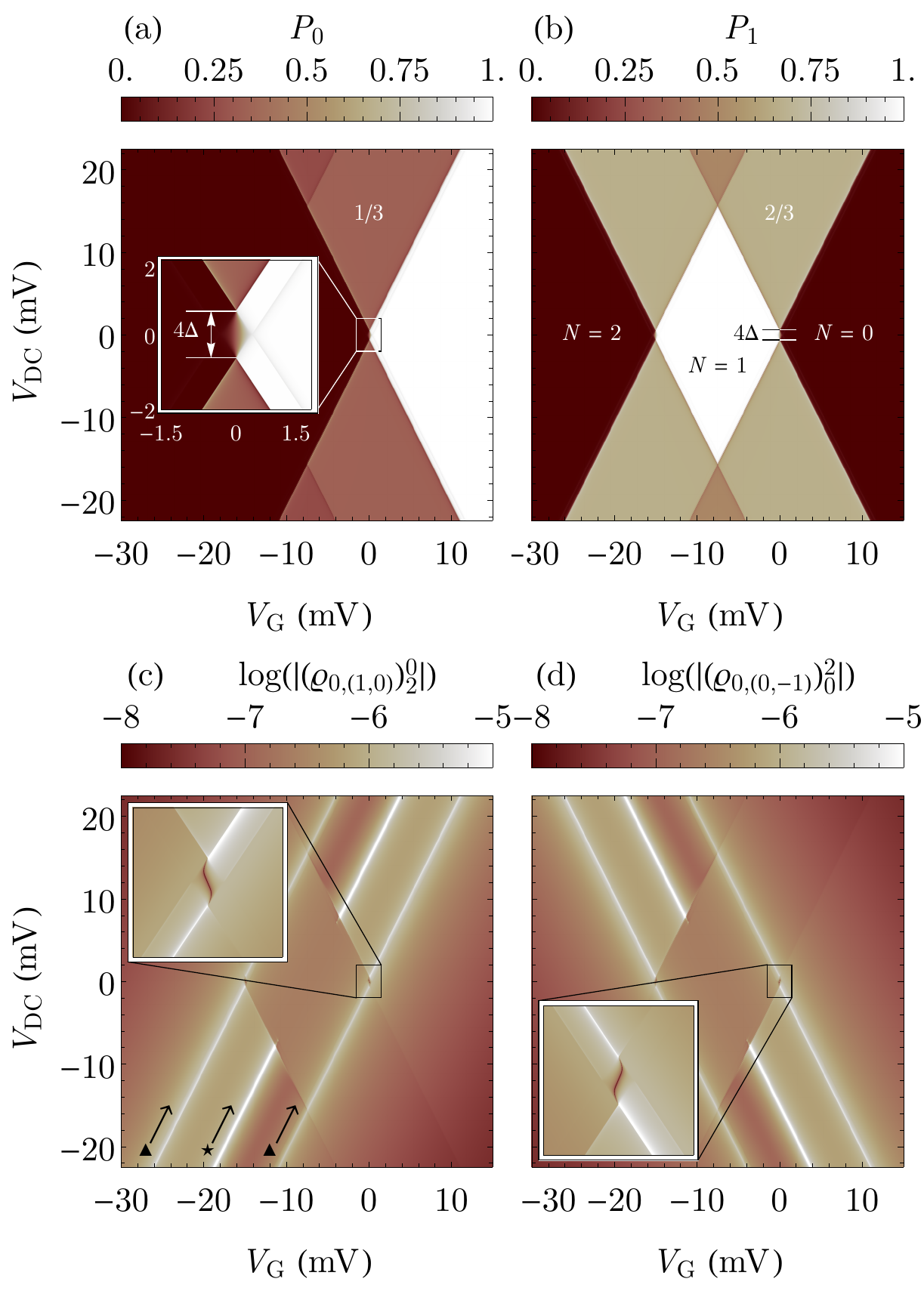}
    \caption{Coulomb blockade and strength of the proximity induced correlations on the dot. Shown are the population $P_0$ of the empty (a) and singly occupied $P_1=P_\uparrow+P_\downarrow$ (b) states of the dot, where $P_\chi:= (\hat{\varrho}_{0,0})^\chi_{\chi}$. The formation of a Coulomb diamond is clearly visible. The absolute value of the induced correlations with $\bm{m}=(1,0)$ (c) and $\bm{m}=(0,-1)$ (d) indicates a strong suppression of proximity induced coherences inside the Coulomb diamond. The proximity effect is largest along the resonance condition established in \cref{eq:cond-res} ($\bigstar$) with appreciable parallel side resonances ($\blacktriangle$) due to resonant pumping. The box in (d) indicates the region around the 1-0 charge degeneracy point on which we focus in later figures. \textit{Parameters}: $U=15\,\T{meV}$, $T=1.2\,\T{K}$, $\Delta_{\tL}=\Delta_{\tR}=0.32\,\T{meV}$, $\gamma=100\,\T{neV}$ and $2\pi D^{0}|t_{\tL}|^{2}=2\pi D^{0}|t_{\tR}|^{2}=93\,\T{neV}$, which results in $2\pi D^{0}|t_{\tL}|^{2}\beta=0.9$.
    \label{fig: coherences}}
\end{figure}

Figs.~\ref{fig: coherences}\,(a) and \ref{fig: coherences}\,(b) display the probability of the dot being empty and with a single charge denoted by $P_1=P_\up + P_\down$, respectively. To not needlessly complicate the figures, we show in the following the case of identical gaps $\Delta_l=\Delta$ and symmetric coupling $a_{\T{L}}=1/2$.
Coulomb blockade features, where the charge is fixed, $P_1\simeq1$, and no current can flow are clearly seen. For a QD-based junction in the weak coupling limit, DC current flows provided that, at the chemical potential of the dot, there are occupied states in one lead (the source) and empty states in the other (the drain). For identical gaps  current can only flow for $|V_{\T{DC}}|\ge 2\Delta$, as marked in both figures. The blow up shows the vicinity of the 1-0 charge degeneracy point ($V_{\T{G}}=0$). The regions of finite DC current correspond to the plateaus with $P_0=1/3$ and $P_1=2/3$ above and below the 1-0 charge degeneracy point, as observed in Figs.~\ref{fig: coherences}\,(a) and \ref{fig: coherences}\,(b). The different populations are due to the two fold spin degeneracy of the single occupied state.

We consider next the terms with $\bm{m} \neq0$. Solving \cref{eq: GME-final} for $\hat{\varrho}_{0,\bm{m}}$, we find the expression
\begin{align}
    \hat{\varrho}_{0,\bm{m}} = &  \frac{1}{i\bm{m}\cdot\bm{\omega}_\T{DC} - \LQD - \widetilde{k}_{\T{T},0,0}(i\bm{m}\cdot\bm{\omega}_\T{DC}) } \nonumber \\
    \times & \sum_{p,l} \widetilde{k}_{\T{T},0,\bar{p}\bm{u}_l} (i\bm{m}\cdot\bm{\omega}_\T{DC}) \hat{\varrho}_{0,\bm{m} +p\bm{u}_l},
    \label{eq: 104}
\end{align}
where we used that the kernels vary $\bm{m}$ by at most $\pm \bm{u}_l$ in the weak coupling limit. \cref{eq: 104} represents the steady state occupancy of a particular Fourier mode of the dot operator in terms of pumping from other modes. The resulting dependence of a given term $\hat{\varrho}_{0,\bm{m}}$ on $\Gamma_l$ is remarkably complicated. Hence, in the spirit of the secular approximation~\cite{Koller2010}, we consider first the parameter regions for which
\begin{equation}
    |\bm{m}\cdot\bm{\omega}_\T{DC}+\omega_\T{QD}| \gtrsim \Gamma_l,\,\,\,\,\,l=\T{L},\T{R}.
    \label{eq:cond-res}
\end{equation}
where $\omega_\T{QD}$ is a generic Rabi frequency associated with the action of $\LQD$ in the denominator of \cref{eq: 104}. I.e. for a term $\ketbra{\chi}{\chi'}$, it is given by
\begin{equation}
    \mathcal{L}_{\T{QD}}\ketbra{\chi}{\chi'}=-i\omega_\T{QD}\ketbra{\chi}{\chi'},
\end{equation}
which, in the absence of a Zeeman splitting, results in $\hbar\omega_\T{QD}=0$ for the populations $P_{\chi}$ and ${\hbar\omega_\T{QD}=}{\pm( 2eV_\T{G}+U)}$ for the coherences $\ketbra{2}{0}$ and $\ketbra{0}{2}$, respectively.

For any two modes $\hat{\varrho}_{0,\bm{m}'}$ and $\hat{\varrho}_{0,\bm{m}}$ with $\sum_l|m'_l|=\sum_l|m_l|+1$, $\hat{\varrho}_{0,\bm{m}'}$ will be smaller than $\hat{\varrho}_{0,\bm{m}}$ by a factor $\sim {\Gamma_l / |\bm{m}'\cdot\bm{\omega}_\T{DC} + \omega_\T{QD}|}$. This imposes a hierarchy of harmonic modes, in which $\hat{\varrho}_{0,\bm{m}}$ become less important as $\bm{m}$ moves away from $\bm{m} = 0$. In particular, in order to calculate the reduced dot operator and the current up to order $\Gamma_l$, we need only the contributions $\bm{m} = 0,\pm \bm{u}_l$, provided that \cref{eq:cond-res} holds.

For $\bm{m}\neq 0$ the region where \cref{eq:cond-res} is not satisfied defines a resonance, where $\hat{\varrho}_{0,\bm{m}}$ is of one order less in $\Gamma_l$ than outside of it. Therefore, the contributions $\bm{m}=\pm \bm{u}_l$ are of the same order in $\Gamma_l$ as $\hat{\varrho}_{0,0}$ along these resonances. 

Previous works on this model identified the coherences in $\bm{m}=\pm\bm{u}_l$ as a proximity induced pair amplitude on the dot \cite{Pala2007}. To understand how this proximity effect impacts transport in the weak coupling limit, we show the absolute value of the induced coherences next to the stability diagram of the junction in {Figs.~\ref{fig: coherences}\,(c)} and \ref{fig: coherences}\,(d), where we denote the entries of the dot operators as
\begin{align}
    (\hat{\varrho}_{n,\bm{m}})^{\chi}_{\chi'} = \bra{\chi}\hat{\varrho}_{n,\bm{m}}  \ket{\chi'}.
\end{align}
We find that the pair amplitude is most prominent along the resonance conditions established in \cref{eq:cond-res} and along parallel features that we identify as the consequence of resonant pumping. Nonetheless, we find them to be significantly diminished even on resonance.

Hermiticity of the density matrix imposes that
\begin{equation}
    |(\hat{\varrho}_{0,(1,0)})^0_2| = |(\hat{\varrho}_{0,(-1,0)})^2_0|,
\end{equation}
but note that in \cref{fig: coherences} (c,d) we have in general
\begin{equation}
    |(\hat{\varrho}_{0,(-1,0)})^2_0|\neq|(\hat{\varrho}_{0,(0,-1)})^2_0|.
\end{equation}
That is: At finite bias, coherences with different $\bm{m}$ can differ even if they share the same states of the quantum dot. This highlights the need to keep track of the state of the condensates when discussing the proximity effect on the dot for non-equilibrium situations.

To explain the absence of an appreciable effect of the coherences on the populations even on resonance and their near perfect suppression inside the Coulomb diamond, we next investigate their size on the resonance. More specifically, we focus on the upper right quadrant of the stability diagram, with the rest following by the particle-hole and left-right symmetries of the problem. We start by considering the coherence $(\hat{\varrho}_{0,(1,0)})^0_2$. The exemplary point $eV_{\T{G}}=0$, $eV_{\T{DC}}=U$ fulfills the resonance condition as $|(1,0)\cdot\bm{\omega}_\T{DC} -U/\hbar|=0$ there. It lies outside of the Coulomb diamond on the plateau region above the 1-0 charge degeneracy point. To leading order in $\Gamma$, we may write
\begin{align}
    (\hat{\varrho}_{0,(1,0)})^0_2=&-\frac{1}{(K_n)^{0,0}_{2,2}}\sum_{\chi}(K_a)^{0,\chi}_{2,\chi}P_{\chi},
    \label{eq: explicit coherence}
\end{align}
with ${K_n=\widetilde{k}_{\T{T},0,0}(iU/\hbar)}$ and ${K_a=\widetilde{k}_{\T{T},0,\bm{u}_{\T{L}}}(iU/\hbar)}$ and where we denoted the matrix elements of the kernel by
\begin{align}
    (\widetilde{k}_{\T{T},l,n,\bm{m}}(\lambda))^{\chi,\chi''}_{\chi',\chi'''} = \bra{\chi} [\widetilde{k}_{\T{T},l,n,\bm{m}}(\lambda) \ketbra{\chi''}{\chi'''}] \ket{\chi'}.
\end{align}
The relevant populations $P_0$ and $P_1$ on the r.h.s. of \cref{eq: explicit coherence} were already discussed for \cref{fig: coherences}. To deduce the absolute value of $(\hat{\varrho}_{0,(1,0)})^0_2$ from these, we still need the kernel elements $(K_n)^{0,0}_{2,2}$ and $(K_a)^{0,\sigma}_{2,\sigma}$, $(K_a)^{0,0}_{2,0}$ with the latter rates representing pumping from the populations. By construction the kernels in \cref{eq: explicit coherence} do not change as we move along the resonance. It is thus sufficient to calculate all the rates for one point on the resonance. The latter observation already indicates that the strong suppression of the coherences \textit{inside} the Coulomb diamond must be purely due to the change in the static populations. In fact, comparing the pumping rates for the even particle number sector of the density matrix
\begin{align}
    (K_a)^{0,0}_{2,0} =({\Gamma_{\T{L}}/D^0})[&f^+(-U/2)A_{\T{L}}(-U/2)\nonumber\\& + i \widetilde{S}^{(2)}_{\T{L}}(-U/2)/\pi],
    \label{eq: even}
\end{align}
($\widetilde{S}^{(2)}_l$ is the real part of the anomalous sequential tunneling integral introduced in \cref{App: sequential tunneling integrals})  and for the odd particle number sector
\begin{align}
    (K_a)^{0,\sigma}_{2,\sigma} = {(2\Gamma_{\T{L}}/D^0)} f^{+}(U/2)A_{\T{L}}(U/2),
    \label{eq: pumping odd}
\end{align}
we find the latter to be exponentially suppressed in $\beta U$ if $U\gg\Delta_l,\beta^{-1}$. This reflects the antagonist role interaction plays to proximitized pairing on the dot. The need for a finite population of the even charge sector of the dot has already been noted in the literature~\cite{Pala2007}.

To explain why the coherences remain suppressed \textit{even on resonance and outside} of the Coulomb diamond, we first note that $|(K_n)^{0,0}_{2,2}|\approx 3\Gamma$. Clearly, the suppression must therefore be due to $(K_a)^{0,0}_{2,0}\ll\Gamma_{\T{L}}$. The latter is a consequence of the anomalous density of states vanishing above the gap. We conclude that the proximity induced coherences on the dot are suppressed as $\Delta/U$ throughout the entire stability diagram for $U\gg\Delta$.

The latter condition is natural e.g. for molecular junctions, where the interaction is typically of the order of multiple $\T{eV}$~\cite{siegert_effects_2015}. For leads manufactured from conventional BCS superconductors as we consider here, the gap is typically of the order of a few $\T{meV}$ at most~\cite{Bardeen1957}. We expect this suppression to also be present, albeit weaker, for extended quantum dots, e.g. carbon nanotubes, where the interaction is still typically an order of magnitude larger than the gap~\cite{Gaass2014}. For weakly interacting junctions, where effects due to the proximity induced coherences on the dot may be sizeable, we predict the dot pair amplitude to be largest on the resonances inside the plateau regions. The latter observation is counter intuitive as quasiparticle transport is usually considered detrimental to coherent phenomena in superconducting junctions~\cite{Frombach2020}. For zero bias, and to sequential tunneling order, we identify the regions around the charge degeneracy points as the most strongly proximitized. However, we note that a treatment at least to cotunneling order is needed to rule out strong proximity effect inside the Coulomb diamond.

\subsection{Current-voltage characteristics}

We turn now to the calculation of the DC-current. If the dot is described by the SIAM of \cref{eq: 1}, the current is given, in general, by the expression
\begin{widetext}
\begin{align}
	I_{\tL}=-e \bigg\lbrace \sum_{\sigma} \bigg[ \Gamma_{\tL}^{\sigma,0} P_{0} +( \Gamma_{\tL}^{2,\sigma} - \Gamma_{\tL}^{0,\sigma}) P_{\sigma} - \Gamma_{\tL}^{\sigma,2} P_{2} \bigg] + 2 \Gamma_{\tL}^{2,0} P_{0} - 2\Gamma_{\tL}^{0,2} P_{2} \bigg\rbrace,
    \label{eq: 29}
\end{align}
\end{widetext}
where $\Gamma_l^{\chi',\chi}$ are the tunneling rates from population $\chi$ to population $\chi'$, mediated by lead $l$. \Cref{eq: 29} has a natural interpretation as the difference of rates by which charges enter and exit the left lead. In the normal case, these rates can be related to elements of the tunneling kernel by ${\Gamma_l^{\chi',\chi} = (\widetilde{k}_{\T{T},l,0,0}(0^+))^{\chi',\chi}_{\chi',\chi}}$, where the $l$ indicates terms stemming from the tunneling to lead $l$. In the case of normal leads the terms $\Gamma_{\T{L}}^{0,2}$ and $\Gamma_{\T{L}}^{2,0}$ do not contribute in lowest order in $\Gamma$, since only one charge at a time can be transferred. In the superconducting case the latter terms emerge even in lowest order, if we account for the effect coherences have on the rates between the populations. Furthermore, a renormalization of the sequential tunneling rates occurs.

Since we only retain the contributions $\bm{m} = 0,\pm \bm{u}_l$, we can express the components $\hat{\varrho}_{0,\pm\bm{u}_l} $ in terms of the populations in  $\hat{\varrho}_{0,0}$ via \cref{eq: 104}. With this, the current becomes a function of the populations as in \cref{eq: 29}, with $\Gamma_l^{\chi',\chi}$ the "secular" rates
\begin{widetext}
\begin{align}
	\Gamma_{l}^{\chi',\chi}= \Gamma_l^{(2),\chi',\chi} - 2\T{Re}\bigg[ \frac{i\hbar}{2eV_\T{G}+U-\hbar\bm{u}_l \cdot\bm{\omega}_{\T{DC}}+i\hbar(\widetilde{k}_{\T{T},0,0}(-i\bm{u}_l \cdot\bm{\omega}_{\T{DC}}))^{2,2}_{0,0}} (\widetilde{k}_{\T{T},0,\bm{u}_l}(0))^{\chi',2}_{\chi',0} (\widetilde{k}_{\T{T},0,-\bm{u}_l}(-i\bm{u}_l \cdot\bm{\omega}_{\T{DC}}))^{2,\chi}_{0,\chi} \bigg].
    \label{eq: 26}
\end{align}
\end{widetext}
The first term is of second order in the tunneling amplitude. The second term in \cref{eq: 26} is $\propto |t_l|^4$ provided that \cref{eq:cond-res} is satisfied, but is relevant inside the resonances. Hence, we see that the main effect of the coherences in the DC current is to renormalize the normal tunneling rates and introduce a new Cooper-pair enabled pair tunneling process. Moreover, the populations $P_\chi$ can be obtained explicitly in terms of the secular rates of \cref{eq: 26}. The corresponding expressions for the populations are given in \cref{App: Analytic solution for the populations}.

\begin{figure}
    \includegraphics[width=\columnwidth]{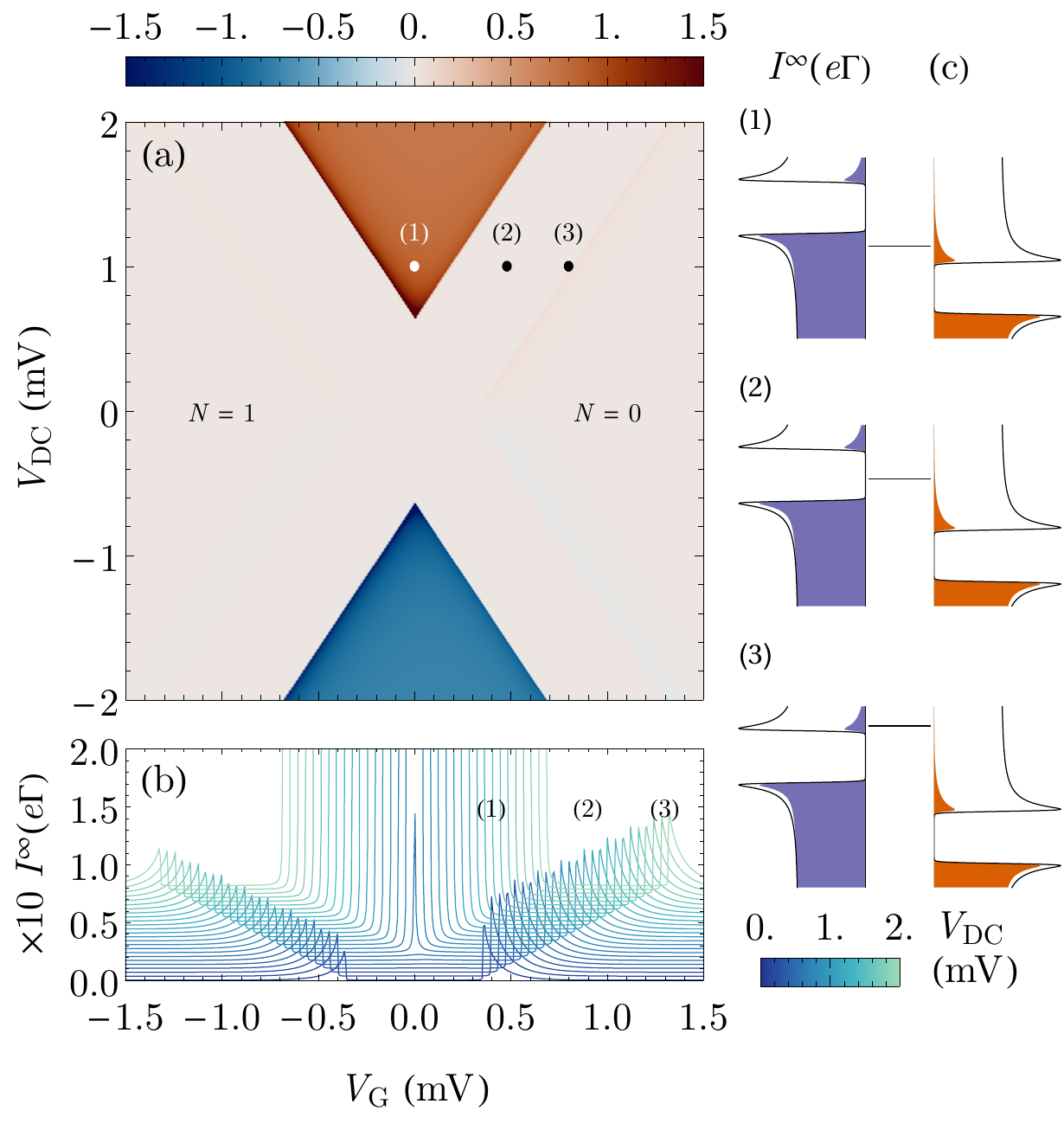}
    \caption{Transport in the DC case. (a) Current near the the 1-0 charge degeneracy point for the DC case ($V_\T{AC}=0$), showing areas of Coulomb blockade (clear colors) and of current flow in both directions. (b) Cascade plot for the same parameters and different values of the DC voltage amplitude, from $V_\T{DC}=0.08$ (blue) to $2\,\T{meV}$ (green), showing clearly current peaks due to thermally excited quasiparticles. (c) Diagrams summarizing the three common transport situations in this setup: (1) Normal transport, where $eV_\T{DC}$ is larger than $2\Delta$ and the chemical potential of the dot is aligned so that a current can flow. (2) Transport blockade inside the gap, where the state of the dot lies inside the superconducting gap. (3) Transport is mediated by thermally excited quasiparticles above the gap. Parameters are the same as in \cref{fig: coherences}.
    \label{fig:DC-case}}
\end{figure}

In \cref{fig:DC-case}~(a), we have represented the current as a function of the gate voltage $V_\T{G}$ and the DC component of the bias voltage $V_\T{DC}$ (the stability diagram of the dot) near $eV_\T{G}=0$, where the empty and the single occupied states of the dot are degenerate (i.e. the 1-0 charge degeneracy point).

As seen in \cref{fig: coherences}, in absence of thermal fluctuations current flow is possible for $|V_\T{DC}|\ge2\Delta$. This situation is present for point~(1) in \cref{fig:DC-case}~(a).
A diagram illustrating the level alignment at this point is represented in \cref{fig:DC-case}~(c.1). At non-zero temperature, transport can also occur by transferring thermally excited quasiparticles that occupy states above the gap~\cite{Pfaller2013,Ratz2014}. At the low temperatures considered here, this subgap thermal current is small compared to the contribution from quasiparticles above the gap, and is hard to observe in the stability diagram of \cref{fig:DC-case}~(a). For that reason, we have also represented in \cref{fig:DC-case}~(b) a cascade plot of the current for different values of $V_\T{DC}=0.08,\ldots,2\,\T{mV}$ capped at a relatively small value of the current, showing clearly the appearance of a set of peaks (marked as (3) both here and in \cref{fig:DC-case}~(a)). An asymmetry due to the different degeneracies of the empty and singly occupied states of the quantum dot is appreciable. The energy level diagram corresponding to this process is represented in \cref{fig:DC-case}~(c.3). Within these two points, there is a region where current does not flow, of size $2\Delta$. This situation is represented schematically in \cref{fig:DC-case}~(c.2). 

\section{Transport characteristics of an AC-DC-biased Josephson quantum dot\label{sec: I-V characteristics}}

We turn now to the general situation of $V_\T{AC}\neq0$, where we are investigating the DC current in presence of simultaneous AC and DC bias as well as the first harmonic of the current. The latter is related to the dynamic nonlinear susceptibility.

\subsection{The average current}

We want to again restrict the range of $n$ we have to consider as we did in \cref{sec: DC case} for $\bm{m}$. While the sequential tunneling kernels could only change $\bm{m}$ by at most $\pm \bm{u}_l$, the kernels connect all $\hat{\varrho}_{n,\bm{m}}$ in \cref{eq: GME-final}. Let us consider first the extension of \cref{eq:cond-res} to non-zero AC voltages, namely
\begin{equation}
    |\bm{m}\cdot\bm{\omega}_\T{DC}+n\omega_\T{AC}+\omega_\T{QD}| \gtrsim \Gamma_l,\,\,\,\,\,l=\T{L},\T{R},
    \label{eq:cond-res-ac}
\end{equation}
which amounts to staying sufficiently far from any photon assisted resonance between the unoccupied and the doubly occupied states (mediated by Cooper pairs). Studying the DC current, we can immediately make the observation that, given \cref{eq:cond-res-ac}, any contribution coming from $\hat{\varrho}_{n,\bm{m}}$ with both $n,\bm{m}\neq 0$ will be at least $\propto|t_l|^4\sim\Gamma^2_l$ and can therefore be neglected in the sequential tunneling approximation. Combined with the discussion in \cref{sec: DC case}, this enables us to focus on $\bm{m}=0,n\in\mathbb{Z}$. Due to the selection rules discussed in \cref{App: Selection rules}, these indices allow only for populations, such that $\omega_\T{QD}=0$. Following the treatment in the DC case, we find
\begin{align}
    \hat{\varrho}_{n,0}= & \frac{1}{in\omega_\T{AC}-\widetilde{k}_{\T{T},0,0}(in\omega_\T{AC})} \sum_{n'}\widetilde{k}_{\T{T},n',0}(in\omega_\T{AC})\hat{\varrho}_{n-n',0}.
    \label{eq: hierarchy AC}
\end{align}
The tunneling kernel can connect Fourier modes with arbitrary difference in $n$. Thus, the hierarchy given by \cref{eq: hierarchy AC} in the region of validity of \cref{eq:cond-res-ac} is flat with all terms $n\neq0$ of order $\Gamma/n\omega_{\T{AC}}$ and $\hat{\varrho}_{0,0}$ again of order 1. Provided \cref{eq:cond-res-ac} holds, this enables us to neglect all $n\neq0$ contributions when discussing the DC current. This is the \emph{high frequency approximation} on which we will focus in the following. Note that this approximation requires only that the photon energy is large compared to $\Gamma_l$, not to all other energy scales of the problem.

For the remaining non-vanishing contribution, $n=0$, we can expand
\begin{align}
	J_{0} \big[ a_l \epsilon_\text{AC} \sin\big(\omega_{\T{AC}}t/2\big) \big] = \sum_{k=-\infty}^{\infty} J_{k}^{2} \big( a_l \epsilon_\text{AC} / 2 \big) e^{ik\omega_{\T{AC}}t}.
	\label{eq: 92}
\end{align}
Using this, we find that the required integrals in \cref{eq: normal integral Laplace} are given by
\begin{align}
	\widetilde{Y}^{q}_{l,0}(\nu) \rightarrow \sum_{k=-\infty}^{\infty} J_{k}^{2} \big( a_l\epsilon_\text{AC} / 2 \big) \widetilde{Y}^{q}_{l,\T{DC}}(\nu+k\hbar\omega_{\T{AC}}).
	\label{eq: 93}
\end{align}
Here, each term corresponds to a photon assisted rate weighted by a factor $J_{k}^{2}\big(\epsilon_\text{AC}/2\big)$ representing the process associated with the absorption or emission of $k$ photons.

\cref{eq: 93} hints at a connection of the resulting expressions for high frequency to the well known Tien-Gordon theory of photon assisted sidebands in tunneling~\cite{Tien1963,Kouwenhoven1994,Kouwenhoven1994b,Whan1996}. Note however, that the original derivation by Tien and Gordon considered a simple tunnel junction, where there is no additional energy scale of the central system to compare against~\cite{Tien1963}. In previous works on photon-assisted tunneling in quantum dots connected to superconducting leads~\cite{Kouwenhoven1994,Kouwenhoven1994b,Whan1996} the tunneling rates themselves were phenomenalogically replaced with a Tien-Gordon like expression. The imaginary part of \cref{eq: 93}, which governs the sequential tunneling rate, recovers the expressions of these earlier works from a microscopic derivation. Meanwhile, our result extends the expressions by encapsulating e.g. the Lamb shift caused by the real part of \cref{eq: 93}, which was omitted in these earlier works. Note that the change in the rates \cref{eq: 93} not only affects the current, but also the steady state solution, via \cref{eq: 74,eq: 75} \cite{Whan1996}. As such, a naive version of the Tien-Gordon model for the current yields qualitatively different results from the correct expression. See \cref{App: Comparison with Tien-Gordon} for details.

\begin{figure}
    \includegraphics[width=\columnwidth]{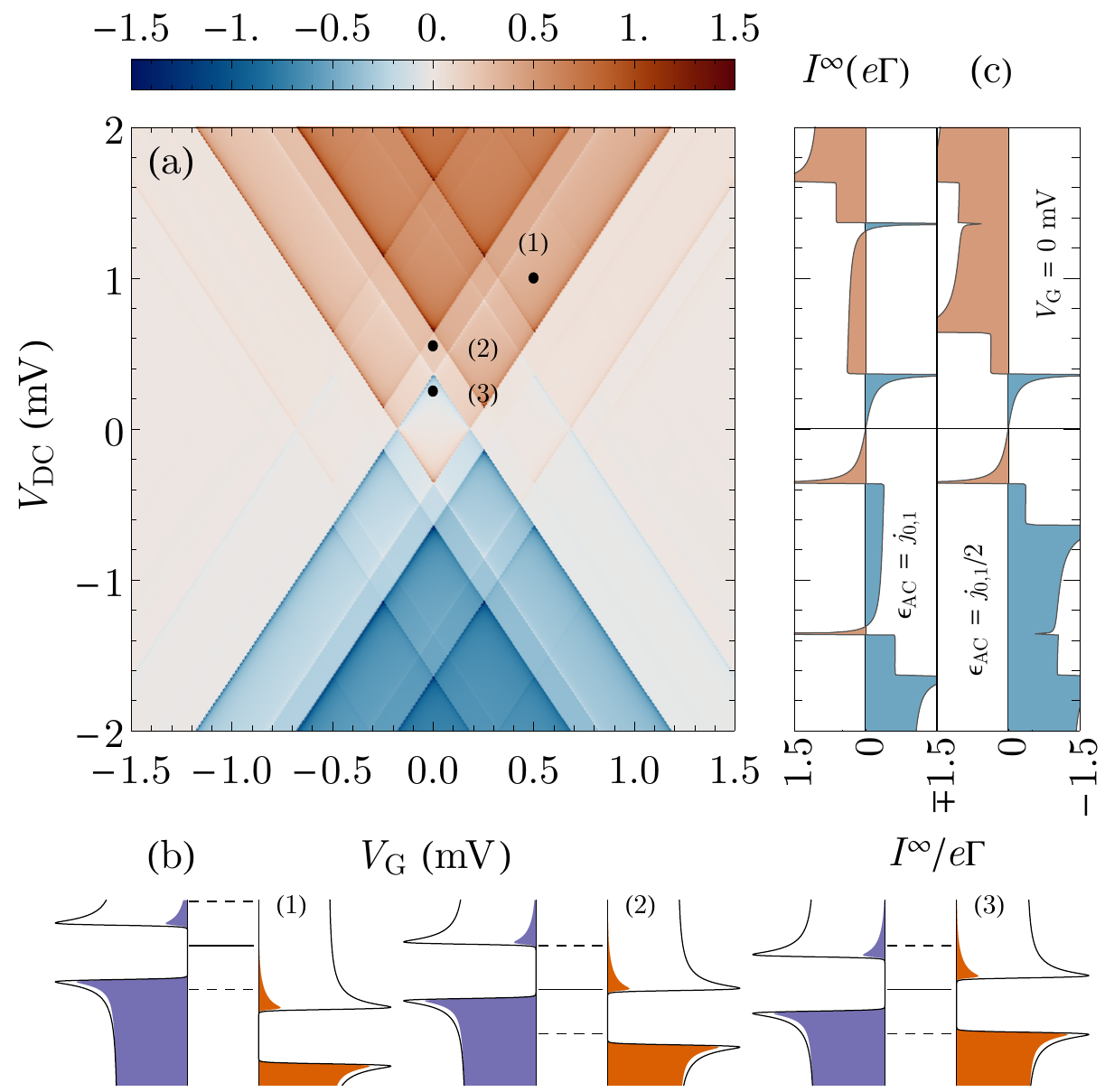}
    \caption{DC Transport in the AC case. (a) Current near the the 1-0 charge degeneracy point for the AC case with $\epsilon_\text{AC}=2b_{0,1}$, where $b_{m,n}$ is the $n$th zero of the $m$th Bessel function of the first kind. New areas of current flow appear here as compared to \protect\cref{fig:DC-case} due to the possibility of photon assisted transport. (b) Diagrams exemplifying several transport situations in this setup: (1) Sideband transport, where current flows even if the chemical potential of the dot lies within the gap of any of the leads.  (2) Subgap transport, where current flows despite $V_\T{DC}<2\Delta$. (3) Current inversion, where the backwards rates are larger than the forwards rates, and the net current flows from the lead at lower to the one with the higher (average) chemical potential (i.e. from drain to source). (c) Current as a function of the DC bias $V_\T{DC}$ for $V_\T{G}=0$, $\epsilon_\text{AC}=2b_{0,1}$ (top) and $\epsilon_\text{AC}=b_{0,1}$ (bottom), corresponding to the cases where two and one photon assisted processes show current inversion, respectively. The color filling indicates the sign of the current (blue for negative, red for positive sign).
    \textit{Parameters}: Same as in \protect\cref{fig:DC-case}. The AC energy $\hbar\omega_{\T{AC}}$ is set to $0.5\,\T{m}e\T{V}$ and the considered range of $n=-30,...,30$ with $m=0,\pm\bm{u}_l$ is in accordance with the discussion in \protect\cref{sec: DC case}.
    \label{fig:AC-case}}
\end{figure}

\begin{figure*}
    \includegraphics[width=1.74\columnwidth]{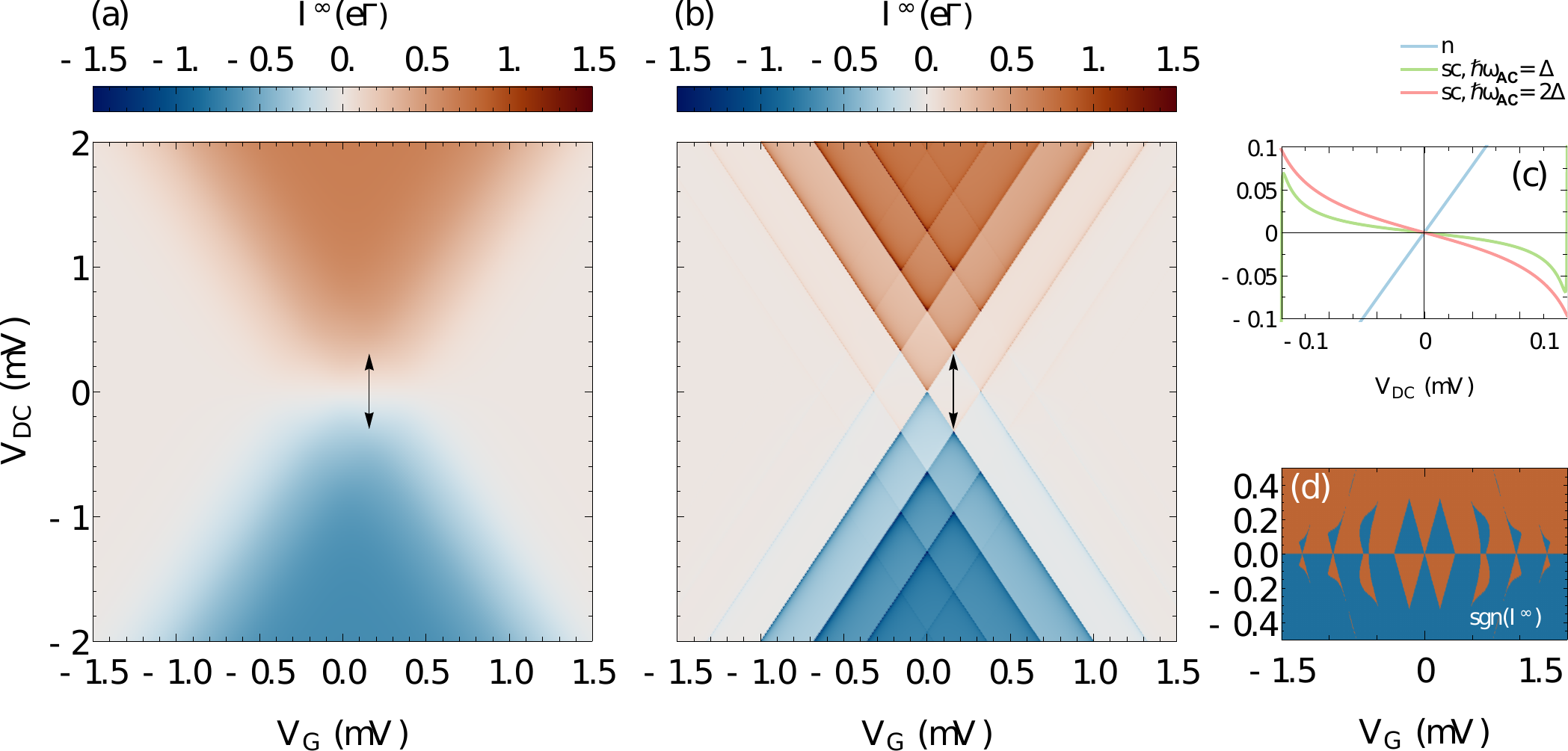}
    \caption{Comparison of currents for AC-DC-driven normal and superconducting junctions. For the latter the current is sharper and can become negative at positive applied DC-bias. (a) Current near the the 1-0 charge degeneracy point for the case of normal conducting leads with $\epsilon_\text{AC}=2b_{0,1}$ and $\hbar\omega_\T{AC}=0.32\,\T{meV}$. (b) Same, for the superconducting case, with $\hbar\omega_\T{AC}=\Delta$. (c) Cut for $V_\T{G}=0.15\,\T{mV}$, along the arrows marked in (a) and~(b), showing current inversion. We have further represented the case $\hbar\omega_{\T{AC}}=2\Delta$ for the same cut, showing an even larger signature of inversion. (d) Sign of the current for the same parameters as Fig.~(b).
    \textit{Parameters}: Same as in \protect\cref{fig:AC-case}.}
    \label{fig:2Col}
\end{figure*}

In general, the high frequency approximation breaks down for a bichromatic drive if the \textit{difference} of the frequencies becomes comparable to the time scale of the dynamics of the system~\cite{Ho1983,GomezLeon2020}. With $\omega_{\T{DC}}$ depending on the position in the stability diagram, this condition of non-degeneracy translates into \cref{eq:cond-res-ac}. The respective integrals as defined according to \cref{eq: normal integral Laplace} are solved analytically in \cref{App: sequential tunneling integrals}.

In \cref{fig:AC-case}~(a) we have represented the current as a function of the gate voltage $V_\T{G}$ and the DC component of the bias voltage $V_\T{DC}$ near the 1-0 degeneracy point for a strength of the AC bias of $\epsilon_\T{AC}=2b_{0,1}$, where $b_{m,n}$ is the $n$th zero of the $m$th Bessel function of the first kind. The resulting stability diagram exhibits features reminiscent of the DC case of \cref{fig:DC-case} (parameters are otherwise the same) but replicated and displaced by integers of $\hbar\omega_\T{AC}/e$. These replicas arise due to the photon assisted rates in \cref{eq: 93}. Their non-trivial nature is expected, since the rates enter in a decisively non-linear manner in the GME. See, for instance, the analytic result of \cref{App: Analytic solution for the populations}, which corresponds to the much simpler DC case. 

We have indicated in \cref{fig:AC-case}~(a) exemplary points which reflect the effect of the AC bias. Point~(1) corresponds to a situation in which the DC voltage is larger than $2\Delta$ but the chemical potential of the dot is not aligned as to lead to current flow. Nonetheless, photon assisted transitions still allow for a non-zero current. This is represented schematically in \cref{fig:AC-case}~(b.1). Here, dashed lines separated by $\pm\hbar\omega_\T{AC}$ from the chemical potential of the dot, represented by a full line, illustrate a tunneling process accompanied by the absorption/emission of a photon. Note that this representation is only for illustrative purposes, since the AC voltage is applied to the leads and not to the dot. On the other hand, point~(2) corresponds to subgap transport, in which the DC bias voltage is not large enough to overcome the superconducting gap, but a current still flows due to AC-induced sidebands, as can be seen in the diagram of \cref{fig:AC-case}~(b.2). Such photon assisted sequential tunneling DC currents beneath the gap are $\sim \epsilon_{\T{AC}}^2$. For weak driving, such that $\epsilon_{\T{AC}}^2<\Gamma\hbar/\Delta$, higher order processes in the tunnel coupling, which are associated with Cooper pair transport, are expected to remain the dominant contribution to the DC current below the gap.

The superconducting case differs from the normal conducting case by also containing regions of current inversion, where the current flows in the opposite direction of the DC bias. This occurs, for instance, at the point labeled (3) in \cref{fig:AC-case}~(a). This  effect is known to appear as a result of a non-flat DOS~\cite{Kouwenhoven1994,Kouwenhoven1994b,Kostur2008}. For point~(3), transport without photon absorption nor emission is suppressed since the chemical potential of the dot lies in the gap of both leads, while photon assisted processes are allowed. In this particular configuration, the backward photon assisted rate transfers charges from the peak in the DOS of the right lead, while the forward rate transfers charges from the flat region of the DOS of the left lead (as represented schematically in \cref{fig:AC-case}~(b.3)). As such, the backward rate is larger than the forward one (by a factor $\sim\Delta/2\gamma$, at most), resulting in a net current flow against the applied DC-bias. \cref{fig:AC-case}~(c) showcases this at two values of the AC bias amplitude, $\epsilon_\text{AC}=2b_{0,1}$ (top) and $\epsilon_\text{AC}=b_{0,1}$ (bottom). For the latter, current inversion occurs only near $eV_\T{DC}=\pm\Delta$, following the same process as described above. For the former, a further zone of current inversion occurs near $eV_\T{DC}=\pm(2\hbar\omega_{\T{AC}}+\Delta)$ well inside the regions of current flow in the DC case.

The asymmetry with respect to the gate in \cref{fig:AC-case}~(a) arises from the spin-degeneracy of the single-occupied states similar to the one in \cref{fig:DC-case}. Note that said asymmetry cannot be understood within a Tien-Gordon-like ansatz for the current~\cite{Tien1963}, but is a consequence of the expression for the rates in \cref{eq: 93}. Overall particle hole symmetry is conserved as the stability diagram at the 1-2 degeneracy point (i.e. the point where $\ket{\sigma}$ and $\ket{2}$ are degenerate at $eV_\T{G} = U$) is a mirror copy of this one.

The complex nature of the stability diagram of \cref{fig:AC-case} is a result of having two incommensurate energy scales. The stability diagram under an AC voltage for the normal case is comparatively simple, as can be seen in \cref{fig:2Col}~(a) for $\hbar\omega_\T{AC}=0.32\,\text{meV}$, since only the AC frequency comes into play. Similarly, for superconducting leads with $\hbar\omega_\T{AC}=\Delta=0.32\,\text{meV}$, as represented in \cref{fig:2Col}~(b), the current also exhibits a simpler structure as compared to \cref{fig:AC-case}, since the two energy scales are commensurate. In this situation, the AC bias pumps quasiparticles from below the gap so that a non-zero current can flow even for for arbitrarily small $V_\T{AC},V_\T{DC}$. For the normal case, the DOS is flat and there is no possibility for current inversion, as the backwards and forward photon assisted rates will be equal. As a result, the current always flows in the direction of the DC bias. \Cref{fig:2Col}~(c) shows a small vertical cut of the stability diagram for both the normal (blue) and the superconducting case (green, the lighter color corresponding to $\hbar\omega_\T{AC}=\Delta$ and the darker color corresponding to  $\hbar\omega_\T{AC}=2\Delta$), showing current inversion in the latter.

In \cref{fig:2Col}~(d) we have represented the sign of the current for the same parameters as in \cref{fig:2Col}~(b), in the region close to $V_\T{DC}=0$, in order to showcase more clearly the pattern resulting from current inversion. In the two diamond-like current inversion regions near $V_\T{G}=0$, both the $k=0$ and the $k=\pm1$ photon processes are blocked by the gap and the $k=\pm2$ rates are dominant. Current inversion then corresponds to the region where the $k=2$ rate is large due to the peak of the DOS and the $k=-2$ is smaller as it samples the flat region of the DOS (and vice versa). Apart from these diamonds, current inversion occurs along $V_\T{G}=0$ in intervals separated by $\hbar\omega_\T{AC}=\Delta$ but now has a conic shape. The explanation of their origin is nonetheless the same (i.e. higher $|k|$ rates being dominant over the ones with lower $k$).

\begin{figure}
    \includegraphics[width=\columnwidth]{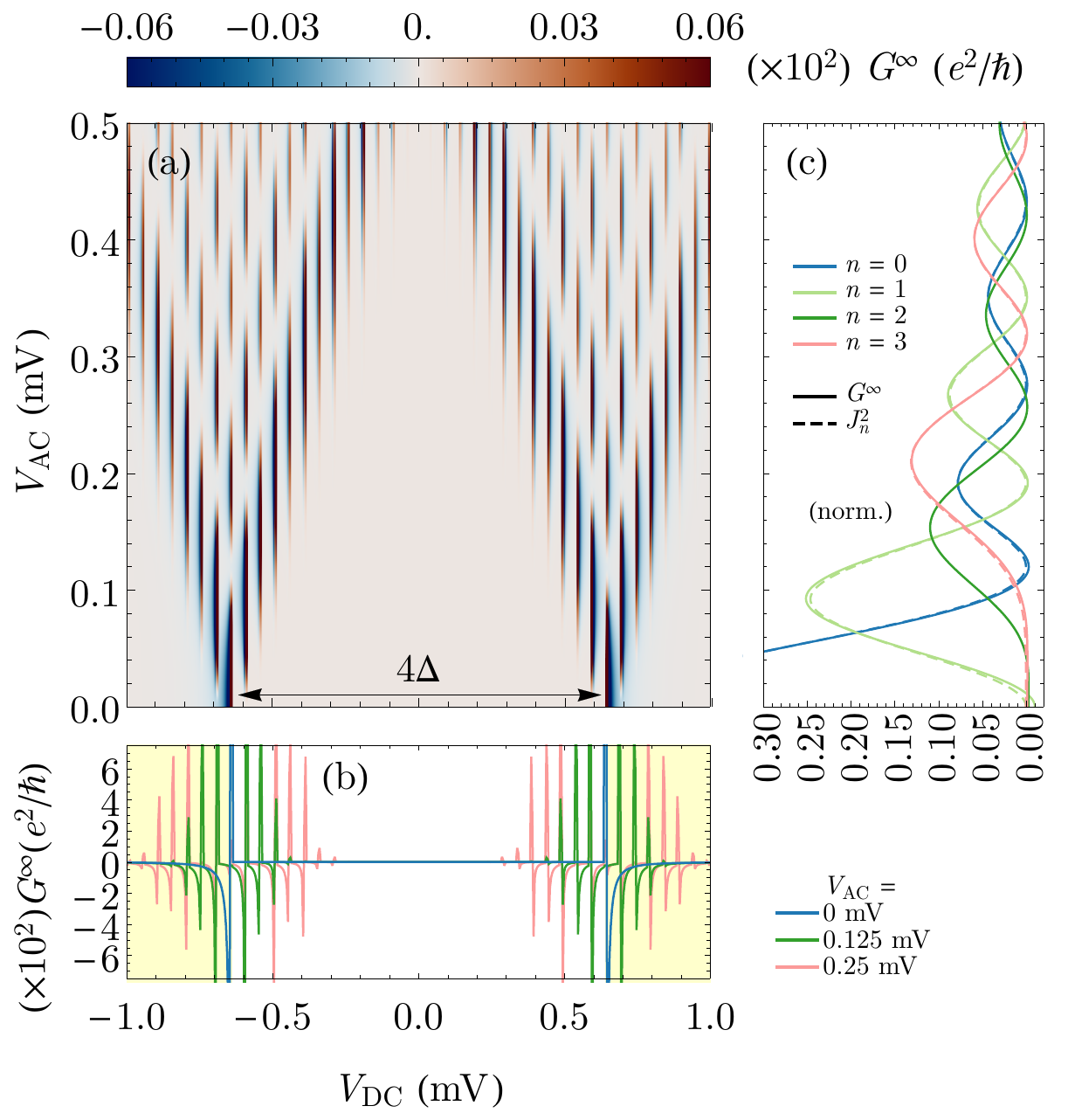}
    \caption{Photon assisted side-bands. (a) Differential conductance as a function of the DC and AC amplitudes of the bias voltage, $V_\T{DC}$ and $V_\T{AC}$, respectively, for $V_\T{G}=0$. The fan-like pattern reflects the appearance of multiple sidebands as $V_\T{AC}$ increases. (b) Cuts of Fig.~(a) for $V_\T{AC}=0$ (blue), 0.125 (green) and $0.25\,\T{mV}$ (light red). (c) Full lines: Cuts of Fig.~(a) for values of $V_\T{DC}$ corresponding to the best matches for the resonances with $n=0-3$ photons (blue, light green, green and light red, respectively). Dashed lines: squared Bessel functions $J_n^2(\epsilon_\text{AC}/2)$ for the same values of $n$. Note the close match for most of the parameter range.
    \textit{Parameters}: Same as in \protect\cref{fig:AC-case} but with $\hbar\omega_{\T{AC}}=25\,\mu eV$.
    \label{fig:AC-fan}}
\end{figure}

\cref{fig:AC-fan} shows the emergence of the photon assisted sidebands as a function of $V_{\T{AC}}$ for $\hbar\omega_{\T{AC}}=25\,\mu eV$. The resulting fan like pattern has a spacing of $2\hbar\omega_{\T{AC}}/e$ in $V_\T{DC}$ between the individual peaks, in agreement with \cref{eq: 93}. As the AC voltage increases, AC induced subgap transport at lower voltages becomes possible. Compared to the normal (i.e. non-superconducting) case, here we obtain \textit{two} fans corresponding to the states at the two sides of the gap. The conductance itself changes sign at the gap edges, as can be seen in \Cref{fig:AC-fan}~(a) and more clearly in \Cref{fig:AC-fan}~(b). This is a well-known result of the peaked DOS of superconductors~\cite{Doh2008}. The resulting conductance peaks at the different resonances follow nonetheless a Bessel-like pattern, shown in \Cref{fig:AC-fan}~(c). We have represented, together with the conductance, the associated squared Bessel function $J_n^2(\epsilon_\text{AC}/2)$ (employing dashed lines). The evaluation of the conductance along the peak of the respective rate contribution results in it dominating the other rates. Hence, the result expected from the Tien-Gordon model is recovered partially, specially at large values of $V_\T{AC}$. Similar features have been observed in recent experiments in scanning tunneling microscopy with superconducting tip and substrate~\cite{Kot2020,Peters2020}.

\begin{figure}
    \includegraphics[width=\columnwidth]{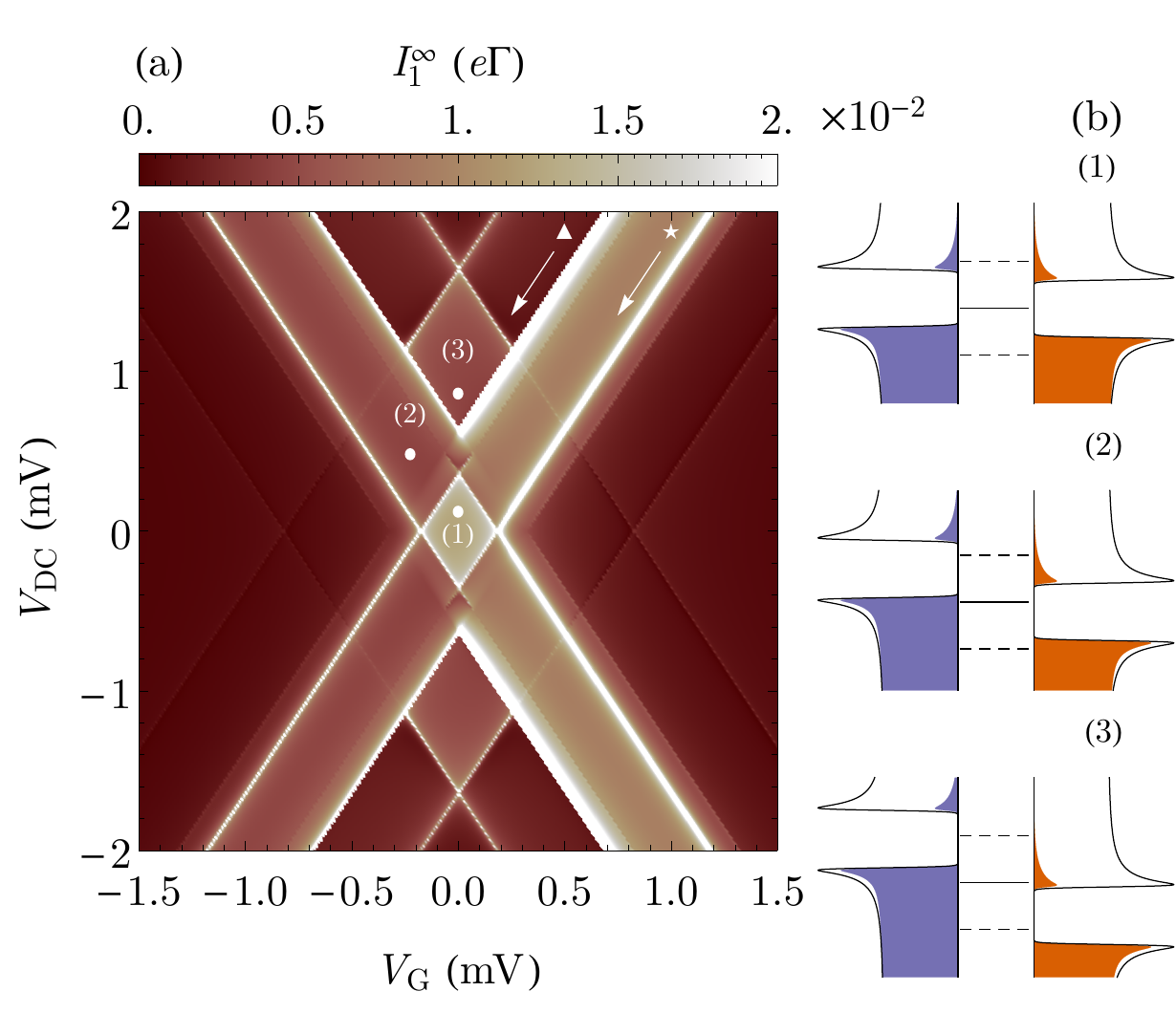}
    \caption{First harmonic of the current. (a) Shown is the absolute value of the first harmonic $I_1^\infty$ for $\epsilon_{\T{AC}}=0.1$ near the 1-0 charge degeneracy point. The other parameters are as in \cref{fig:AC-case}.
    The response is strongest along resonances ($\bigstar$) parallel to the Coulomb resonances ($\blacktriangle$) and offset by $\pm2\hbar\omega_{\T{AC}}$ in $V_{\T{DC}}$. (b) Level alignment of some exemplary points at low bias.
    \label{fig: 1st Harm}}
\end{figure}

\subsection{First current harmonic}

The formalism presented so far can recover also the $n\neq0$ Fourier components of the current. Next to the static response, the DC current, also the dynamic response at the frequency of the drive, the first harmonic $I_1^\infty$ in the current, is of interest. It is related to the nonlinear dynamic susceptibility $\chi(\omega_{\T{AC}},\epsilon_{\T{AC}})$ of the junction via ${I_1^\infty = \epsilon_{\T{AC}} \chi(\omega_{\T{AC}},\epsilon_{\T{AC}})}$. We consider the response of the junction at finite $\epsilon_{\T{AC}}$. For the charge transport through a junction in presence of AC drive it is convenient to use the symmetrized expression $I_1^\infty=\frac{1}{2}(j_{\T{L},1,0}-j_{\T{R},1,0})$. This convention is helpful when discussing AC phenomena as it restores the symmetries of the stability diagram lost due to the presence of displacement currents. The absolute value of $I_1^\infty$ is depicted in \cref{fig: 1st Harm} again for the vicinity of the 1-0 charge degeneracy point. We consider $\epsilon_{\T{AC}}=0.1$ (i.e. relatively weak drive) for clarity, as the general features of the susceptibility are already present at this drive strength. The other parameters were taken from \cref{fig:AC-case}.

To lowest order in $\Gamma$, the first harmonic of the current is given by
\begin{equation}
   I_1^\infty=\text{Tr}_{\text{QD}}\bigg\lbrace \frac{1}{2}\left( \tilde{k}_{\text{I,L};1,0}\left(i\omega_{\T{AC}}\right)-\tilde{k}_{\text{I,R};1,0}\left(i\omega_{\T{AC}}\right) \right)\hat{\varrho}_{0,0}\bigg\rbrace.
\end{equation}
This expression is analogous to the one for the DC current, where now the first harmonic of the current kernels has to be considered, together with the static populations of the dot. The former vanishes far from the Coulomb resonances. Physically, this reflects the condition of the dot's chemical potential being aligned within an energy window of $\hbar\omega_{\T{AC}}$ to either of the leads' chemical potential. The peaked density of states results in replicas of the Coulomb resonance at the boundaries of this energy window at
\begin{equation}
{\pm V_{\T{DC}}=2|V_{\T{G}}|-2V_{0}},
\end{equation}
with $V_{0}=\hbar\omega_{\T{AC}}-\Delta$ ( $\bigstar$ in \cref{fig: 1st Harm}), and
\begin{equation}
{\pm V_{\T{DC}}=2|V_{\T{G}}|+4\hbar\omega_{\T{AC}}}-2V_{0}.
\end{equation}
These resonances merge into the DC Coulomb resonance ($\blacktriangle$) in the limit of vanishing driving frequency, but for finite driving frequency the response inside the energy window is more complex.
Close to the charge degeneracy point the quantum dot's chemical potential is aligned with both leads simultaneously. The diagrams in \cref{fig: 1st Harm}~(b) show the level alignment for some exemplary points marked in (a).

At both (1) and (3) the populations in $\hat{\varrho}_{0,0}$ are $P_0=P_\uparrow=P_\downarrow=1/3$, such that both in and out tunneling rates contribute to the current. The exact form of the rates participating in the first harmonic of the current are given in \cref{eq: AC norm integral from DC}. For the low value of $\epsilon_{\T{AC}}$ we consider, it is sufficient to expand
\begin{equation}
    \widetilde{Y}^q_{l,1}(\nu)\propto a_l \epsilon_{\T{AC}} \left( \widetilde{Y}^q_{l,\T{DC}}(\nu+\hbar\omega_{\T{AC}}) - \widetilde{Y}^q_{l,\T{DC}}(\nu) \right) + \mathcal{O}(\epsilon_{\T{AC}}^2).
\end{equation}
Then the rates in the first harmonic of the current kernel become differences of the DC sequential tunneling integral. At low bias, corresponding to region (1), the second term vanishes for both $l=\tL,\tR$ due to presence of the gap. For the case $\hbar\omega_{\T{AC}}>\Delta$ considered here, the first term is finite and due to $a_\tL=-a_\tR=1/2$ additive in the lead index. Then, the first harmonic in the region around the charge degeneracy points is dominated by photon assisted sequential tunneling and finite even at zero bias. As such near the charge degeneracy point the AC and DC response display different behavior. At point (2) we have $P_1=1$. Looking at the level alignment in \cref{fig: 1st Harm}~(b) would suggest that the out-tunneling rate to the left lead would vanish. Indeed the imaginary parts of the respective DC sequential tunneling integrals either vanish or sum to zero. However, for dynamic properties also the real part contributes, causing a finite response to develop. In point (3) the levels are particle-hole symmetric. Here, the in-tunneling from the source and the out-tunneling into the drain both contribute to the response.

\section{Conclusion\label{sec: Conclusion}}

In this work we provided a microscopic formulation of the transport through an interacting quantum dot Josephson junction in simultaneous presence of a periodic driving and DC-bias. To this extend we generalized the Nakajima-Zwanzig projector operator approach to transport to the case of superconducting contacts and multiple driving frequencies. The formalism provides an exact generalized master equation (GME) for the reduced density operator of the quantum dot and Cooper pair condensate and an integral expression for the current. At long times, the current naturally exhibits a bichromatic periodicity, as a result of the transfer of Cooper pairs across the junction and the AC-drive. In the weak tunneling limit, the theory allows a straightforward calculation of the tunneling kernel of the GME in lowest order in the coupling to the leads. In this regime, the dynamics is dominated by the sequential tunneling of quasiparticles and anomalous processes involving the coherent transfer of Cooper pairs between condensate and quantum dot. The latter are responsible for proximity induced superconducting correlations which manifest in coherences in the charge sector of the quantum dot. Noticeably, the coherences naturally arise in our particle conserving formulation, without the need of invoking $U(1)$ symmetry breaking.

The existence of an Anderson pseudo-spin, associated to such coherences, was already noticed in the literature, and investigated at zero bias and in the infinite gap limit~\cite{Pala2007,Kamp2019,Kamp2021}. Our theory extends those results to a generic finite bias and finite gap situation. In particular, we demonstrate that the coherences become of particular relevance along some resonant lines in the bias voltage - gate voltage plane, cf. \cref{eq:cond-res}.

The theory is non-perturbative in the amplitude and frequency of the AC-drive whilst giving access to all harmonics of both the current and the reduced dot operator. As they encapsulate the static and dynamic response of the current, both the zeroth and first harmonics of the current are of particular experimental and theoretical interest. In the regime of high frequency, the sequential tunneling rates exhibit photon assisted tunneling sidebands, which shape the DC-current and all the harmonics in a highly non-linear manner. For example, a characteristic Bessel pattern is easily identified when considering the DC-current as a function of the amplitude of the external drive. The understanding of other features though requires a microscopic analysis, as the effect of the AC-drive is less trivial. For example, we predict the emergence of total current inversion, in which the current flows in the opposite direction of the DC voltage bias for certain regions of the stability diagram. We explained its origin as being due to the strongly peaked density of states of the superconducting leads and the subsequent dominance of backward photon assisted tunneling rates. 

We find that the same strongly figured density of states also permeates the dynamic response of the junction. In particular, we find that photon assisted sequential tunneling can become the dominant contribution to the AC response at zero DC-bias.

The formalism presented here serves as a counterpoint to non-equilibrium Green's function techniques~\cite{martin-rodero_josephson_2011}. As shown in this work, density operator methods allow a systematic expansion of the current and the tunneling kernels in the tunneling amplitudes. We focused here on the weak coupling limit, but the extensions to the intermediate coupling regime are possible via diagrammatic summation techniques~\cite{Konig1996b,Kern2013}. For superconducting junctions a diagrammatic resummation has been implemented for the case of infinite gap~\cite{Pala2007,Governale2008}. In that direction, future research may relax the weak coupling condition employed throughout this work and consider higher orders in the tunneling. With the supercurrent arising at the next higher order~\cite{Glazman1989}, this lays the groundwork for a microscopic theory of the physics of AC driven Josephson junctions.

Beyond the weak coupling limit, such setups are known to host Shapiro steps, which are commonly investigated within the semiclassical picture~\cite{Shapiro1963,Kautz1996,Dominguez2012,PicoCortes2017,Park2021}. As recent experiments on topological Josephson junctions hint at signatures of Majorana zero modes in the pattern of Shapiro steps~\cite{Kwon2004,Wiedenmann2016,Fischer2022}, a microscopic description of such setups~\cite{Virtanen2013,Li2018,Galaktionov2021} is highly desirable.

\begin{acknowledgments}
We thank G. Platero and A. Donarini for fruitful discussions and acknowledge DFG funding through projects B04 and B09 of SFB 1277 Emerging Relativistic Phenomena in Condensed \mbox{Matter} and support from CSIC Research Platform PTI-001 as well as (MICINN) via Grant No. PID2020-117787GB-100.
\end{acknowledgments}

\appendix

\section{Derivation of the kernels\label{App: Derivatio of the kernels}}
In \cref{eq: def rho} we defined the reduced density operator $\oprho(t)$ for which we now derive the GME governing its time evolution. We employ the Nakajima-Zwanzig formalism~\cite{Nakajima1958,Zwanzig1960} and extend previous treatments on its applications to transport topics (see e.g.~\cite{Saptsov2012,Rohrmeier2021}). The Nakajima-Zwanzig projection operator $\mathcal{P}$ includes a partial trace over a subset of the Hilbert space called the bath, which we identify as the quasiparticles in either lead in agreement with earlier works on the system~\cite{Pfaller2013}. The reduced density matrix can be given in terms of this projection operator as
\begin{align}
    \Pm\rhot (t):=\T{Tr}_{\T{QP}}\lbrace\rhot (t)\rbrace\otimes\hat{\rho}_{\T{QP}}=\oprho(t)\otimes\hat{\rho}_{\T{QP}}.
\end{align}
The Nakajima-Zwanzig formalism requires defining a reference density operator $\hat{\rho}_{\T{QP}}$ for the bath, which we choose to be the grand canonical equilibrium density operator of the quasiparticle sectors of the leads, given by
\begin{align}
	\oprho_{\T{QP}} = \frac{1}{Z_{\T{QP}}} \exp\bigg[ -\beta\bigg( \sum_{l} \hat{H}_{\T{QP},l} - \sum_{l,\bm{k},\sigma}\mu_{l} \hat{\gamma}^{\dagger}_{l,\bm{k},\sigma} \hat{\gamma}_{l,\bm{k},\sigma} \bigg) \bigg].
	\label{eq: thermal qp density op}
\end{align}
Note that the time dependence of the chemical potential drops in \cref{eq: thermal qp density op}, resulting in a static $\oprho_{\T{QP}}$ even in presence of an AC-drive. This choice is valid since in our mean field formulation the grand canonical Hamiltonian of the Cooper pairs vanishes. Thus, the grand canonical equilibrium density operator $\oprho_{\T{G}}$ of the leads factorizes into quasiparticle and Cooper pair sectors. Explicitly, it has the form
\begin{multline}
	\oprho_{\T{G}}=\frac{1}{Z_{\T{G}}}\exp\bigg[-\beta\bigg(\sum_{l}\hat{H}_{\T{CP},l}+\hat{H}_{\T{QP},l}-\mu_{l}\hat{N}_{l}\bigg)\bigg],
	\label{eq: 78}
\end{multline}
where $Z_{\T{G}}$ denotes the partition function in the grand canonical ensemble. By construction it holds
\begin{equation}
	\oprho_{\T{G}}=\oprho_{\T{QP}}\otimes\oprho_{\T{CP}},
	\label{eq: 80}
\end{equation}
where $\oprho_{\T{CP}}=\mathds{1}_{\T{CP}}/Z_{\T{CP}}$. This enables us to evaluate the partition function as
\begin{align}
    Z_{\T{G}}=:Z_{\T{QP}}Z_{\T{CP}},
    \label{eq: 85}
\end{align}
where $Z_{\T{CP}}$ is formally divergent. The latter is a consequence of the Cooper pair Hamiltonian being linear in the particle number. Considering e.g. a charging energy term would result in finite $Z_{\T{CP}}$ and reproduces the above expression in the limit of vanishing capacitance, i.e. for macroscopic leads.

The system part of the setup is given by the Cooper pairs and the dot. We work in the transformed frame given by \cref{eq: 9}. Since the Cooper pair Liouvillian is removed by this transformation, the system Liouvillian consists only of the Liouvillian of the dot. Introducing the orthogonal projector $\Qm:=1-\Pm$, the time-ordering superoperator $\mathcal{T}$ and assuming the leads to be in equilibrium at initial time, such that $\Qm\rhot'(0)=0$ the Nakajima-Zwanzig equation for our system reads
\begin{align}
    \begin{split}
    \Pm\dot{\hat{\rho}}'_{\text{tot}}(t) = & \mathcal{L}_{\T{QD}} \Pm \rhot'(t)\\
    +& \int_{0}^{t}ds \Pm \LT'(t) \mathcal{G}'(t,s) \LT'(s) \Pm \rhot'(s),
    \end{split}
    \label{eq: Nakajima-Zwanzig Projector}
\end{align}
where we used  the propagator
\begin{align}
    \mathcal{G}'(t,s)=\mathcal{T} \exp(\int_s^t ds' \mathcal{L}_{\T{QD}}(s') + \LQP(s') +\LT'(s')\Qm).
    \label{eq: Propagator}
\end{align}
Expanding \cref{eq: Propagator} in powers of $\LT'$ and exploiting the fact that $\Pm\LT'^{2n+1}\Pm=0$ for all integers $n$ due to particle conservation, we find to first non vanishing order in $\LT'$
\begin{align}
    \Km'^{(2)}_{\T{T}}(t,s)\oprho(s) = \T{Tr}_{\T{QP}}\lbrace \LT'(t) \mathcal{G}'_0(t,s) \LT'(s) \oprho'(s) \otimes \oprho_{\T{QP}} \rbrace.
    \label{eq: Tunneling Kernel App}
\end{align}
Here we introduced the propagator in absence of tunneling
\begin{align}
    \mathcal{G}'_0(t,s) = \exp(\int_s^t ds' \mathcal{L}_{\T{QD}}(s') + \LQP(s')).
\end{align}
Enforcing particle conservation in \cref{eq: exp val of I} yields
\begin{equation}
    I_{\tL}(t) = \T{Tr}\lbrace \hat{I}_{\tL}(t) (\Qm+\Pm) \rhot'(t) \rbrace = \T{Tr}\lbrace \hat{I}_{\tL}(t) \Qm \rhot'(t) \rbrace.
\end{equation}
Inserting the formal solution for $\Qm\rhot$~\cite{Breuer2007} and comparing with \cref{eq: 49}, we can identify
\begin{align}
   \Km'_{\T{I}}(t,s) \oprho'(s) = \T{Tr}\lbrace \hat{I}'_{\T{L}} \mathcal{G}'(t,s) \LT'(s) \oprho'(s) \otimes \oprho_{\T{QP}} \rbrace.
\end{align}
To lowest  order, we find
\begin{align}
   \Km'^{(2)}_{\T{I}}(t,s) \oprho(s) = \T{Tr}\lbrace \hat{I}'_{\T{L}} \mathcal{G}'_0(t,s) \LT'(s) \oprho'(s) \otimes \oprho_{\T{QP}} \rbrace.
\end{align}

\section{Phase representation\label{App: initial conditions}}

In this appendix we briefly discuss the origin of the $F(\bm{\varphi})$ term in the solution of the reduced dot operator, \cref{eq: Fourier-comp-DO-phase}. For the reduced density operator, the trace property complements the GME, resulting in an inhomogenous set of equations (see e.g. \cref{App: Analytic solution for the populations}).  For the generalized partial trace we employ, the trace property inherited by $\hat{\varrho}'(0;t)$ only constrains the average value
\begin{equation}
    \int_{\square}d\bm{\varphi}F\left(\bm{\varphi}\right)=1,
    \label{eq:trace-f}
\end{equation}
which is not sufficient to uniquely determine the solution to the GME. Thus, we need to supplement the GME with a further condition, i.e. matching to the initial preparation. For our calculations we choose a factorized initial condition, whereby the dot and the superconductors are decoupled. Thus, the dot and the Cooper pairs are initially diagonal in their respective particle numbers, which results in $F(\bm{\varphi})=1$.

The origin of this dependence lies in the fact that the phase variable associated with the Cooper pairs is uncoupled to the quasiparticles in an ideal superconductor. Cooper pairs and quasiparticles appear together only in the anomalous term of the tunneling Hamiltonian. However, since the Cooper pair operators are diagonal in the phase representation, the phase operator still commutes with the full Hamiltonian in the transformed frame and remains uncoupled to the quasiparticle bath. The time evolution of the phase operator is only determined by the voltage bias, which can be gauged away by going to the transformed frame. As a result, the phase operator in the transformed frame yields a constant of motion. In particular
\begin{align}
    \left\langle \hat{\bm{\varphi}} \right\rangle'(t) =\text{Tr}\{ \hat{\bm{\varphi}} \hat{\rho}'_{\T{tot}}\left(t\right)\} =\text{Tr}\{ \hat{U}^{\prime\dagger}\left(t\right) \hat{\bm{\varphi}} \hat{U}^{\prime}\left(t\right) \hat{\rho}_{\T{tot}}\left(0\right)\},
\end{align}
where $\hat{U}^{\prime}\left(t\right)=\mathcal{T}\exp[-i\int_{0}^{t}ds\hat{H}^{\prime}\left(s\right)]$ and $\hat{H}^{\prime}\left(s\right)$ is the total Hamiltonian in the transformed frame. Since this Hamiltonian commutes with the phase operator at all times, we obtain
\begin{equation}
   \left\langle \hat{\bm{\varphi}} \right\rangle'(t) = \text{Tr}\{ \hat{\bm{\varphi}} \hat{\rho}_{\T{tot}}\left(0\right)\}=\left\langle \hat{\bm{\varphi}} \right\rangle'(0).
    \label{eq: constant of motion}
\end{equation}
\Cref{eq: constant of motion} is then the condition required to supplement the GME. Throughout this work, we consider the zeroth harmonic $j_{\tL,n,0}$ of the current in the Josephson frequency. For it $\bm{m}=0$ and only the average value of $F(\bm{\varphi})$ enters. As this is already fixed by \cref{eq:trace-f}, our results presented here remain independent of the particular choice of initial condition. For $j_{\tL,n,\bm{m}}$ with $\bm{m}\neq0$, it remains necessary to supplement the GME with the proper initial state. This dependence can be remedied by accounting for effects (e.g. charging effects or voltage fluctuations) that couple the phase to dissipative degrees of freedom.

\section{Selection rules \label{App: Selection rules}}

The evaluation of the behavior of the junction at long times is simplified by symmetries and superselection rules of the problem we consider. We start by noting that for any self-adjoint operator $\hat{X}$, which fulfills $\Lm(t)\hat{X}=0\,\forall t$, we can choose a simultaneous eigenbasis of the Hamiltonian and $\hat{X}$. One finds that, in the steady state, coherences that involve states with two different eigenvalues of a conserved operator vanish. I.e. $\hat{\rho}'(t)$ can be block-diagonalized into blocks of constant eigenvalues of $\hat{X}$. A well known example of such superselection rule is the absence of coherences between different spin states for a dot connected to spin unpolarized leads as we consider throughout this work. The most relevant superselection rule for us will be the one with respect to the overall particle number, which ensures that coherences between states with different particle numbers are zero. Hence, the superselection rule indicates that the elements $\varrho^{\prime}_{\chi,\chi'}(\Delta\bm{M},\bm{M};t)$ in \cref{eq: structure of rho} are constrained by
\begin{equation}
    N_\chi+N_{\bm{M}+\Delta\bm{M}}=N_{\chi'}+N_{\bm{M}},
    \label{eq: Num constraint}
\end{equation}
where $N_\chi$ measures the number of particles on the dot for state $\ket{\chi}$ and ${N_{\bm{M}}=2\sum_l M_{l}}$. Since the dot can host at most a particle content of two electrons, we find that all $\varrho^{\prime}_{\chi,\chi'}(\Delta\bm{M},\bm{M};t)$ must vanish except where ${\sum_l \Delta M_{\T{l}}=0,\pm1}$. ${\bm{\Delta M}=0}$ allows only populations, while ${\bm{\Delta M}=\pm1}$ allows only for coherences of the type $\ketbra{0}{2}$ and $\ketbra{2}{0}$ respectively. While the range of  $\Delta\bm{M}$ is $\mathbb{Z}^2$, all nonzero $\varrho^{\prime}_{\chi,\chi'}(\Delta\bm{M},\bm{M};t)$ lie along the one dimensional subset $\Delta\bm{M}=n\bm{u}_{\tL}-n\bm{u}_{\tR}+r\bm{u}_l$ with $r=0,\pm1$. These superselection rules significantly reduce the number of entries we have to account for~\cite{Hiltscher2012}. They further ensure that no defects in current conservation can occur in the approach presented here as compared to approaches not explicitly conserving the number of particles~\cite{zonda_perturbation_2016}.

Finally, symmetry under hermitian conjugation is inherited by the reduced dot operators $\hat{\varrho},\hat{\varrho}^\circ$ from the reduced density operator upon exchange ${\Delta\bm{M},\bm{\varphi}\rightarrow-\Delta\bm{M},-\bm{\varphi}}$. As a result, the reduced operators satisfy
\begin{equation}
	\hat{\varrho}_{n,\bm{m}}^{\dagger}=\hat{\varrho}_{-n,-\bm{m}},
	\label{eq: 23}
\end{equation}
while the kernels satisfy
\begin{equation}
	(\widetilde{k}_{\T{T},n,\bm{m}}(\lambda))^{\dagger}=\widetilde{k}_{\T{T},n,\bm{m}}^{\dagger}(\lambda^{*}),
	\label{eq: 24}
\end{equation}
further simplifying the calculation.

\section{Sequential tunneling integrals\label{App: sequential tunneling integrals}}

In this appendix, we give analytic results for the normal and anomalous sequential tunneling integrals defined according to \cref{eq: normal integral Laplace} for the case of vanishing Dynes parameter $\gamma$. We start by considering the simple case of $V_{\T{AC}}=0$, for which the time integral can easily be executed. As a result, the term $n=0$ is the only non vanishing contribution and we find
\begin{align}
    \widetilde{Y}^{q}_{l,\T{DC}}(\nu) = \lim_{W\rightarrow\infty} \int_{-\infty}^{\infty}dE \frac{D_{l}(E)f^{q}(E)}{i0^{+}+E+\nu} L_W(E),
    \label{eq: 87}
\end{align}
and
\begin{align}
    \widetilde{Z}^{q}_{l,\T{DC}}(\nu) = \int_{-\infty}^{\infty}dE \frac{A_{l}(E)f^{q}(E)\text{sign}(E)}{i0^{+}+E+\nu},
    \label{eq: 88}
\end{align}
where we added the label $\T{DC}$ and removed the subscript $0$ to avoid confusion with the general $\widetilde{Y}^{q}_{l,0}$ in the AC case. Moreover, we have introduced a Lorentzian ${L_W(E)=W^{2}/(W^{2}+E^2)}$ with bandwidth $W$ in \cref{eq: 87} in order to regularize the integral. In physical terms, it corrects the ultraviolet divergence due to the non-vanishing density of states at large energies in the wide band limit.

\begin{figure}[t]
	\centering
    \includegraphics{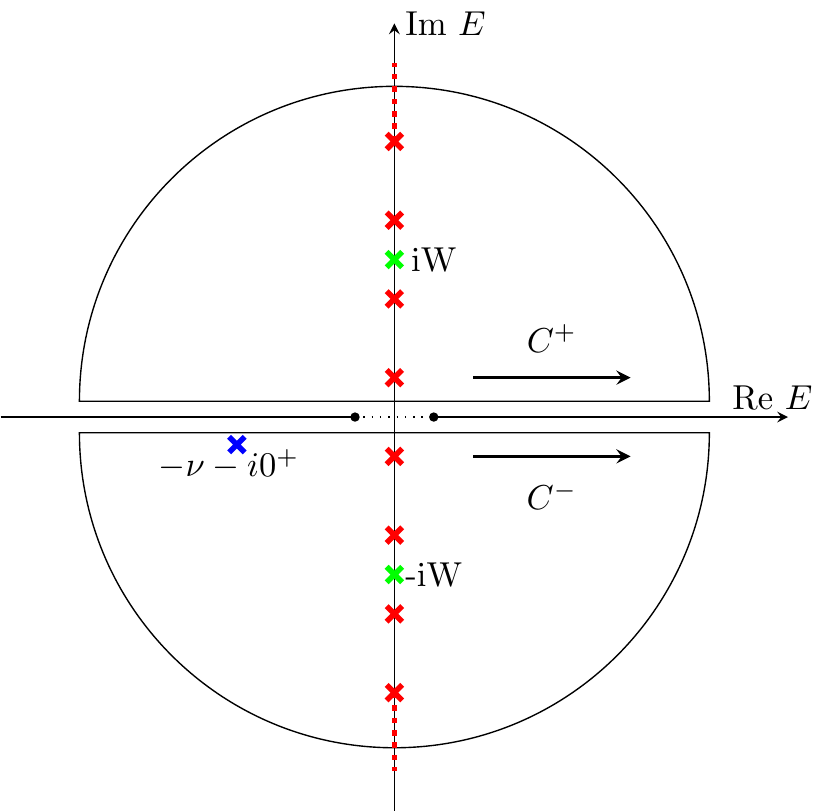}
	\caption{Integration contours for the normal sequential tunneling integral. The semicircles extend to infinite radius. The branch cut $\left[-\Delta_{l},\Delta_{l}\right]$ cancels the contribution along this interval if both contours are summed up with the orientations given by the arrows. The poles at the Matsubara frequencies $i\omega_{k}$ are shown as red crosses, the poles due to the Lorentzian and the denominator of the integrand are indicated in green and blue, respectively.}
	\label{fig: Contour}
\end{figure}

For the normal integral, the DOS function is the real part of a complex function
\begin{align}
    g_l(z)=D^0\sqrt{\frac{z^2}{z^2-|\Delta_l|^2}},\label{eq:g-function}
\end{align}
in the limit $\gamma\to0$. We define its analytic continuation $g_l(z)$ so that it has its branch cut along $\Re\left\{ z\right\} \in\left[-\left|\Delta_{l}\right|,\left|\Delta_{l}\right|\right]$. The real part corresponds to the segments outside of the branch cut.  Then, one can perform an integration along the contours shown in \cref{fig: Contour} in order to solve the integral~\cite{PicoCortesTh2021}. The contour is divided into two parts covering the upper and lower complex half-plane, leaving out the real axis. Doing so avoids the branch cut in the interval $[-|\Delta_l|,|\Delta_l|$ which originates from the density of states and cancels this contribution, due to the opposite signs along the two contours, precisely in the region $\Re\left\{ z\right\} \in\left[-\left|\Delta_{l}\right|,\left|\Delta_{l}\right|\right]$. Thus, we have
\begin{align}
    &\int_{-\infty}^{\infty}dE \frac{f^{q}\left(E\right)D_{l}\left(E\right)}{E+\nu+i0^{+}}L_W\left(E\right)\nonumber \\
    & = \lim_{R\to\infty} \sum_{s=\pm} \frac{s}{2} \int_{C^{s}} dz \frac{f^{q}\left(z\right)g_{l}\left(z\right)}{z+\nu+i0^{+}}L_W\left(z\right).
\end{align}
The resulting contour integrals can then be executed employing the residue theorem. For $f^{q}(z)$ the poles are the Matsubara frequencies $i\omega_{k}=2\pi i\beta^{-1}\left(k+1/2\right)$, $k\in\mathbb{Z}$, with residue $-q\beta^{-1}$, while the Lorentzian has poles at $z=\pm iW$ with residuum $\left(1/2\right)\left(\mp iW\right)$. Taking the limits, the integral yields
\begin{align}
    \widetilde{Y}^{q}_{l,\T{DC}}(\nu) & = -i\pi f^{\bar{q}}\left(\nu\right) g_{l}\left(\nu-i0^+\right)
    \label{eq:seq-tun-int}\\
    & -q \left[ \mathcal{S}_{l}^{\left(2\right)}\left(\nu\right) - \lim_{W\to\infty}\mathcal{C}_{l}^{\left(2\right)}\left(W\right) \right],\nonumber \\
    \mathcal{S}_{l}^{\left(2\right)}\left(\nu\right) & = \frac{2\pi}{\beta} \sum_{k=0}^{\infty} \frac{\omega_{k}g_{l}\left(i\omega_{k}\right)}{\omega_{k}^{2}+\nu^{2}},
    \label{eq:Matsubara-sum-2}\\
    \mathcal{C}_{l}^{\left(2\right)}\left(W\right) & = \frac{2\pi}{\beta} \sum_{k=0}^{\infty}\frac{1}{2} \left[ \frac{g_{l}\left(i\omega_{k}\right) + g_{l}\left(iW\right)} {\omega_{k}+W} \right]\nonumber \\
    & + \frac{2\pi}{\beta} \sum_{k=0}^{\infty} \frac{1}{2} \left[\frac{g_{l}\left(i\omega_{k}\right) - g_{l}\left(iW\right)}{\omega_{k}-W} \right].
    \label{eq:Divergence-constant}
\end{align}
It satisfies the property $\widetilde{Y}^{q}_{l,\T{DC}}(\nu) = -\widetilde{Y}^{\bar{q}*}_{l,\T{DC}}(-\nu)$ which guarantees the Hermiticity of the reduced dot operator. The last can be written for large $\omega_{k}\sim W$ as
\begin{align}
    \frac{g_{l}\left(iW\right)-g_{l}\left(i\omega_{k}\right)}{\omega_{k}-W}\simeq-D^0\frac{\left|\Delta_{l}\right|^{2}}{2\omega_{k}^{2}W^{2}}\left(\omega_{k}+W\right)\sim D^0\frac{\left|\Delta_{l}\right|^{2}}{W^{3}},\nonumber
\end{align}
and hence neglected when the limit $W\to\infty$ is taken. The real valued function $g_{l}\left(iE\right)$ is mostly flat except for the region $\left|E\right|\lesssim\left|\Delta_{l}\right|$ around zero, where it exhibits a dip. For small gap (or high temperature), the sampling at the Matsubara frequencies ignores the dip and the density of states is effectively flat. In that case, we recover the formula for the case of normal leads
\begin{align}
    \lim_{\beta\left|\Delta_{l}\right|\to0}
    \frac{\mathcal{S}_{l}^{\left(2\right)}\left(\nu\right)}{D^0}= & \Re\left\{ \Psi^{\left(0\right)}\left(\frac{1}{2}+\frac{i\beta\nu}{2\pi}\right)\right\} +\gamma_{\Psi},\\
    \lim_{\beta\left|\Delta_{l}\right|\to0}\frac{\mathcal{C}_{l}^{\left(2\right)}\left(W\right)}{D^0}= & \Re\left\{ \Psi^{\left(0\right)}\left(\frac{1}{2}+\frac{\beta W}{2\pi}\right)\right\} +\gamma_{\Psi},
\end{align}
where $\gamma_{\Psi}=\sum_{k=1}^{\infty}\log\left(1+\frac{1}{k}\right).$ This expressions can be employed to approximate the Matsubara sums for finite $\beta|\Delta_l|$.

For the anomalous kernel, let us consider the function
\begin{align}
    h^\rightleftharpoons_l(z) = D^0 \sqrt{\frac{|\Delta_l|^2}{z^2-|\Delta_l|^2}}.
    \label{eq:g-function-anom-wrong}
\end{align}
With the regular branch cut chosen to lie in $\Re\left\{ z\right\} \in\left[-\left|\Delta_{l}\right|,\left|\Delta_{l}\right|\right]$, $h^\rightleftharpoons_{l}\left(z\right)$ has an extra branch cut in the imaginary axis. Considering the continuation of $A_l(E)$ to the complex plane as $\Re{h_l(E)}$, where
\begin{align}
    h_l(z)=D^0\sqrt{\frac{|\Delta_l|^2}{z^2-|\Delta_l|^2}}\text{sgn}(\Re{z}+0^+),
    \label{eq:g-function-anom}
\end{align}
removes the branch cut along the imaginary axis. The integral can then be calculated employing the same contour as for the normal integral, yielding
\begin{align}
    \widetilde{Z}^{q}_{l,\T{DC}}(\nu) & = i\pi f^{\bar{q}}\left(\nu\right) h_{l}\left(\nu\right) + q \tilde{\mathcal{S}}_{l}^{\left(2\right)}\left(\nu\right),
    \label{eq:seq-tun-anom-int}\\
    \tilde{\mathcal{S}}_{l}^{\left(2\right)}\left(\nu\right) & = \frac{2\pi}{\beta} \sum_{k=0}^{\infty} \frac{i\nu h_{l}\left(i\omega_{k}\right)}{\omega_{k}^{2}+\nu^{2}}.
    \label{eq:anomalous-matsubara-S2}
\end{align}
It satisfies the property $\widetilde{Z}^{q}_{l,\T{DC}}(\nu)=\widetilde{Z}^{\bar{q}*}_{l,\T{DC}}(-\nu)$. Note that the function $\tilde{\mathcal{S}}_{l}^{\left(2\right)}\left(\nu\right)$ is real, since $h_{l}\left(i\nu\right)$ is purely imaginary. In a similar manner to the Matsubara sum found in the normal case, $\tilde{\mathcal{S}}_{l}^{\left(2\right)}\left(\nu\right)$ can be truncated for finite $\beta|\Delta_l|$, as it decays for $k\to\infty$ as $\sim\omega_k^{-3}$.

We note that in both integrals it is the $g_l(z),h_l(z)$ functions that appear in the results and not the densities of states defined in \cref{eq:DOS-n,eq:DOS-sc}. This is similar to how the imaginary part of a function arises from a real-valued function in Kramers-Kronig relations and can be obtained by extending said theorem to functions with poles in the upper and lower complex half-planes through contour integration. For the DC current to sequential tunneling order symmetry properties force the real parts of these integrals to cancel in the evaluation of tunneling rates between populations, thus reproducing the known "Fermi's golden rule" expressions albeit with a modified BCS like density of states. Considering higher orders in the expansion of the kernel, the finite real parts can become important~\cite{Rohrmeier2021}. For coherences and higher harmonics in the drive they contribute a Lamb shift.

The sequential tunneling integrals in presence of an AC-drive, \cref{eq: normal integral Laplace,lap1,lap2}, are difficult to evaluate numerically. However, using the definition of the $n-$th Bessel function,
\begin{align}
    J_n(x)=\sum_{r=0}^\infty \frac{(-1)^r(\frac{x}{2})^{2r+n}}{(r+n)!r!},
\end{align}
where we consider $n\ge0$ ($J_{-n}(x)=(-1)^n J_{n}(x)$), we can solve the integral over time as
\begin{widetext}
\begin{align}
    \int_0^{\infty}d\tau & \exp\bigg[ \frac{i\tau}{\hbar}\bigg(\nu+E+\frac{n\hbar\omega_{\T{AC}}}{2}+i0^+\bigg)  \bigg] J_n\bigg[ \epsilon_\text{AC} \sin(\frac{\omega_\T{AC}\tau}{2}) \bigg] \nonumber\\
    & = i\hbar \sum_{r=0}^{\infty} \sum_{k=0}^{2r+n} \frac{(-1)^{r+k}(\frac{a_l\epsilon_\text{AC}}{4i})^{2r+n}}{(r+n)!r!} \begin{pmatrix}2r+n\\k\end{pmatrix} \frac{1}{i0^+ +E+\nu+(r+n-k)\hbar\omega_{\T{AC}} }.
\end{align}
Comparing to \cref{eq: 87,eq: 88}, we find
\begin{align}
    \widetilde{Y}^{q}_{l,n}(\nu)= & (-1)^{n} \sum_{r=0}^{\infty} \sum_{k=0}^{2r+n} \frac{(-1)^{r+k}(\frac{a_l\epsilon_\text{AC}}{4i})^{2r+n}}{(r+n)!r!} \begin{pmatrix}2r+n\\k\end{pmatrix} \widetilde{Y}^{q}_{l,\T{DC}} \bigg(\nu+(r+n-k)\hbar\omega_{\T{AC}} \bigg),
    \label{eq: AC norm integral from DC}\\
    \widetilde{Z}^{q}_{l,n}(\nu) = & (i)^{n} \sum_{r=0}^{\infty} \sum_{k=0}^{2r+n} \frac{(-1)^{r}(\frac{a_l\epsilon_\text{AC}}{4})^{2r+n}}{(r+n)!r!} \begin{pmatrix}2r+n\\k\end{pmatrix} \widetilde{Z}^{q}_{l,\T{DC}}\bigg(\nu+(r+n-k)\hbar\omega_{\T{AC}} \bigg).
    \label{eq: AC anom integral from DC}
\end{align}
\end{widetext}
From \cref{eq: AC norm integral from DC,eq: AC anom integral from DC} we see that the AC integrals are sums over the DC integrals, with their arguments offset by multiples of the photon energy associated to the drive. The factorials in the denominator of either summand enable a straightforward truncation of the sums for numerical evaluation. We further remark that for all $n$ and $r$, i.e. all orders in $\epsilon_{\T{AC}}$, the imaginary part of the normal integral vanish for large $|\nu|\gg |r+n|\omega_{\T{AC}}+\beta^{-1}$. There the Fermi functions in the DC integrals saturate and the inner sum over $k$ becomes an alternating sum over binomial coefficients which vanishes by a simple combinatorical argument. The real part of the normal integral and the anomalous integral trivially vanish for large arguments. This ensures the physical behavior of the AC-drive only inducing a significant response of the system if the dot's chemical potential is aligned within an energy window of $n\omega_{\T{AC}}$ of the respective lead for a $n-$photon assisted tunneling process.

\section{Analytic solution for the populations\label{App: Analytic solution for the populations}}

For the DC driven case, we define $\Gamma^{\chi,\chi'} =\sum_{l} \Gamma^{\chi,\chi'}_{l}$ as the total flow out of the population of the dot state $\chi'$ to population $\chi$. Then, the rate equation one has to solve for the steady state can be brought into the form of a linear inhomogeneous system
\begin{equation}
	\begin{pmatrix}
		\Gamma^{0,0}	&	\Gamma^{0,\up}	    &	\Gamma^{0,\down}	&	\Gamma^{0,2}\\
		\Gamma^{\up,0}	&	\Gamma^{\up,\up}	&	\Gamma^{\up,\down}	&	\Gamma^{\up,2}\\
		\Gamma^{\down,0}&	\Gamma^{\down,\up}	&	\Gamma^{\down,\down}&	\Gamma^{\down,2}\\
		1 	            &	                1	&	                1	&	1
	\end{pmatrix}\cdot
	\begin{pmatrix}
		P_{0}\\P_{\up}\\P_{\down}\\P_{2}
	\end{pmatrix}=
	\begin{pmatrix}
		0\\0\\0\\1
	\end{pmatrix}.
\label{eq: 28}
\end{equation}
The populations can then be given analytically as a function of the rates $\Gamma^{\chi,\chi'}$ as
\begin{align}
	P_{0} = & \frac{1}{N} \sum_{\sigma} \lbrace \Gamma^{0,\sigma} \Gamma^{\sigma,2} ( \Gamma^{0,\bar{\sigma}} + \Gamma^{2,\bar{\sigma}} ) \nonumber \\
	+& \Gamma^{\sigma,\bar{\sigma}} [ \Gamma^{0,\sigma} ( \Gamma^{\up,2} + \Gamma^{\down,2} ) + \Gamma^{0,2}( \Gamma^{0,\sigma} + \Gamma^{2,\sigma} ) ] \nonumber \\
	+ &	\Gamma^{0,2} \Gamma^{2,\sigma} \Gamma^{0,\bar{\sigma}} \rbrace + \Gamma^{0,2} (\Gamma^{0,\up} \Gamma^{0,\down} + \Gamma^{2,\up} \Gamma^{2,\down} ),
	\label{eq: QD1.3.2.3}
\end{align}
\begin{align}
	P_{\up} = & \frac{1}{N}	\sum_{\sigma} \lbrace \Gamma^{\up,0} \Gamma^{0,\down} \Gamma^{\sigma,2} + \Gamma^{\up,2} \Gamma^{2,\down} \Gamma^{\sigma,0} \nonumber \\
	+ &	\Gamma^{\up,\down} [ \Gamma^{0,2} \Gamma^{\sigma,0} + \Gamma^{2,0} \Gamma^{\sigma,2} + \Gamma^{\sigma,2} ( \Gamma^{\up,0} + \Gamma^{\down,0})] \rbrace \nonumber \\
	+ &	(\Gamma^{0,\down} + \Gamma^{2,\down} )( \Gamma^{0,2} \Gamma^{\up,0} + \Gamma^{2,0} \Gamma^{\up,2} ),
\end{align}
\begin{align}
	P_{\down}=&\frac{1}{N} \sum_{\sigma} \lbrace \Gamma^{\down,0} \Gamma^{0,\up} \Gamma^{\sigma,2} + \Gamma^{\down,2} \Gamma^{2,\up} \Gamma^{\sigma,0} \nonumber \\
	+ & \Gamma^{\down,\up} [ \Gamma^{0,2} \Gamma^{\sigma,0} + \Gamma^{2,0} \Gamma^{\sigma,2} + \Gamma^{\sigma,2} ( \Gamma^{\down,0} + \Gamma^{\up,0})]
	\rbrace \nonumber \\
	+ & ( \Gamma^{0,\up} + \Gamma^{2,\up} )( \Gamma^{0,2} \Gamma^{\down,0} + \Gamma^{2,0} \Gamma^{\down,2} ),
	\label{eq: QD1.3.2.4}
	\end{align}
and
\begin{align}
	P_{2} = & \frac{1}{N} \sum_{\sigma} \lbrace	\Gamma^{2,\bar{\sigma}} \Gamma^{\bar{\sigma},0} ( \Gamma^{0,\sigma} + \Gamma^{2,\sigma} )\nonumber \\
	+ & \Gamma^{\sigma,\bar{\sigma}} [ \Gamma^{2,\sigma} ( \Gamma^{\sigma,0} + \Gamma^{\bar{\sigma},0} ) + \Gamma^{2,0} ( \Gamma^{0,\sigma} + \Gamma^{2,\sigma} ) ] \nonumber \\
	+ & \Gamma^{2,0} \Gamma^{0,\sigma} \Gamma^{2,\bar{\sigma}} \rbrace + \Gamma^{2,0} ( \Gamma^{0,\up} \Gamma^{0,\down} +\Gamma^{2,\up} \Gamma^{2,\down} ),
	\label{eq: QD1.3.2.5}
	\end{align}
where $N$ is an appropriate normalization constant, such that $P_0+P_\uparrow +P_\downarrow+P_2=1$.

\section{Comparison with Tien-Gordon theory\label{App: Comparison with Tien-Gordon}}

\begin{figure}
    \includegraphics[width=0.7\columnwidth]{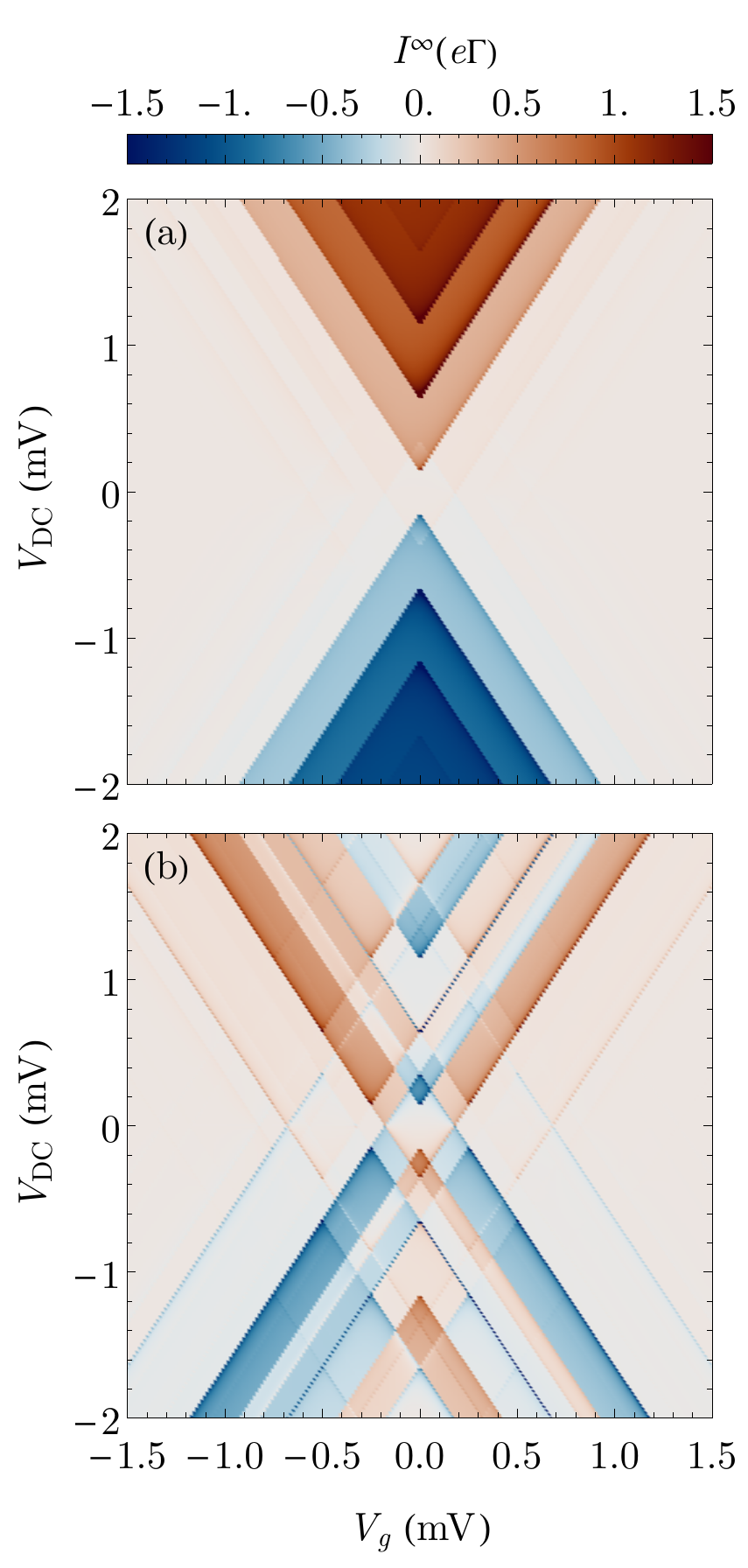}
    \caption{Stability diagram within a simple Tien-Gordon approach for the current. (a) Current near the the 1-0 charge degeneracy point obtained employing the Tien-Gordon expression for the current, \protect\cref{eq: naive TG Ansatz}, showing the simpler structure of replicas compared to \protect\cref{fig:AC-case}. (b) Difference between the full result of \protect\cref{fig:AC-case}~(a) and the Tien-Gordon approximation.
    \textit{Parameters}: Same as in \protect\cref{fig:AC-case}.
    \label{fig:TG}}
\end{figure}

In this appendix, we elaborate on some aspects of the photon assisted sidebands exhibited by the current for high frequency driving. The main difference in our result as compared to a direct application of Tien-Gordon theory to the DC current, cf. \cref{eq: naive TG Ansatz} is that the photon assisted replicas of the rates do not cause an identical behavior of the current~\cite{Whan1996}. In \cref{fig:TG}~(a) we have represented the results of applying the Tien-Gordon like equation
\begin{align}
	I(V_{\T{DC}},V_\T{g}) \rightarrow \sum_{k=-\infty}^{\infty} J_{k}^{2}\big(\epsilon_\text{AC}/2\big) I(V_{\T{DC}}-k\hbar\omega_{\T{AC}}/e,V_\T{g}),
	\label{eq: naive TG Ansatz}
\end{align}
for the current with the same parameters as in \cref{fig:AC-case}. It can be seen, that the two results are markedly different, as expected due to the non-linear way in which the rates enter the current. The difference between the two situations is represented in \cref{fig:TG}~(b).

%


\begin{thebibliography}{102}%
\makeatletter
\providecommand \@ifxundefined [1]{%
 \@ifx{#1\undefined}
}%
\providecommand \@ifnum [1]{%
 \ifnum #1\expandafter \@firstoftwo
 \else \expandafter \@secondoftwo
 \fi
}%
\providecommand \@ifx [1]{%
 \ifx #1\expandafter \@firstoftwo
 \else \expandafter \@secondoftwo
 \fi
}%
\providecommand \natexlab [1]{#1}%
\providecommand \enquote  [1]{``#1''}%
\providecommand \bibnamefont  [1]{#1}%
\providecommand \bibfnamefont [1]{#1}%
\providecommand \citenamefont [1]{#1}%
\providecommand \href@noop [0]{\@secondoftwo}%
\providecommand \href [0]{\begingroup \@sanitize@url \@href}%
\providecommand \@href[1]{\@@startlink{#1}\@@href}%
\providecommand \@@href[1]{\endgroup#1\@@endlink}%
\providecommand \@sanitize@url [0]{\catcode `\\12\catcode `\$12\catcode
  `\&12\catcode `\#12\catcode `\^12\catcode `\_12\catcode `\%12\relax}%
\providecommand \@@startlink[1]{}%
\providecommand \@@endlink[0]{}%
\providecommand \url  [0]{\begingroup\@sanitize@url \@url }%
\providecommand \@url [1]{\endgroup\@href {#1}{\urlprefix }}%
\providecommand \urlprefix  [0]{URL }%
\providecommand \Eprint [0]{\href }%
\providecommand \doibase [0]{https://doi.org/}%
\providecommand \selectlanguage [0]{\@gobble}%
\providecommand \bibinfo  [0]{\@secondoftwo}%
\providecommand \bibfield  [0]{\@secondoftwo}%
\providecommand \translation [1]{[#1]}%
\providecommand \BibitemOpen [0]{}%
\providecommand \bibitemStop [0]{}%
\providecommand \bibitemNoStop [0]{.\EOS\space}%
\providecommand \EOS [0]{\spacefactor3000\relax}%
\providecommand \BibitemShut  [1]{\csname bibitem#1\endcsname}%
\let\auto@bib@innerbib\@empty
\bibitem [{\citenamefont {Makhlin}\ \emph {et~al.}(2001)\citenamefont
  {Makhlin}, \citenamefont {{Sch\"on}},\ and\ \citenamefont
  {Shnirman}}]{Makhlin2010}%
  \BibitemOpen
  \bibfield  {author} {\bibinfo {author} {\bibfnamefont {Y.}~\bibnamefont
  {Makhlin}}, \bibinfo {author} {\bibfnamefont {G.}~\bibnamefont {{Sch\"on}}},\
  and\ \bibinfo {author} {\bibfnamefont {A.}~\bibnamefont {Shnirman}},\
  }\bibfield  {title} {\bibinfo {title} {{Quantum-state engineering with
  Josephson-junction devices}},\ }\href
  {https://doi.org/10.1103/RevModPhys.73.357} {\bibfield  {journal} {\bibinfo
  {journal} {Rev. Mod. Phys.}\ }\textbf {\bibinfo {volume} {73}},\ \bibinfo
  {pages} {357} (\bibinfo {year} {2001})}\BibitemShut {NoStop}%
\bibitem [{\citenamefont {Ladd}\ \emph {et~al.}(2010)\citenamefont {Ladd},
  \citenamefont {Jelezko}, \citenamefont {Laflamme}, \citenamefont {Nakamura},
  \citenamefont {Monroe},\ and\ \citenamefont {O’Brien}}]{Ladd2010}%
  \BibitemOpen
  \bibfield  {author} {\bibinfo {author} {\bibfnamefont {T.~D.}\ \bibnamefont
  {Ladd}}, \bibinfo {author} {\bibfnamefont {F.}~\bibnamefont {Jelezko}},
  \bibinfo {author} {\bibfnamefont {R.}~\bibnamefont {Laflamme}}, \bibinfo
  {author} {\bibfnamefont {Y.}~\bibnamefont {Nakamura}}, \bibinfo {author}
  {\bibfnamefont {C.}~\bibnamefont {Monroe}},\ and\ \bibinfo {author}
  {\bibfnamefont {J.~L.}\ \bibnamefont {O’Brien}},\ }\bibfield  {title}
  {\bibinfo {title} {{Quantum computers}},\ }\href
  {https://doi.org/10.1038/nature08812} {\bibfield  {journal} {\bibinfo
  {journal} {Nature}\ }\textbf {\bibinfo {volume} {464}},\ \bibinfo {pages}
  {45} (\bibinfo {year} {2010})}\BibitemShut {NoStop}%
\bibitem [{\citenamefont {Wendin}(2017)}]{Wendin2017}%
  \BibitemOpen
  \bibfield  {author} {\bibinfo {author} {\bibfnamefont {G.}~\bibnamefont
  {Wendin}},\ }\bibfield  {title} {\bibinfo {title} {{Quantum information
  processing with superconducting circuits: a review}},\ }\href
  {https://doi.org/10.1088/1361-6633/aa7e1a} {\bibfield  {journal} {\bibinfo
  {journal} {Rep. Prog. Phys.}\ }\textbf {\bibinfo {volume} {80}},\ \bibinfo
  {pages} {106001} (\bibinfo {year} {2017})}\BibitemShut {NoStop}%
\bibitem [{\citenamefont {Giaever}(1960)}]{Giaever1960}%
  \BibitemOpen
  \bibfield  {author} {\bibinfo {author} {\bibfnamefont {I.}~\bibnamefont
  {Giaever}},\ }\bibfield  {title} {\bibinfo {title} {{Electron Tunneling
  Between Two Superconductors}},\ }\href
  {https://doi.org/10.1103/PhysRevLett.5.464} {\bibfield  {journal} {\bibinfo
  {journal} {Phys. Rev. Lett.}\ }\textbf {\bibinfo {volume} {5}},\ \bibinfo
  {pages} {464} (\bibinfo {year} {1960})}\BibitemShut {NoStop}%
\bibitem [{\citenamefont {Grove-Rasmussen}\ \emph {et~al.}(2009)\citenamefont
  {Grove-Rasmussen}, \citenamefont {Jørgensen}, \citenamefont {Andersen},
  \citenamefont {Paaske}, \citenamefont {Jespersen}, \citenamefont {Nygård},
  \citenamefont {Flensberg},\ and\ \citenamefont {Lindelof}}]{Grove2009}%
  \BibitemOpen
  \bibfield  {author} {\bibinfo {author} {\bibfnamefont {K.}~\bibnamefont
  {Grove-Rasmussen}}, \bibinfo {author} {\bibfnamefont {H.~I.}\ \bibnamefont
  {Jørgensen}}, \bibinfo {author} {\bibfnamefont {B.~M.}\ \bibnamefont
  {Andersen}}, \bibinfo {author} {\bibfnamefont {J.}~\bibnamefont {Paaske}},
  \bibinfo {author} {\bibfnamefont {T.~S.}\ \bibnamefont {Jespersen}}, \bibinfo
  {author} {\bibfnamefont {J.}~\bibnamefont {Nygård}}, \bibinfo {author}
  {\bibfnamefont {K.}~\bibnamefont {Flensberg}},\ and\ \bibinfo {author}
  {\bibfnamefont {P.~E.}\ \bibnamefont {Lindelof}},\ }\bibfield  {title}
  {\bibinfo {title} {{Superconductivity-enhanced bias spectroscopy in carbon
  nanotube quantum dots}},\ }\href {https://doi.org/10.1103/PhysRevB.79.134518}
  {\bibfield  {journal} {\bibinfo  {journal} {Phys. Rev. B}\ }\textbf {\bibinfo
  {volume} {79}},\ \bibinfo {pages} {134518} (\bibinfo {year}
  {2009})}\BibitemShut {NoStop}%
\bibitem [{\citenamefont {Dirks}\ \emph {et~al.}(2009)\citenamefont {Dirks},
  \citenamefont {Chen}, \citenamefont {Birge},\ and\ \citenamefont
  {Mason}}]{Dirks2009}%
  \BibitemOpen
  \bibfield  {author} {\bibinfo {author} {\bibfnamefont {T.}~\bibnamefont
  {Dirks}}, \bibinfo {author} {\bibfnamefont {Y.-F.}\ \bibnamefont {Chen}},
  \bibinfo {author} {\bibfnamefont {N.~O.}\ \bibnamefont {Birge}},\ and\
  \bibinfo {author} {\bibfnamefont {N.}~\bibnamefont {Mason}},\ }\bibfield
  {title} {\bibinfo {title} {{Superconducting tunneling spectroscopy of a
  carbon nanotube quantum dot}},\ }\href {https://doi.org/10.1063/1.3253705}
  {\bibfield  {journal} {\bibinfo  {journal} {Appl. Phys. Lett.}\ }\textbf
  {\bibinfo {volume} {95}},\ \bibinfo {pages} {192103} (\bibinfo {year}
  {2009})}\BibitemShut {NoStop}%
\bibitem [{\citenamefont {Bespalov}\ \emph {et~al.}(2016)\citenamefont
  {Bespalov}, \citenamefont {Houzet}, \citenamefont {Meyer},\ and\
  \citenamefont {Nazarov}}]{Bespalov2016}%
  \BibitemOpen
  \bibfield  {author} {\bibinfo {author} {\bibfnamefont {A.}~\bibnamefont
  {Bespalov}}, \bibinfo {author} {\bibfnamefont {M.}~\bibnamefont {Houzet}},
  \bibinfo {author} {\bibfnamefont {J.~S.}\ \bibnamefont {Meyer}},\ and\
  \bibinfo {author} {\bibfnamefont {Y.~V.}\ \bibnamefont {Nazarov}},\
  }\bibfield  {title} {\bibinfo {title} {{Theoretical Model to Explain Excess
  of Quasiparticles in Superconductors}},\ }\href
  {https://doi.org/10.1103/PhysRevLett.117.117002} {\bibfield  {journal}
  {\bibinfo  {journal} {Phys. Rev. Lett.}\ }\textbf {\bibinfo {volume} {117}},\
  \bibinfo {pages} {117002} (\bibinfo {year} {2016})}\BibitemShut {NoStop}%
\bibitem [{\citenamefont {Frombach}\ and\ \citenamefont
  {Recher}(2020)}]{Frombach2020}%
  \BibitemOpen
  \bibfield  {author} {\bibinfo {author} {\bibfnamefont {D.}~\bibnamefont
  {Frombach}}\ and\ \bibinfo {author} {\bibfnamefont {P.}~\bibnamefont
  {Recher}},\ }\bibfield  {title} {\bibinfo {title} {{Quasiparticle poisoning
  effects on the dynamics of topological Josephson junctions}},\ }\href
  {https://doi.org/10.1103/PhysRevB.101.115304} {\bibfield  {journal} {\bibinfo
   {journal} {Phys. Rev. B}\ }\textbf {\bibinfo {volume} {101}},\ \bibinfo
  {pages} {115304} (\bibinfo {year} {2020})}\BibitemShut {NoStop}%
\bibitem [{\citenamefont {Whan}\ and\ \citenamefont
  {Orlando}(1996)}]{Whan1996}%
  \BibitemOpen
  \bibfield  {author} {\bibinfo {author} {\bibfnamefont {C.}~\bibnamefont
  {Whan}}\ and\ \bibinfo {author} {\bibfnamefont {T.}~\bibnamefont {Orlando}},\
  }\bibfield  {title} {\bibinfo {title} {{Transport properties of a quantum dot
  with superconducting leads}},\ }\href
  {https://doi.org/10.1103/PhysRevB.54.R5255} {\bibfield  {journal} {\bibinfo
  {journal} {Phy. Rev. B}\ }\textbf {\bibinfo {volume} {54}},\ \bibinfo {pages}
  {R5255} (\bibinfo {year} {1996})}\BibitemShut {NoStop}%
\bibitem [{\citenamefont {Pfaller}\ \emph {et~al.}(2013)\citenamefont
  {Pfaller}, \citenamefont {Donarini},\ and\ \citenamefont
  {Grifoni}}]{Pfaller2013}%
  \BibitemOpen
  \bibfield  {author} {\bibinfo {author} {\bibfnamefont {S.}~\bibnamefont
  {Pfaller}}, \bibinfo {author} {\bibfnamefont {A.}~\bibnamefont {Donarini}},\
  and\ \bibinfo {author} {\bibfnamefont {M.}~\bibnamefont {Grifoni}},\
  }\bibfield  {title} {\bibinfo {title} {{Subgap features due to quasiparticle
  tunneling in quantum dots coupled to superconducting leads}},\ }\href
  {https://doi.org/10.1103/PhysRevB.87.155439} {\bibfield  {journal} {\bibinfo
  {journal} {Phys. Rev. B}\ }\textbf {\bibinfo {volume} {87}},\ \bibinfo
  {pages} {155439} (\bibinfo {year} {2013})}\BibitemShut {NoStop}%
\bibitem [{\citenamefont {Ratz}\ \emph {et~al.}(2014)\citenamefont {Ratz},
  \citenamefont {Donarini}, \citenamefont {Steininger}, \citenamefont {Geiger},
  \citenamefont {Kumar}, \citenamefont {Hüttel}, \citenamefont {Strunk},\ and\
  \citenamefont {Grifoni}}]{Ratz2014}%
  \BibitemOpen
  \bibfield  {author} {\bibinfo {author} {\bibfnamefont {S.}~\bibnamefont
  {Ratz}}, \bibinfo {author} {\bibfnamefont {A.}~\bibnamefont {Donarini}},
  \bibinfo {author} {\bibfnamefont {D.}~\bibnamefont {Steininger}}, \bibinfo
  {author} {\bibfnamefont {T.}~\bibnamefont {Geiger}}, \bibinfo {author}
  {\bibfnamefont {A.}~\bibnamefont {Kumar}}, \bibinfo {author} {\bibfnamefont
  {A.~K.}\ \bibnamefont {Hüttel}}, \bibinfo {author} {\bibfnamefont
  {C.}~\bibnamefont {Strunk}},\ and\ \bibinfo {author} {\bibfnamefont
  {M.}~\bibnamefont {Grifoni}},\ }\bibfield  {title} {\bibinfo {title}
  {{Thermally induced subgap features in the cotunneling spectroscopy of a
  carbon nanotube}},\ }\href {https://doi.org/10.1088/1367-2630/16/12/123040}
  {\bibfield  {journal} {\bibinfo  {journal} {New J. Phys.}\ }\textbf {\bibinfo
  {volume} {16}},\ \bibinfo {pages} {123040} (\bibinfo {year}
  {2014})}\BibitemShut {NoStop}%
\bibitem [{\citenamefont {Gaass}\ \emph {et~al.}(2014)\citenamefont {Gaass},
  \citenamefont {Pfaller}, \citenamefont {Geiger}, \citenamefont {Donarini},
  \citenamefont {Grifoni}, \citenamefont {Hüttel},\ and\ \citenamefont
  {Strunk}}]{Gaass2014}%
  \BibitemOpen
  \bibfield  {author} {\bibinfo {author} {\bibfnamefont {M.}~\bibnamefont
  {Gaass}}, \bibinfo {author} {\bibfnamefont {S.}~\bibnamefont {Pfaller}},
  \bibinfo {author} {\bibfnamefont {T.}~\bibnamefont {Geiger}}, \bibinfo
  {author} {\bibfnamefont {A.}~\bibnamefont {Donarini}}, \bibinfo {author}
  {\bibfnamefont {M.}~\bibnamefont {Grifoni}}, \bibinfo {author} {\bibfnamefont
  {A.~K.}\ \bibnamefont {Hüttel}},\ and\ \bibinfo {author} {\bibfnamefont
  {C.}~\bibnamefont {Strunk}},\ }\bibfield  {title} {\bibinfo {title} {{Subgap
  spectroscopy of thermally excited quasiparticles in a Nb-contacted carbon
  nanotube quantum dot}},\ }\href {https://doi.org/10.1103/PhysRevB.89.241405}
  {\bibfield  {journal} {\bibinfo  {journal} {Phys. Rev. B}\ }\textbf {\bibinfo
  {volume} {89}},\ \bibinfo {pages} {241405} (\bibinfo {year}
  {2014})}\BibitemShut {NoStop}%
\bibitem [{\citenamefont {Gramich}\ \emph {et~al.}(2016)\citenamefont
  {Gramich}, \citenamefont {Baumgartner},\ and\ \citenamefont
  {{{Sch\"onenberger}}}}]{Gramich2016}%
  \BibitemOpen
  \bibfield  {author} {\bibinfo {author} {\bibfnamefont {J.}~\bibnamefont
  {Gramich}}, \bibinfo {author} {\bibfnamefont {A.}~\bibnamefont
  {Baumgartner}},\ and\ \bibinfo {author} {\bibfnamefont {C.}~\bibnamefont
  {{{Sch\"onenberger}}}},\ }\bibfield  {title} {\bibinfo {title} {{Subgap
  resonant quasiparticle transport in normal-superconductor quantum dot
  devices}},\ }\href {https://doi.org/10.1063/1.4948352} {\bibfield  {journal}
  {\bibinfo  {journal} {Appl. Phys. Lett.}\ }\textbf {\bibinfo {volume}
  {108}},\ \bibinfo {pages} {172604} (\bibinfo {year} {2016})}\BibitemShut
  {NoStop}%
\bibitem [{\citenamefont {Andreev}(1964)}]{Andreev1964}%
  \BibitemOpen
  \bibfield  {author} {\bibinfo {author} {\bibfnamefont {A.~F.}\ \bibnamefont
  {Andreev}},\ }\bibfield  {title} {\bibinfo {title} {{The Thermal Conductivity
  of the Intermediate State in Superconductors}},\ }\href
  {http://www.jetp.ras.ru/cgi-bin/e/index/e/19/5/p1228?a=list} {\bibfield
  {journal} {\bibinfo  {journal} {J. Exp. Theor. Phys.}\ }\textbf {\bibinfo
  {volume} {19}},\ \bibinfo {pages} {1228} (\bibinfo {year}
  {1964})}\BibitemShut {NoStop}%
\bibitem [{\citenamefont {Flensberg}\ \emph {et~al.}(1988)\citenamefont
  {Flensberg}, \citenamefont {Hansen},\ and\ \citenamefont
  {Octavio}}]{Flensberg1988}%
  \BibitemOpen
  \bibfield  {author} {\bibinfo {author} {\bibfnamefont {K.}~\bibnamefont
  {Flensberg}}, \bibinfo {author} {\bibfnamefont {J.~B.}\ \bibnamefont
  {Hansen}},\ and\ \bibinfo {author} {\bibfnamefont {M.}~\bibnamefont
  {Octavio}},\ }\bibfield  {title} {\bibinfo {title} {{Subharmonic energy-gap
  structure in superconducting weak links}},\ }\href
  {https://doi.org/10.1103/PhysRevB.38.8707} {\bibfield  {journal} {\bibinfo
  {journal} {Phys. Rev. B}\ }\textbf {\bibinfo {volume} {38}},\ \bibinfo
  {pages} {8707} (\bibinfo {year} {1988})}\BibitemShut {NoStop}%
\bibitem [{\citenamefont {Grifoni}\ and\ \citenamefont
  {{H\"{a}nggi}}(1998)}]{Grifoni1998}%
  \BibitemOpen
  \bibfield  {author} {\bibinfo {author} {\bibfnamefont {M.}~\bibnamefont
  {Grifoni}}\ and\ \bibinfo {author} {\bibfnamefont {P.}~\bibnamefont
  {{H\"{a}nggi}}},\ }\bibfield  {title} {\bibinfo {title} {{Driven quantum
  tunneling}},\ }\href
  {https://doi.org/https://doi.org/10.1016/S0370-1573(98)00022-2} {\bibfield
  {journal} {\bibinfo  {journal} {Phys. Rep.}\ }\textbf {\bibinfo {volume}
  {304}},\ \bibinfo {pages} {229} (\bibinfo {year} {1998})}\BibitemShut
  {NoStop}%
\bibitem [{\citenamefont {Platero}\ and\ \citenamefont
  {Aguado}(2004)}]{Platero2004}%
  \BibitemOpen
  \bibfield  {author} {\bibinfo {author} {\bibfnamefont {G.}~\bibnamefont
  {Platero}}\ and\ \bibinfo {author} {\bibfnamefont {R.}~\bibnamefont
  {Aguado}},\ }\bibfield  {title} {\bibinfo {title} {{Photon-assisted transport
  in semiconductor nanostructures}},\ }\href
  {https://doi.org/10.1016/j.physrep.2004.01.004} {\bibfield  {journal}
  {\bibinfo  {journal} {Phys. Rep.}\ }\textbf {\bibinfo {volume} {395}},\
  \bibinfo {pages} {1} (\bibinfo {year} {2004})}\BibitemShut {NoStop}%
\bibitem [{\citenamefont {Shirley}(1965)}]{Shirley1965}%
  \BibitemOpen
  \bibfield  {author} {\bibinfo {author} {\bibfnamefont {J.~H.}\ \bibnamefont
  {Shirley}},\ }\bibfield  {title} {\bibinfo {title} {{Solution of the
  {Schr\"odinger} Equation with a Hamiltonian Periodic in Time}},\ }\href
  {https://doi.org/10.1103/PhysRev.138.B979} {\bibfield  {journal} {\bibinfo
  {journal} {Phys. Rev.}\ }\textbf {\bibinfo {volume} {138}},\ \bibinfo {pages}
  {B979} (\bibinfo {year} {1965})}\BibitemShut {NoStop}%
\bibitem [{\citenamefont {Oka}\ and\ \citenamefont {Kitamura}(2019)}]{Oka2019}%
  \BibitemOpen
  \bibfield  {author} {\bibinfo {author} {\bibfnamefont {T.}~\bibnamefont
  {Oka}}\ and\ \bibinfo {author} {\bibfnamefont {S.}~\bibnamefont {Kitamura}},\
  }\bibfield  {title} {\bibinfo {title} {{Floquet Engineering of Quantum
  Materials}},\ }\href
  {https://doi.org/10.1146/annurev-conmatphys-031218-013423} {\bibfield
  {journal} {\bibinfo  {journal} {Annu. Rev. Condens. Matter Phys.}\ }\textbf
  {\bibinfo {volume} {10}},\ \bibinfo {pages} {387} (\bibinfo {year}
  {2019})}\BibitemShut {NoStop}%
\bibitem [{\citenamefont {Tien}\ and\ \citenamefont {Gordon}(1963)}]{Tien1963}%
  \BibitemOpen
  \bibfield  {author} {\bibinfo {author} {\bibfnamefont {P.~K.}\ \bibnamefont
  {Tien}}\ and\ \bibinfo {author} {\bibfnamefont {J.~P.}\ \bibnamefont
  {Gordon}},\ }\bibfield  {title} {\bibinfo {title} {{Multiphoton process
  observed in the interaction of microwave fields with the tunneling between
  superconductor films}},\ }\href {https://doi.org/10.1103/PhysRev.129.647}
  {\bibfield  {journal} {\bibinfo  {journal} {Phys. Rev.}\ }\textbf {\bibinfo
  {volume} {129}},\ \bibinfo {pages} {647} (\bibinfo {year}
  {1963})}\BibitemShut {NoStop}%
\bibitem [{\citenamefont {Kouwenhoven}\ \emph
  {et~al.}(1994{\natexlab{a}})\citenamefont {Kouwenhoven}, \citenamefont
  {Jauhar}, \citenamefont {Orenstein}, \citenamefont {McEuen}, \citenamefont
  {Nagamune}, \citenamefont {Motohisa},\ and\ \citenamefont
  {Sakaki}}]{Kouwenhoven1994}%
  \BibitemOpen
  \bibfield  {author} {\bibinfo {author} {\bibfnamefont {L.~P.}\ \bibnamefont
  {Kouwenhoven}}, \bibinfo {author} {\bibfnamefont {S.}~\bibnamefont {Jauhar}},
  \bibinfo {author} {\bibfnamefont {J.}~\bibnamefont {Orenstein}}, \bibinfo
  {author} {\bibfnamefont {P.~L.}\ \bibnamefont {McEuen}}, \bibinfo {author}
  {\bibfnamefont {Y.}~\bibnamefont {Nagamune}}, \bibinfo {author}
  {\bibfnamefont {J.}~\bibnamefont {Motohisa}},\ and\ \bibinfo {author}
  {\bibfnamefont {H.}~\bibnamefont {Sakaki}},\ }\bibfield  {title} {\bibinfo
  {title} {{Observation of Photon-Assisted Tunneling through a Quantum Dot}},\
  }\href {https://doi.org/10.1103/PhysRevLett.73.3443} {\bibfield  {journal}
  {\bibinfo  {journal} {Phys. Rev. Lett.}\ }\textbf {\bibinfo {volume} {73}},\
  \bibinfo {pages} {3443} (\bibinfo {year} {1994}{\natexlab{a}})}\BibitemShut
  {NoStop}%
\bibitem [{\citenamefont {Kouwenhoven}\ \emph
  {et~al.}(1994{\natexlab{b}})\citenamefont {Kouwenhoven}, \citenamefont
  {Jauhar}, \citenamefont {McCormick}, \citenamefont {Dixon}, \citenamefont
  {McEuen}, \citenamefont {Nazarov}, \citenamefont {van~der Vaart},\ and\
  \citenamefont {Foxon}}]{Kouwenhoven1994b}%
  \BibitemOpen
  \bibfield  {author} {\bibinfo {author} {\bibfnamefont {L.~P.}\ \bibnamefont
  {Kouwenhoven}}, \bibinfo {author} {\bibfnamefont {S.}~\bibnamefont {Jauhar}},
  \bibinfo {author} {\bibfnamefont {K.}~\bibnamefont {McCormick}}, \bibinfo
  {author} {\bibfnamefont {D.}~\bibnamefont {Dixon}}, \bibinfo {author}
  {\bibfnamefont {P.~L.}\ \bibnamefont {McEuen}}, \bibinfo {author}
  {\bibfnamefont {Y.~V.}\ \bibnamefont {Nazarov}}, \bibinfo {author}
  {\bibfnamefont {N.~C.}\ \bibnamefont {van~der Vaart}},\ and\ \bibinfo
  {author} {\bibfnamefont {C.~T.}\ \bibnamefont {Foxon}},\ }\bibfield  {title}
  {\bibinfo {title} {{Photon-assisted tunneling through a quantum dot}},\
  }\href {https://doi.org/10.1103/PhysRevB.50.2019} {\bibfield  {journal}
  {\bibinfo  {journal} {Phys. Rev. B}\ }\textbf {\bibinfo {volume} {50}},\
  \bibinfo {pages} {2019} (\bibinfo {year} {1994}{\natexlab{b}})}\BibitemShut
  {NoStop}%
\bibitem [{\citenamefont {Roychowdhury}\ \emph {et~al.}(2015)\citenamefont
  {Roychowdhury}, \citenamefont {Dreyer}, \citenamefont {Anderson},
  \citenamefont {Lobb},\ and\ \citenamefont {Wellstood}}]{Roychowdhury2015}%
  \BibitemOpen
  \bibfield  {author} {\bibinfo {author} {\bibfnamefont {A.}~\bibnamefont
  {Roychowdhury}}, \bibinfo {author} {\bibfnamefont {M.}~\bibnamefont
  {Dreyer}}, \bibinfo {author} {\bibfnamefont {J.~R.}\ \bibnamefont
  {Anderson}}, \bibinfo {author} {\bibfnamefont {C.~J.}\ \bibnamefont {Lobb}},\
  and\ \bibinfo {author} {\bibfnamefont {F.~C.}\ \bibnamefont {Wellstood}},\
  }\bibfield  {title} {\bibinfo {title} {{Microwave Photon-Assisted Incoherent
  Cooper-Pair Tunneling in a Josephson STM}},\ }\href
  {https://doi.org/10.1103/PhysRevApplied.4.034011} {\bibfield  {journal}
  {\bibinfo  {journal} {Phys. Rev. Appl.}\ }\textbf {\bibinfo {volume} {4}},\
  \bibinfo {pages} {034011} (\bibinfo {year} {2015})}\BibitemShut {NoStop}%
\bibitem [{\citenamefont {Kot}\ \emph {et~al.}(2020)\citenamefont {Kot},
  \citenamefont {Drost}, \citenamefont {Uhl}, \citenamefont {Ankerhold},
  \citenamefont {Cuevas},\ and\ \citenamefont {Ast}}]{Kot2020}%
  \BibitemOpen
  \bibfield  {author} {\bibinfo {author} {\bibfnamefont {P.}~\bibnamefont
  {Kot}}, \bibinfo {author} {\bibfnamefont {R.}~\bibnamefont {Drost}}, \bibinfo
  {author} {\bibfnamefont {M.}~\bibnamefont {Uhl}}, \bibinfo {author}
  {\bibfnamefont {J.}~\bibnamefont {Ankerhold}}, \bibinfo {author}
  {\bibfnamefont {J.~C.}\ \bibnamefont {Cuevas}},\ and\ \bibinfo {author}
  {\bibfnamefont {C.~R.}\ \bibnamefont {Ast}},\ }\bibfield  {title} {\bibinfo
  {title} {{Microwave-assisted tunneling and interference effects in
  superconducting junctions under fast driving signals}},\ }\href
  {https://doi.org/10.1103/PhysRevB.101.134507} {\bibfield  {journal} {\bibinfo
   {journal} {Phys. Rev. B}\ }\textbf {\bibinfo {volume} {101}},\ \bibinfo
  {pages} {134507} (\bibinfo {year} {2020})}\BibitemShut {NoStop}%
\bibitem [{\citenamefont {Peters}\ \emph {et~al.}(2020)\citenamefont {Peters},
  \citenamefont {Bogdanoff}, \citenamefont {González}, \citenamefont
  {Melischek}, \citenamefont {Simon}, \citenamefont {Reecht}, \citenamefont
  {Winkelmann}, \citenamefont {von Oppen},\ and\ \citenamefont
  {Franke}}]{Peters2020}%
  \BibitemOpen
  \bibfield  {author} {\bibinfo {author} {\bibfnamefont {O.}~\bibnamefont
  {Peters}}, \bibinfo {author} {\bibfnamefont {N.}~\bibnamefont {Bogdanoff}},
  \bibinfo {author} {\bibfnamefont {S.~A.}\ \bibnamefont {González}}, \bibinfo
  {author} {\bibfnamefont {L.}~\bibnamefont {Melischek}}, \bibinfo {author}
  {\bibfnamefont {J.~R.}\ \bibnamefont {Simon}}, \bibinfo {author}
  {\bibfnamefont {G.}~\bibnamefont {Reecht}}, \bibinfo {author} {\bibfnamefont
  {C.~B.}\ \bibnamefont {Winkelmann}}, \bibinfo {author} {\bibfnamefont
  {F.}~\bibnamefont {von Oppen}},\ and\ \bibinfo {author} {\bibfnamefont
  {K.~J.}\ \bibnamefont {Franke}},\ }\bibfield  {title} {\bibinfo {title}
  {{Resonant Andreev reflections probed by photon-assisted tunnelling at the
  atomic scale}},\ }\href {https://doi.org/10.1038/s41567-020-0972-z}
  {\bibfield  {journal} {\bibinfo  {journal} {Nat. Phys.}\ }\textbf {\bibinfo
  {volume} {16}},\ \bibinfo {pages} {1222} (\bibinfo {year}
  {2020})}\BibitemShut {NoStop}%
\bibitem [{\citenamefont {Shapiro}(1963)}]{Shapiro1963}%
  \BibitemOpen
  \bibfield  {author} {\bibinfo {author} {\bibfnamefont {S.}~\bibnamefont
  {Shapiro}},\ }\bibfield  {title} {\bibinfo {title} {{Josephson currents in
  superconducting tunneling: The effect of microwaves and other
  observations}},\ }\href {https://doi.org/10.1103/PhysRevLett.11.80}
  {\bibfield  {journal} {\bibinfo  {journal} {Phys. Rev. Lett.}\ }\textbf
  {\bibinfo {volume} {11}},\ \bibinfo {pages} {80} (\bibinfo {year}
  {1963})}\BibitemShut {NoStop}%
\bibitem [{\citenamefont {Dubos}\ \emph {et~al.}(2001)\citenamefont {Dubos},
  \citenamefont {Courtois}, \citenamefont {Buisson},\ and\ \citenamefont
  {Pannetier}}]{Dubos2001}%
  \BibitemOpen
  \bibfield  {author} {\bibinfo {author} {\bibfnamefont {P.}~\bibnamefont
  {Dubos}}, \bibinfo {author} {\bibfnamefont {H.}~\bibnamefont {Courtois}},
  \bibinfo {author} {\bibfnamefont {O.}~\bibnamefont {Buisson}},\ and\ \bibinfo
  {author} {\bibfnamefont {B.}~\bibnamefont {Pannetier}},\ }\bibfield  {title}
  {\bibinfo {title} {{Coherent Low-Energy Charge Transport in a Diffusive S-N-S
  Junction}},\ }\href {https://doi.org/10.1103/PhysRevLett.87.206801}
  {\bibfield  {journal} {\bibinfo  {journal} {Phys. Rev. Lett.}\ }\textbf
  {\bibinfo {volume} {87}},\ \bibinfo {pages} {206801} (\bibinfo {year}
  {2001})}\BibitemShut {NoStop}%
\bibitem [{\citenamefont {Cuevas}\ \emph {et~al.}(2002)\citenamefont {Cuevas},
  \citenamefont {Heurich}, \citenamefont {Mart\'{\i}n-Rodero}, \citenamefont
  {Levy~Yeyati},\ and\ \citenamefont {Sch\"on}}]{Cuevas2002}%
  \BibitemOpen
  \bibfield  {author} {\bibinfo {author} {\bibfnamefont {J.~C.}\ \bibnamefont
  {Cuevas}}, \bibinfo {author} {\bibfnamefont {J.}~\bibnamefont {Heurich}},
  \bibinfo {author} {\bibfnamefont {A.}~\bibnamefont {Mart\'{\i}n-Rodero}},
  \bibinfo {author} {\bibfnamefont {A.}~\bibnamefont {Levy~Yeyati}},\ and\
  \bibinfo {author} {\bibfnamefont {G.}~\bibnamefont {Sch\"on}},\ }\bibfield
  {title} {\bibinfo {title} {{Subharmonic Shapiro Steps and Assisted Tunneling
  in Superconducting Point Contacts}},\ }\href
  {https://doi.org/10.1103/PhysRevLett.88.157001} {\bibfield  {journal}
  {\bibinfo  {journal} {Phys. Rev. Lett.}\ }\textbf {\bibinfo {volume} {88}},\
  \bibinfo {pages} {157001} (\bibinfo {year} {2002})}\BibitemShut {NoStop}%
\bibitem [{\citenamefont {Rokhinson}\ and\ \citenamefont
  {Furdyna}(2012)}]{Rokhinson2012}%
  \BibitemOpen
  \bibfield  {author} {\bibinfo {author} {\bibfnamefont {X.}~\bibnamefont
  {Rokhinson}, \bibfnamefont {L.~P.and~Liu}}\ and\ \bibinfo {author}
  {\bibfnamefont {J.~K.}\ \bibnamefont {Furdyna}},\ }\bibfield  {title}
  {\bibinfo {title} {{The fractional a.c. Josephson effect in a
  semiconductor{\textendash}superconductor nanowire as a signature of Majorana
  particles}},\ }\href {https://doi.org/10.1038/nphys2429} {\bibfield
  {journal} {\bibinfo  {journal} {Nat. Phys.}\ }\textbf {\bibinfo {volume}
  {8}},\ \bibinfo {pages} {795} (\bibinfo {year} {2012})}\BibitemShut {NoStop}%
\bibitem [{\citenamefont {Wiedenmann}\ \emph {et~al.}(2016)\citenamefont
  {Wiedenmann}, \citenamefont {Bocquillon}, \citenamefont {Deacon},
  \citenamefont {Hartinger}, \citenamefont {Herrmann}, \citenamefont
  {Klapwijk}, \citenamefont {Maier}, \citenamefont {Ames}, \citenamefont
  {{{Br{\"{u}}ne}}}, \citenamefont {Gould}, \citenamefont {Oiwa}, \citenamefont
  {Ishibashi}, \citenamefont {Tarucha}, \citenamefont {Buhmann},\ and\
  \citenamefont {Molenkamp}}]{Wiedenmann2016}%
  \BibitemOpen
  \bibfield  {author} {\bibinfo {author} {\bibfnamefont {J.}~\bibnamefont
  {Wiedenmann}}, \bibinfo {author} {\bibfnamefont {E.}~\bibnamefont
  {Bocquillon}}, \bibinfo {author} {\bibfnamefont {R.~S.}\ \bibnamefont
  {Deacon}}, \bibinfo {author} {\bibfnamefont {S.}~\bibnamefont {Hartinger}},
  \bibinfo {author} {\bibfnamefont {O.}~\bibnamefont {Herrmann}}, \bibinfo
  {author} {\bibfnamefont {T.~M.}\ \bibnamefont {Klapwijk}}, \bibinfo {author}
  {\bibfnamefont {L.}~\bibnamefont {Maier}}, \bibinfo {author} {\bibfnamefont
  {C.}~\bibnamefont {Ames}}, \bibinfo {author} {\bibfnamefont {C.}~\bibnamefont
  {{{Br{\"{u}}ne}}}}, \bibinfo {author} {\bibfnamefont {C.}~\bibnamefont
  {Gould}}, \bibinfo {author} {\bibfnamefont {A.}~\bibnamefont {Oiwa}},
  \bibinfo {author} {\bibfnamefont {K.}~\bibnamefont {Ishibashi}}, \bibinfo
  {author} {\bibfnamefont {S.}~\bibnamefont {Tarucha}}, \bibinfo {author}
  {\bibfnamefont {H.}~\bibnamefont {Buhmann}},\ and\ \bibinfo {author}
  {\bibfnamefont {L.~W.}\ \bibnamefont {Molenkamp}},\ }\bibfield  {title}
  {\bibinfo {title} {{4$\pi$-periodic Josephson supercurrent in HgTe-based
  topological Josephson junctions}},\ }\href
  {https://doi.org/10.1038/ncomms10303} {\bibfield  {journal} {\bibinfo
  {journal} {Nat. Commun.}\ }\textbf {\bibinfo {volume} {7}},\ \bibinfo {pages}
  {10303} (\bibinfo {year} {2016})}\BibitemShut {NoStop}%
\bibitem [{\citenamefont {Laroche}\ \emph {et~al.}(2019)\citenamefont
  {Laroche}, \citenamefont {Bouman}, \citenamefont {van Woerkom}, \citenamefont
  {Proutski}, \citenamefont {Murthy}, \citenamefont {Pikulin}, \citenamefont
  {Nayak}, \citenamefont {van Gulik}, \citenamefont {{Nyg{\aa}rd}},
  \citenamefont {Krogstrup}, \citenamefont {Kouwenhoven},\ and\ \citenamefont
  {Geresdi}}]{Laroche2019}%
  \BibitemOpen
  \bibfield  {author} {\bibinfo {author} {\bibfnamefont {D.}~\bibnamefont
  {Laroche}}, \bibinfo {author} {\bibfnamefont {D.}~\bibnamefont {Bouman}},
  \bibinfo {author} {\bibfnamefont {D.~J.}\ \bibnamefont {van Woerkom}},
  \bibinfo {author} {\bibfnamefont {A.}~\bibnamefont {Proutski}}, \bibinfo
  {author} {\bibfnamefont {C.}~\bibnamefont {Murthy}}, \bibinfo {author}
  {\bibfnamefont {D.~I.}\ \bibnamefont {Pikulin}}, \bibinfo {author}
  {\bibfnamefont {C.}~\bibnamefont {Nayak}}, \bibinfo {author} {\bibfnamefont
  {R.~J.~J.}\ \bibnamefont {van Gulik}}, \bibinfo {author} {\bibfnamefont
  {J.}~\bibnamefont {{Nyg{\aa}rd}}}, \bibinfo {author} {\bibfnamefont
  {P.}~\bibnamefont {Krogstrup}}, \bibinfo {author} {\bibfnamefont {L.~P.}\
  \bibnamefont {Kouwenhoven}},\ and\ \bibinfo {author} {\bibfnamefont
  {A.}~\bibnamefont {Geresdi}},\ }\bibfield  {title} {\bibinfo {title}
  {{Observation of the $4\ensuremath{\pi}$-periodic Josephson effect in indium
  arsenide nanowires}},\ }\href {https://doi.org/10.1038/s41467-018-08161-2}
  {\bibfield  {journal} {\bibinfo  {journal} {Nat. Comm.}\ }\textbf {\bibinfo
  {volume} {10}},\ \bibinfo {pages} {245} (\bibinfo {year} {2019})}\BibitemShut
  {NoStop}%
\bibitem [{\citenamefont {Fischer}\ \emph {et~al.}(2022)\citenamefont
  {Fischer}, \citenamefont {{Pic\'o-Cort\'es}}, \citenamefont {Himmler},
  \citenamefont {Platero}, \citenamefont {Grifoni}, \citenamefont {Kozlov},
  \citenamefont {Mikhailov}, \citenamefont {Dvoretsky}, \citenamefont
  {Strunk},\ and\ \citenamefont {Weiss}}]{Fischer2022}%
  \BibitemOpen
  \bibfield  {author} {\bibinfo {author} {\bibfnamefont {R.}~\bibnamefont
  {Fischer}}, \bibinfo {author} {\bibfnamefont {J.}~\bibnamefont
  {{Pic\'o-Cort\'es}}}, \bibinfo {author} {\bibfnamefont {W.}~\bibnamefont
  {Himmler}}, \bibinfo {author} {\bibfnamefont {G.}~\bibnamefont {Platero}},
  \bibinfo {author} {\bibfnamefont {M.}~\bibnamefont {Grifoni}}, \bibinfo
  {author} {\bibfnamefont {D.~A.}\ \bibnamefont {Kozlov}}, \bibinfo {author}
  {\bibfnamefont {N.~N.}\ \bibnamefont {Mikhailov}}, \bibinfo {author}
  {\bibfnamefont {S.~A.}\ \bibnamefont {Dvoretsky}}, \bibinfo {author}
  {\bibfnamefont {C.}~\bibnamefont {Strunk}},\ and\ \bibinfo {author}
  {\bibfnamefont {D.}~\bibnamefont {Weiss}},\ }\bibfield  {title} {\bibinfo
  {title} {{$4\pi$-periodic supercurrent tuned by an axial magnetic flux in
  topological insulator nanowires}},\ }\href
  {https://doi.org/10.1103/PhysRevResearch.4.013087} {\bibfield  {journal}
  {\bibinfo  {journal} {Phys. Rev. Res.}\ }\textbf {\bibinfo {volume} {4}},\
  \bibinfo {pages} {013087} (\bibinfo {year} {2022})}\BibitemShut {NoStop}%
\bibitem [{\citenamefont {Clerk}\ and\ \citenamefont
  {Ambegaokar}(2000)}]{Clerk2000}%
  \BibitemOpen
  \bibfield  {author} {\bibinfo {author} {\bibfnamefont {A.~A.}\ \bibnamefont
  {Clerk}}\ and\ \bibinfo {author} {\bibfnamefont {V.}~\bibnamefont
  {Ambegaokar}},\ }\bibfield  {title} {\bibinfo {title} {{Loss of
  $\pi$-junction behavior in an interacting impurity Josephson junction}},\
  }\href {https://doi.org/10.1103/PhysRevB.61.9109} {\bibfield  {journal}
  {\bibinfo  {journal} {Phys. Rev. B}\ }\textbf {\bibinfo {volume} {61}},\
  \bibinfo {pages} {9109} (\bibinfo {year} {2000})}\BibitemShut {NoStop}%
\bibitem [{\citenamefont {{J{\o}rgensen}}\ \emph {et~al.}(2007)\citenamefont
  {{J{\o}rgensen}}, \citenamefont {{Novotn{\'{y}}}}, \citenamefont
  {{Grove-Rasmussen}}, \citenamefont {Flensberg},\ and\ \citenamefont
  {Lindelof}}]{Jrgensen2007}%
  \BibitemOpen
  \bibfield  {author} {\bibinfo {author} {\bibfnamefont {H.~I.}\ \bibnamefont
  {{J{\o}rgensen}}}, \bibinfo {author} {\bibfnamefont {T.}~\bibnamefont
  {{Novotn{\'{y}}}}}, \bibinfo {author} {\bibfnamefont {K.}~\bibnamefont
  {{Grove-Rasmussen}}}, \bibinfo {author} {\bibfnamefont {K.}~\bibnamefont
  {Flensberg}},\ and\ \bibinfo {author} {\bibfnamefont {P.~E.}\ \bibnamefont
  {Lindelof}},\ }\bibfield  {title} {\bibinfo {title} {{Critical Current
  0-$\pi$ Transition in Designed Josephson Quantum Dot Junctions}},\ }\href
  {https://doi.org/10.1021/nl071152w} {\bibfield  {journal} {\bibinfo
  {journal} {Nano Lett.}\ }\textbf {\bibinfo {volume} {7}},\ \bibinfo {pages}
  {2441} (\bibinfo {year} {2007})}\BibitemShut {NoStop}%
\bibitem [{\citenamefont {{Mart{\'{\i}}n-Rodero}}\ and\ \citenamefont
  {Yeyati}(2011)}]{MartinRodero2011}%
  \BibitemOpen
  \bibfield  {author} {\bibinfo {author} {\bibfnamefont {A.}~\bibnamefont
  {{Mart{\'{\i}}n-Rodero}}}\ and\ \bibinfo {author} {\bibfnamefont {A.~L.}\
  \bibnamefont {Yeyati}},\ }\bibfield  {title} {\bibinfo {title} {{Josephson
  and Andreev transport through quantum dots}},\ }\href
  {https://doi.org/10.1080/00018732.2011.624266} {\bibfield  {journal}
  {\bibinfo  {journal} {Adv. Phys}\ }\textbf {\bibinfo {volume} {60}},\
  \bibinfo {pages} {899} (\bibinfo {year} {2011})}\BibitemShut {NoStop}%
\bibitem [{\citenamefont {Pala}\ \emph {et~al.}(2007)\citenamefont {Pala},
  \citenamefont {Governale},\ and\ \citenamefont {König}}]{Pala2007}%
  \BibitemOpen
  \bibfield  {author} {\bibinfo {author} {\bibfnamefont {M.~G.}\ \bibnamefont
  {Pala}}, \bibinfo {author} {\bibfnamefont {M.}~\bibnamefont {Governale}},\
  and\ \bibinfo {author} {\bibfnamefont {J.}~\bibnamefont {König}},\
  }\bibfield  {title} {\bibinfo {title} {{Nonequilibrium Josephson and Andreev
  current through interacting quantum dots}},\ }\href
  {https://doi.org/10.1088/1367-2630/9/8/278} {\bibfield  {journal} {\bibinfo
  {journal} {New J. Phys.}\ }\textbf {\bibinfo {volume} {9}},\ \bibinfo {pages}
  {278} (\bibinfo {year} {2007})}\BibitemShut {NoStop}%
\bibitem [{\citenamefont {Governale}\ \emph {et~al.}(2008)\citenamefont
  {Governale}, \citenamefont {Pala},\ and\ \citenamefont
  {König}}]{Governale2008}%
  \BibitemOpen
  \bibfield  {author} {\bibinfo {author} {\bibfnamefont {M.}~\bibnamefont
  {Governale}}, \bibinfo {author} {\bibfnamefont {M.~G.}\ \bibnamefont
  {Pala}},\ and\ \bibinfo {author} {\bibfnamefont {J.}~\bibnamefont {König}},\
  }\bibfield  {title} {\bibinfo {title} {{Real-time diagrammatic approach to
  transport through interacting quantum dots with normal and superconducting
  leads}},\ }\href {https://doi.org/10.1103/PhysRevB.77.134513} {\bibfield
  {journal} {\bibinfo  {journal} {Phys. Rev. B}\ }\textbf {\bibinfo {volume}
  {77}},\ \bibinfo {pages} {134513} (\bibinfo {year} {2008})}\BibitemShut
  {NoStop}%
\bibitem [{\citenamefont {Hiltscher}\ \emph {et~al.}(2012)\citenamefont
  {Hiltscher}, \citenamefont {Governale},\ and\ \citenamefont
  {König}}]{Hiltscher2012}%
  \BibitemOpen
  \bibfield  {author} {\bibinfo {author} {\bibfnamefont {B.}~\bibnamefont
  {Hiltscher}}, \bibinfo {author} {\bibfnamefont {M.}~\bibnamefont
  {Governale}},\ and\ \bibinfo {author} {\bibfnamefont {J.}~\bibnamefont
  {König}},\ }\bibfield  {title} {\bibinfo {title} {{Ac Josephson transport
  through interacting quantum dots}},\ }\href
  {https://doi.org/10.1103/PhysRevB.86.235427} {\bibfield  {journal} {\bibinfo
  {journal} {Phys. Rev. B}\ }\textbf {\bibinfo {volume} {86}},\ \bibinfo
  {pages} {235427} (\bibinfo {year} {2012})}\BibitemShut {NoStop}%
\bibitem [{\citenamefont {Vecino}\ \emph {et~al.}(2003)\citenamefont {Vecino},
  \citenamefont {Martin-Rodero},\ and\ \citenamefont {Yeyati}}]{Vecino2003}%
  \BibitemOpen
  \bibfield  {author} {\bibinfo {author} {\bibfnamefont {E.}~\bibnamefont
  {Vecino}}, \bibinfo {author} {\bibfnamefont {A.}~\bibnamefont
  {Martin-Rodero}},\ and\ \bibinfo {author} {\bibfnamefont {A.~L.}\
  \bibnamefont {Yeyati}},\ }\bibfield  {title} {\bibinfo {title} {{Josephson
  current through a correlated quantum level}},\ }\href
  {https://doi.org/10.1103/PhysRevB.68.035105} {\bibfield  {journal} {\bibinfo
  {journal} {Phys. Rev. B}\ }\textbf {\bibinfo {volume} {68}},\ \bibinfo
  {pages} {035105} (\bibinfo {year} {2003})}\BibitemShut {NoStop}%
\bibitem [{\citenamefont {{van Dam}}\ \emph {et~al.}(2006)\citenamefont {{van
  Dam}}, \citenamefont {Nazarov}, \citenamefont {Bakkers}, \citenamefont {{De
  Franceschi}},\ and\ \citenamefont {Kouwenhoven}}]{vanDam2006}%
  \BibitemOpen
  \bibfield  {author} {\bibinfo {author} {\bibfnamefont {J.~A.}\ \bibnamefont
  {{van Dam}}}, \bibinfo {author} {\bibfnamefont {Y.~V.}\ \bibnamefont
  {Nazarov}}, \bibinfo {author} {\bibfnamefont {E.~P. A.~M.}\ \bibnamefont
  {Bakkers}}, \bibinfo {author} {\bibfnamefont {S.}~\bibnamefont {{De
  Franceschi}}},\ and\ \bibinfo {author} {\bibfnamefont {L.~P.}\ \bibnamefont
  {Kouwenhoven}},\ }\bibfield  {title} {\bibinfo {title} {{Supercurrent
  reversal in quantum dots}},\ }\href {https://doi.org/10.1038/nature05018}
  {\bibfield  {journal} {\bibinfo  {journal} {Nature}\ }\textbf {\bibinfo
  {volume} {442}},\ \bibinfo {pages} {667} (\bibinfo {year}
  {2006})}\BibitemShut {NoStop}%
\bibitem [{\citenamefont {Buitelaar}\ \emph {et~al.}(2003)\citenamefont
  {Buitelaar}, \citenamefont {Belzig}, \citenamefont {Nussbaumer},
  \citenamefont {Babić}, \citenamefont {Bruder},\ and\ \citenamefont
  {Schönenberger}}]{Buitelaar2003}%
  \BibitemOpen
  \bibfield  {author} {\bibinfo {author} {\bibfnamefont {M.~R.}\ \bibnamefont
  {Buitelaar}}, \bibinfo {author} {\bibfnamefont {W.}~\bibnamefont {Belzig}},
  \bibinfo {author} {\bibfnamefont {T.}~\bibnamefont {Nussbaumer}}, \bibinfo
  {author} {\bibfnamefont {B.}~\bibnamefont {Babić}}, \bibinfo {author}
  {\bibfnamefont {C.}~\bibnamefont {Bruder}},\ and\ \bibinfo {author}
  {\bibfnamefont {C.}~\bibnamefont {Schönenberger}},\ }\bibfield  {title}
  {\bibinfo {title} {{Multiple Andreev Reflections in a Carbon Nanotube Quantum
  Dot}},\ }\href {https://doi.org/10.1103/PhysRevLett.91.057005} {\bibfield
  {journal} {\bibinfo  {journal} {Phys. Rev. Lett.}\ }\textbf {\bibinfo
  {volume} {91}},\ \bibinfo {pages} {057005} (\bibinfo {year}
  {2003})}\BibitemShut {NoStop}%
\bibitem [{\citenamefont {Vecino}\ \emph {et~al.}(2004)\citenamefont {Vecino},
  \citenamefont {Buitelaar}, \citenamefont {{Marti\'{n}-Roder}}, \citenamefont
  {{Sch\"onberger}},\ and\ \citenamefont {{Levy Yeyati}}}]{Vecino2004}%
  \BibitemOpen
  \bibfield  {author} {\bibinfo {author} {\bibfnamefont {E.}~\bibnamefont
  {Vecino}}, \bibinfo {author} {\bibfnamefont {M.~R.}\ \bibnamefont
  {Buitelaar}}, \bibinfo {author} {\bibfnamefont {A.}~\bibnamefont
  {{Marti\'{n}-Roder}}}, \bibinfo {author} {\bibfnamefont {C.}~\bibnamefont
  {{Sch\"onberger}}},\ and\ \bibinfo {author} {\bibfnamefont {A.}~\bibnamefont
  {{Levy Yeyati}}},\ }\bibfield  {title} {\bibinfo {title} {{Conductance
  properties of nanotubes coupled to superconducting leads: signatures of
  Andreev states dynamics}},\ }\href
  {https://doi.org/https://doi.org/10.1016/j.ssc.2004.05.031} {\bibfield
  {journal} {\bibinfo  {journal} {Solid State Commun.}\ }\textbf {\bibinfo
  {volume} {131}},\ \bibinfo {pages} {625} (\bibinfo {year}
  {2004})}\BibitemShut {NoStop}%
\bibitem [{\citenamefont {Cleuziou}\ \emph {et~al.}(2006)\citenamefont
  {Cleuziou}, \citenamefont {Wernsdorfer}, \citenamefont {Bouchiat},
  \citenamefont {{Ondar{\c{c}}uhu}},\ and\ \citenamefont
  {Monthioux}}]{Cleuziou2006}%
  \BibitemOpen
  \bibfield  {author} {\bibinfo {author} {\bibfnamefont {J.}~\bibnamefont
  {Cleuziou}}, \bibinfo {author} {\bibfnamefont {W.}~\bibnamefont
  {Wernsdorfer}}, \bibinfo {author} {\bibfnamefont {V.}~\bibnamefont
  {Bouchiat}}, \bibinfo {author} {\bibfnamefont {T.}~\bibnamefont
  {{Ondar{\c{c}}uhu}}},\ and\ \bibinfo {author} {\bibfnamefont
  {M.}~\bibnamefont {Monthioux}},\ }\bibfield  {title} {\bibinfo {title}
  {{Carbon nanotube superconducting quantum interference device}},\ }\href
  {https://doi.org/10.1038/nnano.2006.54} {\bibfield  {journal} {\bibinfo
  {journal} {Nat. Nanotechnol.}\ }\textbf {\bibinfo {volume} {1}},\ \bibinfo
  {pages} {53} (\bibinfo {year} {2006})}\BibitemShut {NoStop}%
\bibitem [{\citenamefont {Eichler}\ \emph {et~al.}(2007)\citenamefont
  {Eichler}, \citenamefont {Weiss}, \citenamefont {Oberholzer}, \citenamefont
  {{Sch\"onenberger}}, \citenamefont {Levy~Yeyati}, \citenamefont {Cuevas},\
  and\ \citenamefont {{Mart\'{\i}n-Rodero}}}]{Eichler2007}%
  \BibitemOpen
  \bibfield  {author} {\bibinfo {author} {\bibfnamefont {A.}~\bibnamefont
  {Eichler}}, \bibinfo {author} {\bibfnamefont {M.}~\bibnamefont {Weiss}},
  \bibinfo {author} {\bibfnamefont {S.}~\bibnamefont {Oberholzer}}, \bibinfo
  {author} {\bibfnamefont {C.}~\bibnamefont {{Sch\"onenberger}}}, \bibinfo
  {author} {\bibfnamefont {A.}~\bibnamefont {Levy~Yeyati}}, \bibinfo {author}
  {\bibfnamefont {J.~C.}\ \bibnamefont {Cuevas}},\ and\ \bibinfo {author}
  {\bibfnamefont {A.}~\bibnamefont {{Mart\'{\i}n-Rodero}}},\ }\bibfield
  {title} {\bibinfo {title} {{Even-Odd Effect in Andreev Transport through a
  Carbon Nanotube Quantum Dot}},\ }\href
  {https://doi.org/10.1103/PhysRevLett.99.126602} {\bibfield  {journal}
  {\bibinfo  {journal} {Phys. Rev. Lett.}\ }\textbf {\bibinfo {volume} {99}},\
  \bibinfo {pages} {126602} (\bibinfo {year} {2007})}\BibitemShut {NoStop}%
\bibitem [{\citenamefont {Pillet}\ \emph {et~al.}(2010)\citenamefont {Pillet},
  \citenamefont {Quay}, \citenamefont {Morfin}, \citenamefont {Bena},
  \citenamefont {Yeyati},\ and\ \citenamefont {Joyez}}]{Pillet2010}%
  \BibitemOpen
  \bibfield  {author} {\bibinfo {author} {\bibfnamefont {J.-D.}\ \bibnamefont
  {Pillet}}, \bibinfo {author} {\bibfnamefont {C.~H.~L.}\ \bibnamefont {Quay}},
  \bibinfo {author} {\bibfnamefont {P.}~\bibnamefont {Morfin}}, \bibinfo
  {author} {\bibfnamefont {C.}~\bibnamefont {Bena}}, \bibinfo {author}
  {\bibfnamefont {A.~L.}\ \bibnamefont {Yeyati}},\ and\ \bibinfo {author}
  {\bibfnamefont {P.}~\bibnamefont {Joyez}},\ }\bibfield  {title} {\bibinfo
  {title} {{Andreev bound states in supercurrent-carrying carbon nanotubes
  revealed}},\ }\href {https://doi.org/10.1038/nphys1811} {\bibfield  {journal}
  {\bibinfo  {journal} {Nat. Phys.}\ }\textbf {\bibinfo {volume} {6}},\
  \bibinfo {pages} {965} (\bibinfo {year} {2010})}\BibitemShut {NoStop}%
\bibitem [{\citenamefont {Winkelmann}\ \emph {et~al.}(2009)\citenamefont
  {Winkelmann}, \citenamefont {Roch}, \citenamefont {Wernsdorfer},
  \citenamefont {Bouchiat},\ and\ \citenamefont {Balestro}}]{Winkelmann2009}%
  \BibitemOpen
  \bibfield  {author} {\bibinfo {author} {\bibfnamefont {C.~B.}\ \bibnamefont
  {Winkelmann}}, \bibinfo {author} {\bibfnamefont {N.}~\bibnamefont {Roch}},
  \bibinfo {author} {\bibfnamefont {W.}~\bibnamefont {Wernsdorfer}}, \bibinfo
  {author} {\bibfnamefont {V.}~\bibnamefont {Bouchiat}},\ and\ \bibinfo
  {author} {\bibfnamefont {F.}~\bibnamefont {Balestro}},\ }\bibfield  {title}
  {\bibinfo {title} {{Superconductivity in a single-C60 transistor}},\ }\href
  {https://doi.org/10.1038/nphys1433} {\bibfield  {journal} {\bibinfo
  {journal} {Nat. Phys.}\ }\textbf {\bibinfo {volume} {5}},\ \bibinfo {pages}
  {876} (\bibinfo {year} {2009})}\BibitemShut {NoStop}%
\bibitem [{\citenamefont {Dirks}\ \emph {et~al.}(2011)\citenamefont {Dirks},
  \citenamefont {Hughes}, \citenamefont {Lal}, \citenamefont {Uchoa},
  \citenamefont {Chen}, \citenamefont {Chialvo}, \citenamefont {Goldbart},\
  and\ \citenamefont {Mason}}]{Dirks2011}%
  \BibitemOpen
  \bibfield  {author} {\bibinfo {author} {\bibfnamefont {T.}~\bibnamefont
  {Dirks}}, \bibinfo {author} {\bibfnamefont {T.~L.}\ \bibnamefont {Hughes}},
  \bibinfo {author} {\bibfnamefont {S.}~\bibnamefont {Lal}}, \bibinfo {author}
  {\bibfnamefont {B.}~\bibnamefont {Uchoa}}, \bibinfo {author} {\bibfnamefont
  {Y.-F.}\ \bibnamefont {Chen}}, \bibinfo {author} {\bibfnamefont
  {C.}~\bibnamefont {Chialvo}}, \bibinfo {author} {\bibfnamefont {P.~M.}\
  \bibnamefont {Goldbart}},\ and\ \bibinfo {author} {\bibfnamefont
  {N.}~\bibnamefont {Mason}},\ }\bibfield  {title} {\bibinfo {title}
  {{Transport through Andreev bound states in a graphene quantum dot}},\ }\href
  {https://doi.org/10.1038/nphys1911} {\bibfield  {journal} {\bibinfo
  {journal} {Nat. Phys.}\ }\textbf {\bibinfo {volume} {7}},\ \bibinfo {pages}
  {386} (\bibinfo {year} {2011})}\BibitemShut {NoStop}%
\bibitem [{\citenamefont {Padurariu}\ and\ \citenamefont
  {Nazarov}(2010)}]{Padurariu2010}%
  \BibitemOpen
  \bibfield  {author} {\bibinfo {author} {\bibfnamefont {C.}~\bibnamefont
  {Padurariu}}\ and\ \bibinfo {author} {\bibfnamefont {Y.~V.}\ \bibnamefont
  {Nazarov}},\ }\bibfield  {title} {\bibinfo {title} {{Theoretical proposal for
  superconducting spin qubits}},\ }\href
  {https://doi.org/10.1103/PhysRevB.81.144519} {\bibfield  {journal} {\bibinfo
  {journal} {Phys. Rev. B}\ }\textbf {\bibinfo {volume} {81}},\ \bibinfo
  {pages} {144519} (\bibinfo {year} {2010})}\BibitemShut {NoStop}%
\bibitem [{\citenamefont {Park}\ and\ \citenamefont {Yeyati}(2017)}]{Park2017}%
  \BibitemOpen
  \bibfield  {author} {\bibinfo {author} {\bibfnamefont {S.}~\bibnamefont
  {Park}}\ and\ \bibinfo {author} {\bibfnamefont {A.~L.}\ \bibnamefont
  {Yeyati}},\ }\bibfield  {title} {\bibinfo {title} {{Andreev spin qubits in
  multichannel Rashba nanowires}},\ }\href
  {https://doi.org/10.1103/PhysRevB.96.125416} {\bibfield  {journal} {\bibinfo
  {journal} {Phys. Rev. B}\ }\textbf {\bibinfo {volume} {96}},\ \bibinfo
  {pages} {125416} (\bibinfo {year} {2017})}\BibitemShut {NoStop}%
\bibitem [{\citenamefont {Pave\ifmmode \check{s}\else
  \v{s}\fi{}i\ifmmode~\acute{c}\else \'{c}\fi{}}\ and\ \citenamefont
  {\ifmmode~\check{Z}\else \v{Z}\fi{}itko}(2022)}]{Pavesic2022}%
  \BibitemOpen
  \bibfield  {author} {\bibinfo {author} {\bibfnamefont {L.}~\bibnamefont
  {Pave\ifmmode \check{s}\else \v{s}\fi{}i\ifmmode~\acute{c}\else \'{c}\fi{}}}\
  and\ \bibinfo {author} {\bibfnamefont {R.}~\bibnamefont
  {\ifmmode~\check{Z}\else \v{Z}\fi{}itko}},\ }\bibfield  {title} {\bibinfo
  {title} {{Qubit based on spin-singlet Yu-Shiba-Rusinov states}},\ }\href
  {https://doi.org/10.1103/PhysRevB.105.075129} {\bibfield  {journal} {\bibinfo
   {journal} {Phys. Rev. B}\ }\textbf {\bibinfo {volume} {105}},\ \bibinfo
  {pages} {075129} (\bibinfo {year} {2022})}\BibitemShut {NoStop}%
\bibitem [{\citenamefont {Spethmann}\ \emph {et~al.}(2022)\citenamefont
  {Spethmann}, \citenamefont {Zhang}, \citenamefont {Klinovaja},\ and\
  \citenamefont {Loss}}]{2205.03843}%
  \BibitemOpen
  \bibfield  {author} {\bibinfo {author} {\bibfnamefont {M.}~\bibnamefont
  {Spethmann}}, \bibinfo {author} {\bibfnamefont {X.-P.}\ \bibnamefont
  {Zhang}}, \bibinfo {author} {\bibfnamefont {J.}~\bibnamefont {Klinovaja}},\
  and\ \bibinfo {author} {\bibfnamefont {D.}~\bibnamefont {Loss}},\ }\bibfield
  {title} {\bibinfo {title} {{Coupled superconducting spin qubits with
  spin-orbit interaction}},\ }\href
  {https://doi.org/10.1103/PhysRevB.106.115411} {\bibfield  {journal} {\bibinfo
   {journal} {Phys. Rev. B}\ }\textbf {\bibinfo {volume} {106}},\ \bibinfo
  {pages} {115411} (\bibinfo {year} {2022})}\BibitemShut {NoStop}%
\bibitem [{\citenamefont {Bloch}(1957)}]{Bloch1957}%
  \BibitemOpen
  \bibfield  {author} {\bibinfo {author} {\bibfnamefont {F.}~\bibnamefont
  {Bloch}},\ }\bibfield  {title} {\bibinfo {title} {{Generalized Theory of
  Relaxation}},\ }\href {https://doi.org/10.1103/PhysRev.105.1206} {\bibfield
  {journal} {\bibinfo  {journal} {Phys. Rev.}\ }\textbf {\bibinfo {volume}
  {105}},\ \bibinfo {pages} {1206} (\bibinfo {year} {1957})}\BibitemShut
  {NoStop}%
\bibitem [{\citenamefont {Redfield}(1965)}]{Redfield1965}%
  \BibitemOpen
  \bibfield  {author} {\bibinfo {author} {\bibfnamefont {A.~G.}\ \bibnamefont
  {Redfield}},\ }\bibfield  {title} {\bibinfo {title} {{The Theory of
  Relaxation Processes}},\ }in\ \href
  {https://doi.org/10.1016/B978-1-4832-3114-3.50007-6} {\emph {\bibinfo
  {booktitle} {Advances in Magnetic and Optical Resonance}}},\ \bibinfo
  {series} {Advances in Magnetic Resonance}, Vol.~\bibinfo {volume} {1},\
  \bibinfo {editor} {edited by\ \bibinfo {editor} {\bibfnamefont {J.~S.}\
  \bibnamefont {Waugh}}}\ (\bibinfo  {publisher} {Academic Press},\ \bibinfo
  {year} {1965})\ pp.\ \bibinfo {pages} {1--32}\BibitemShut {NoStop}%
\bibitem [{\citenamefont {König}\ \emph {et~al.}(1997)\citenamefont {König},
  \citenamefont {Schoeller},\ and\ \citenamefont {Schön}}]{Konig1997}%
  \BibitemOpen
  \bibfield  {author} {\bibinfo {author} {\bibfnamefont {J.}~\bibnamefont
  {König}}, \bibinfo {author} {\bibfnamefont {H.}~\bibnamefont {Schoeller}},\
  and\ \bibinfo {author} {\bibfnamefont {G.}~\bibnamefont {Schön}},\
  }\bibfield  {title} {\bibinfo {title} {{Cotunneling at Resonance for the
  Single-Electron Transistor}},\ }\href
  {https://doi.org/10.1103/PhysRevLett.78.4482} {\bibfield  {journal} {\bibinfo
   {journal} {Phys. Rev. Lett.}\ }\textbf {\bibinfo {volume} {78}},\ \bibinfo
  {pages} {4482} (\bibinfo {year} {1997})}\BibitemShut {NoStop}%
\bibitem [{\citenamefont {König}\ \emph {et~al.}(1998)\citenamefont {König},
  \citenamefont {Schoeller},\ and\ \citenamefont {Schön}}]{Konig1998}%
  \BibitemOpen
  \bibfield  {author} {\bibinfo {author} {\bibfnamefont {J.}~\bibnamefont
  {König}}, \bibinfo {author} {\bibfnamefont {H.}~\bibnamefont {Schoeller}},\
  and\ \bibinfo {author} {\bibfnamefont {G.}~\bibnamefont {Schön}},\
  }\bibfield  {title} {\bibinfo {title} {{Cotunneling and renormalization
  effects for the single-electron transistor}},\ }\href
  {https://doi.org/10.1103/PhysRevB.58.7882} {\bibfield  {journal} {\bibinfo
  {journal} {Phys. Rev. B}\ }\textbf {\bibinfo {volume} {58}},\ \bibinfo
  {pages} {7882} (\bibinfo {year} {1998})}\BibitemShut {NoStop}%
\bibitem [{\citenamefont {Pedersen}\ and\ \citenamefont
  {Wacker}(2005)}]{Pedersen2005}%
  \BibitemOpen
  \bibfield  {author} {\bibinfo {author} {\bibfnamefont {J.~N.}\ \bibnamefont
  {Pedersen}}\ and\ \bibinfo {author} {\bibfnamefont {A.}~\bibnamefont
  {Wacker}},\ }\bibfield  {title} {\bibinfo {title} {{Tunneling through
  nanosystems: Combining broadening with many-particle states}},\ }\href
  {https://doi.org/10.1103/PhysRevB.72.195330} {\bibfield  {journal} {\bibinfo
  {journal} {Phys. Rev. B}\ }\textbf {\bibinfo {volume} {72}},\ \bibinfo
  {pages} {195330} (\bibinfo {year} {2005})}\BibitemShut {NoStop}%
\bibitem [{\citenamefont {Timm}(2008)}]{Timm2008}%
  \BibitemOpen
  \bibfield  {author} {\bibinfo {author} {\bibfnamefont {C.}~\bibnamefont
  {Timm}},\ }\bibfield  {title} {\bibinfo {title} {{Tunneling through molecules
  and quantum dots: Master-equation approaches}},\ }\href
  {https://doi.org/10.1103/PhysRevB.77.195416} {\bibfield  {journal} {\bibinfo
  {journal} {Phys. Rev. B}\ }\textbf {\bibinfo {volume} {77}},\ \bibinfo
  {pages} {195416} (\bibinfo {year} {2008})}\BibitemShut {NoStop}%
\bibitem [{\citenamefont {Leijnse}\ and\ \citenamefont
  {Wegewijs}(2008)}]{Leijnse2008}%
  \BibitemOpen
  \bibfield  {author} {\bibinfo {author} {\bibfnamefont {M.}~\bibnamefont
  {Leijnse}}\ and\ \bibinfo {author} {\bibfnamefont {M.~R.}\ \bibnamefont
  {Wegewijs}},\ }\bibfield  {title} {\bibinfo {title} {{Kinetic equations for
  transport through single-molecule transistors}},\ }\href
  {https://doi.org/10.1103/PhysRevB.78.235424} {\bibfield  {journal} {\bibinfo
  {journal} {Phys. Rev. B}\ }\textbf {\bibinfo {volume} {78}},\ \bibinfo
  {pages} {235424} (\bibinfo {year} {2008})}\BibitemShut {NoStop}%
\bibitem [{\citenamefont {Koller}\ \emph {et~al.}(2010)\citenamefont {Koller},
  \citenamefont {Grifoni}, \citenamefont {Leijnse},\ and\ \citenamefont
  {Wegewijs}}]{Koller2010}%
  \BibitemOpen
  \bibfield  {author} {\bibinfo {author} {\bibfnamefont {S.}~\bibnamefont
  {Koller}}, \bibinfo {author} {\bibfnamefont {M.}~\bibnamefont {Grifoni}},
  \bibinfo {author} {\bibfnamefont {M.}~\bibnamefont {Leijnse}},\ and\ \bibinfo
  {author} {\bibfnamefont {M.~R.}\ \bibnamefont {Wegewijs}},\ }\bibfield
  {title} {\bibinfo {title} {{Density-operator approaches to transport through
  interacting quantum dots: Simplifications in fourth-order perturbation
  theory}},\ }\href {https://doi.org/10.1103/PhysRevB.82.235307} {\bibfield
  {journal} {\bibinfo  {journal} {Phys. Rev. B}\ }\textbf {\bibinfo {volume}
  {82}},\ \bibinfo {pages} {235307} (\bibinfo {year} {2010})}\BibitemShut
  {NoStop}%
\bibitem [{\citenamefont {Karlström}\ \emph {et~al.}(2013)\citenamefont
  {Karlström}, \citenamefont {Emary}, \citenamefont {Zedler}, \citenamefont
  {Pedersen}, \citenamefont {Bergenfeldt}, \citenamefont {Samuelsson},
  \citenamefont {Brandes},\ and\ \citenamefont {Wacker}}]{Karlstrom2013}%
  \BibitemOpen
  \bibfield  {author} {\bibinfo {author} {\bibfnamefont {O.}~\bibnamefont
  {Karlström}}, \bibinfo {author} {\bibfnamefont {C.}~\bibnamefont {Emary}},
  \bibinfo {author} {\bibfnamefont {P.}~\bibnamefont {Zedler}}, \bibinfo
  {author} {\bibfnamefont {J.~N.}\ \bibnamefont {Pedersen}}, \bibinfo {author}
  {\bibfnamefont {C.}~\bibnamefont {Bergenfeldt}}, \bibinfo {author}
  {\bibfnamefont {P.}~\bibnamefont {Samuelsson}}, \bibinfo {author}
  {\bibfnamefont {T.}~\bibnamefont {Brandes}},\ and\ \bibinfo {author}
  {\bibfnamefont {A.}~\bibnamefont {Wacker}},\ }\bibfield  {title} {\bibinfo
  {title} {{A diagrammatic description of the equations of motion, current and
  noise within the second-order von Neumann approach}},\ }\href
  {https://doi.org/10.1088/1751-8113/46/6/065301} {\bibfield  {journal}
  {\bibinfo  {journal} {J. Phys. A: Math. Theor.}\ }\textbf {\bibinfo {volume}
  {46}},\ \bibinfo {pages} {065301} (\bibinfo {year} {2013})}\BibitemShut
  {NoStop}%
\bibitem [{\citenamefont {Josephson}(1962)}]{Josephson1962}%
  \BibitemOpen
  \bibfield  {author} {\bibinfo {author} {\bibfnamefont {B.~D.}\ \bibnamefont
  {Josephson}},\ }\bibfield  {title} {\bibinfo {title} {{Possible new effects
  in superconductive tunnelling}},\ }\href
  {https://doi.org/10.1016/0031-9163(62)91369-0} {\bibfield  {journal}
  {\bibinfo  {journal} {Phys. Lett.}\ }\textbf {\bibinfo {volume} {1}},\
  \bibinfo {pages} {251} (\bibinfo {year} {1962})}\BibitemShut {NoStop}%
\bibitem [{\citenamefont {Kamp}\ and\ \citenamefont
  {Sothmann}(2019)}]{Kamp2019}%
  \BibitemOpen
  \bibfield  {author} {\bibinfo {author} {\bibfnamefont {M.}~\bibnamefont
  {Kamp}}\ and\ \bibinfo {author} {\bibfnamefont {B.}~\bibnamefont
  {Sothmann}},\ }\bibfield  {title} {\bibinfo {title} {{Phase-dependent heat
  and charge transport through superconductor–quantum dot hybrids}},\ }\href
  {https://doi.org/10.1103/PhysRevB.99.045428} {\bibfield  {journal} {\bibinfo
  {journal} {Phys. Rev. B}\ }\textbf {\bibinfo {volume} {99}},\ \bibinfo
  {pages} {045428} (\bibinfo {year} {2019})}\BibitemShut {NoStop}%
\bibitem [{\citenamefont {Kamp}\ and\ \citenamefont
  {Sothmann}(2021)}]{Kamp2021}%
  \BibitemOpen
  \bibfield  {author} {\bibinfo {author} {\bibfnamefont {M.}~\bibnamefont
  {Kamp}}\ and\ \bibinfo {author} {\bibfnamefont {B.}~\bibnamefont
  {Sothmann}},\ }\bibfield  {title} {\bibinfo {title} {{Higgs-like pair
  amplitude dynamics in superconductor–quantum-dot hybrids}},\ }\href
  {https://doi.org/10.1103/PhysRevB.103.045414} {\bibfield  {journal} {\bibinfo
   {journal} {Phys. Rev. B}\ }\textbf {\bibinfo {volume} {103}},\ \bibinfo
  {pages} {045414} (\bibinfo {year} {2021})}\BibitemShut {NoStop}%
\bibitem [{\citenamefont {Heckschen}\ and\ \citenamefont
  {Sothmann}(2022)}]{Heckschen2022}%
  \BibitemOpen
  \bibfield  {author} {\bibinfo {author} {\bibfnamefont {M.}~\bibnamefont
  {Heckschen}}\ and\ \bibinfo {author} {\bibfnamefont {B.}~\bibnamefont
  {Sothmann}},\ }\bibfield  {title} {\bibinfo {title} {{Pair-amplitude dynamics
  in strongly coupled superconductor--quantum dot hybrids}},\ }\href
  {https://doi.org/10.1103/PhysRevB.105.045420} {\bibfield  {journal} {\bibinfo
   {journal} {Phys. Rev. B}\ }\textbf {\bibinfo {volume} {105}},\ \bibinfo
  {pages} {045420} (\bibinfo {year} {2022})}\BibitemShut {NoStop}%
\bibitem [{\citenamefont {Nakajima}(1958)}]{Nakajima1958}%
  \BibitemOpen
  \bibfield  {author} {\bibinfo {author} {\bibfnamefont {S.}~\bibnamefont
  {Nakajima}},\ }\bibfield  {title} {\bibinfo {title} {{On quantum theory of
  transport phenomena}},\ }\href {https://doi.org/10.1143/ptp.20.948}
  {\bibfield  {journal} {\bibinfo  {journal} {Prog. Theor. Phys.}\ }\textbf
  {\bibinfo {volume} {20}},\ \bibinfo {pages} {948} (\bibinfo {year}
  {1958})}\BibitemShut {NoStop}%
\bibitem [{\citenamefont {Zwanzig}(1960)}]{Zwanzig1960}%
  \BibitemOpen
  \bibfield  {author} {\bibinfo {author} {\bibfnamefont {R.}~\bibnamefont
  {Zwanzig}},\ }\bibfield  {title} {\bibinfo {title} {{Ensemble method in the
  theory of irreversibility}},\ }\href {https://doi.org/10.1063/1.1731409}
  {\bibfield  {journal} {\bibinfo  {journal} {J. Chem. Phys.}\ }\textbf
  {\bibinfo {volume} {33}},\ \bibinfo {pages} {1338} (\bibinfo {year}
  {1960})}\BibitemShut {NoStop}%
\bibitem [{\citenamefont {Anderson}(1961)}]{Anderson1961}%
  \BibitemOpen
  \bibfield  {author} {\bibinfo {author} {\bibfnamefont {P.~W.}\ \bibnamefont
  {Anderson}},\ }\bibfield  {title} {\bibinfo {title} {{Localized magnetic
  states in metals}},\ }\href {https://doi.org/10.1103/PhysRev.124.41}
  {\bibfield  {journal} {\bibinfo  {journal} {Phys. Rev.}\ }\textbf {\bibinfo
  {volume} {124}},\ \bibinfo {pages} {41} (\bibinfo {year} {1961})}\BibitemShut
  {NoStop}%
\bibitem [{\citenamefont {Hubbard}(1963)}]{Hubbard1963}%
  \BibitemOpen
  \bibfield  {author} {\bibinfo {author} {\bibfnamefont {J.}~\bibnamefont
  {Hubbard}},\ }\bibfield  {title} {\bibinfo {title} {{Electron correlations in
  narrow energy bands}},\ }\href {https://doi.org/10.1098/rspa.1963.0204}
  {\bibfield  {journal} {\bibinfo  {journal} {Proc. R. Soc. Lond.}\ }\textbf
  {\bibinfo {volume} {276}},\ \bibinfo {pages} {238–257} (\bibinfo {year}
  {1963})}\BibitemShut {NoStop}%
\bibitem [{\citenamefont {Bardeen}(1962)}]{Bardeen1962}%
  \BibitemOpen
  \bibfield  {author} {\bibinfo {author} {\bibfnamefont {J.}~\bibnamefont
  {Bardeen}},\ }\bibfield  {title} {\bibinfo {title} {{Tunneling Into
  Superconductors}},\ }\href {https://doi.org/10.1103/PhysRevLett.9.147}
  {\bibfield  {journal} {\bibinfo  {journal} {Phys. Rev. Lett.}\ }\textbf
  {\bibinfo {volume} {9}},\ \bibinfo {pages} {147} (\bibinfo {year}
  {1962})}\BibitemShut {NoStop}%
\bibitem [{\citenamefont {Cooper}(1956)}]{Cooper1956}%
  \BibitemOpen
  \bibfield  {author} {\bibinfo {author} {\bibfnamefont {L.~N.}\ \bibnamefont
  {Cooper}},\ }\bibfield  {title} {\bibinfo {title} {{Bound electron pairs in a
  degenerate Fermi gas}},\ }\href {https://doi.org/10.1103/PhysRev.104.1189}
  {\bibfield  {journal} {\bibinfo  {journal} {Phys. Rev.}\ }\textbf {\bibinfo
  {volume} {104}},\ \bibinfo {pages} {1189} (\bibinfo {year}
  {1956})}\BibitemShut {NoStop}%
\bibitem [{\citenamefont {Leggett}(2006)}]{Leggett2008}%
  \BibitemOpen
  \bibfield  {author} {\bibinfo {author} {\bibfnamefont {A.~J.}\ \bibnamefont
  {Leggett}},\ }\href
  {https://doi.org/10.1093/acprof:oso/9780198526438.001.0001} {\emph {\bibinfo
  {title} {{Quantum Liquids: Bose condensation and Cooper pairing in
  condensed-matter systems}}}}\ (\bibinfo  {publisher} {Oxford University
  Press},\ \bibinfo {year} {2006})\BibitemShut {NoStop}%
\bibitem [{Note1()}]{Note1}%
  \BibitemOpen
  \bibinfo {note} {Due to particle conservation in this formalism, any
  displacement currents vanish in the time average. Therefore, alternative
  conventions differ in at most a phase(sign) for the AC(DC) part of the
  current.}\BibitemShut {Stop}%
\bibitem [{\citenamefont {Josephson}(1974)}]{Josephson1974}%
  \BibitemOpen
  \bibfield  {author} {\bibinfo {author} {\bibfnamefont {B.~D.}\ \bibnamefont
  {Josephson}},\ }\bibfield  {title} {\bibinfo {title} {{The discovery of
  tunnelling supercurrents}},\ }\href
  {https://doi.org/10.1103/RevModPhys.46.251} {\bibfield  {journal} {\bibinfo
  {journal} {Rev. Mod. Phys.}\ }\textbf {\bibinfo {volume} {46}},\ \bibinfo
  {pages} {251} (\bibinfo {year} {1974})}\BibitemShut {NoStop}%
\bibitem [{\citenamefont {Abramowitz}\ and\ \citenamefont
  {Stegun}(1964)}]{Abramowitz1965}%
  \BibitemOpen
  \bibfield  {author} {\bibinfo {author} {\bibfnamefont {M.}~\bibnamefont
  {Abramowitz}}\ and\ \bibinfo {author} {\bibfnamefont {I.}~\bibnamefont
  {Stegun}},\ }\href {https://books.google.de/books?id=2FowulitwpUC} {\emph
  {\bibinfo {title} {{Handbook of Mathematical Functions with Formulas, Graphs,
  and Mathematical Tables}}}},\ Applied mathematics series\ (\bibinfo
  {publisher} {U.S. Government Printing Office},\ \bibinfo {year}
  {1964})\BibitemShut {NoStop}%
\bibitem [{Note2()}]{Note2}%
  \BibitemOpen
  \bibinfo {note} {Any other choice is identical up to an additional phase
  factor absorbed into $v_{l,k}$}\BibitemShut {NoStop}%
\bibitem [{\citenamefont {Bardeen}\ \emph {et~al.}(1957)\citenamefont
  {Bardeen}, \citenamefont {Cooper},\ and\ \citenamefont
  {Schrieffer}}]{Bardeen1957}%
  \BibitemOpen
  \bibfield  {author} {\bibinfo {author} {\bibfnamefont {J.}~\bibnamefont
  {Bardeen}}, \bibinfo {author} {\bibfnamefont {L.~N.}\ \bibnamefont
  {Cooper}},\ and\ \bibinfo {author} {\bibfnamefont {J.~R.}\ \bibnamefont
  {Schrieffer}},\ }\bibfield  {title} {\bibinfo {title} {{Theory of
  superconductivity}},\ }\href {https://doi.org/10.1103/PhysRev.108.1175}
  {\bibfield  {journal} {\bibinfo  {journal} {Phys. Rev.}\ }\textbf {\bibinfo
  {volume} {108}},\ \bibinfo {pages} {1175} (\bibinfo {year}
  {1957})}\BibitemShut {NoStop}%
\bibitem [{\citenamefont {Peierls}(1991)}]{Peierls1991}%
  \BibitemOpen
  \bibfield  {author} {\bibinfo {author} {\bibfnamefont {R.}~\bibnamefont
  {Peierls}},\ }\bibfield  {title} {\bibinfo {title} {{Spontaneously broken
  symmetries}},\ }\href
  {https://iopscience.iop.org/article/10.1088/0305-4470/24/22/011/meta}
  {\bibfield  {journal} {\bibinfo  {journal} {J. Phys. A: Math. Gen.}\ ,\
  \bibinfo {pages} {5273}} (\bibinfo {year} {1991})}\BibitemShut {NoStop}%
\bibitem [{Note3()}]{Note3}%
  \BibitemOpen
  \bibinfo {note} {This property also holds to all orders as higher order
  kernels at most contain the convolutions of terms of the form \protect \cref
  {eq: expansions1,eq: expansions2,eq: expansions3} taken at different
  times.}\BibitemShut {Stop}%
\bibitem [{\citenamefont {Yeyati}\ \emph {et~al.}(1997)\citenamefont {Yeyati},
  \citenamefont {Cuevas}, \citenamefont {Lo~Pez-Da},\ and\ \citenamefont {Marti
  N-Rodero}}]{Yeyati1997}%
  \BibitemOpen
  \bibfield  {author} {\bibinfo {author} {\bibfnamefont {A.~L.}\ \bibnamefont
  {Yeyati}}, \bibinfo {author} {\bibfnamefont {J.~C.}\ \bibnamefont {Cuevas}},
  \bibinfo {author} {\bibfnamefont {A.}~\bibnamefont {Lo~Pez-Da}},\ and\
  \bibinfo {author} {\bibfnamefont {A.}~\bibnamefont {Marti N-Rodero}},\
  }\bibfield  {title} {\bibinfo {title} {{Resonant tunneling through a small
  quantum dot coupled to superconducting leads}},\ }\href
  {https://doi.org/https://doi.org/10.1103/PhysRevB.55.R6137} {\bibfield
  {journal} {\bibinfo  {journal} {Phys. Rev. B}\ }\textbf {\bibinfo {volume}
  {55}},\ \bibinfo {pages} {R6137} (\bibinfo {year} {1997})}\BibitemShut
  {NoStop}%
\bibitem [{\citenamefont {Dynes}\ \emph {et~al.}(1978)\citenamefont {Dynes},
  \citenamefont {Narayanamurti},\ and\ \citenamefont {Garno}}]{Dynes1978}%
  \BibitemOpen
  \bibfield  {author} {\bibinfo {author} {\bibfnamefont {R.~C.}\ \bibnamefont
  {Dynes}}, \bibinfo {author} {\bibfnamefont {V.}~\bibnamefont
  {Narayanamurti}},\ and\ \bibinfo {author} {\bibfnamefont {J.~P.}\
  \bibnamefont {Garno}},\ }\bibfield  {title} {\bibinfo {title} {{Direct
  Measurement of Quasiparticle-Lifetime Broadening in a Strong-Coupled
  Superconductor}},\ }\href {https://doi.org/10.1103/PhysRevLett.41.1509}
  {\bibfield  {journal} {\bibinfo  {journal} {Phys. Rev. Lett.}\ }\textbf
  {\bibinfo {volume} {41}},\ \bibinfo {pages} {1509} (\bibinfo {year}
  {1978})}\BibitemShut {NoStop}%
\bibitem [{\citenamefont {Siegert}\ and\ \citenamefont
  {Grifoni}(2015)}]{siegert_effects_2015}%
  \BibitemOpen
  \bibfield  {author} {\bibinfo {author} {\bibfnamefont {A.}~\bibnamefont
  {Siegert}, \bibfnamefont {B.and~Donarini}}\ and\ \bibinfo {author}
  {\bibfnamefont {M.}~\bibnamefont {Grifoni}},\ }\bibfield  {title} {\bibinfo
  {title} {{Effects of spin–orbit coupling and many-body correlations in
  {STM} transport through copper phthalocyanine}},\ }\href
  {https://doi.org/10.3762/bjnano.6.254} {\bibfield  {journal} {\bibinfo
  {journal} {Beilstein J. Nanotechnol.}\ }\textbf {\bibinfo {volume} {6}},\
  \bibinfo {pages} {2452–2462} (\bibinfo {year} {2015})}\BibitemShut
  {NoStop}%
\bibitem [{\citenamefont {Ho}\ \emph {et~al.}(1983)\citenamefont {Ho},
  \citenamefont {Chu},\ and\ \citenamefont {Tietz}}]{Ho1983}%
  \BibitemOpen
  \bibfield  {author} {\bibinfo {author} {\bibfnamefont {T.~S.}\ \bibnamefont
  {Ho}}, \bibinfo {author} {\bibfnamefont {S.~I.}\ \bibnamefont {Chu}},\ and\
  \bibinfo {author} {\bibfnamefont {J.~V.}\ \bibnamefont {Tietz}},\ }\bibfield
  {title} {\bibinfo {title} {{Semiclassical many-mode Floquet theory}},\ }\href
  {https://doi.org/https://doi.org/10.1016/0009-2614(83)80732-5} {\bibfield
  {journal} {\bibinfo  {journal} {Chem. Phys. Lett.}\ }\textbf {\bibinfo
  {volume} {96}},\ \bibinfo {pages} {464} (\bibinfo {year} {1983})}\BibitemShut
  {NoStop}%
\bibitem [{\citenamefont {{G\'omez-Le\'on}}\ and\ \citenamefont
  {Platero}(2020)}]{GomezLeon2020}%
  \BibitemOpen
  \bibfield  {author} {\bibinfo {author} {\bibfnamefont {A.}~\bibnamefont
  {{G\'omez-Le\'on}}}\ and\ \bibinfo {author} {\bibfnamefont {G.}~\bibnamefont
  {Platero}},\ }\bibfield  {title} {\bibinfo {title} {{Designing adiabatic time
  evolution from high-frequency bichromatic sources}},\ }\href
  {https://doi.org/10.1103/PhysRevResearch.2.033412} {\bibfield  {journal}
  {\bibinfo  {journal} {Phys. Rev. Research}\ }\textbf {\bibinfo {volume}
  {2}},\ \bibinfo {pages} {033412} (\bibinfo {year} {2020})}\BibitemShut
  {NoStop}%
\bibitem [{\citenamefont {Kostur}\ \emph {et~al.}(2008)\citenamefont {Kostur},
  \citenamefont {Machura}, \citenamefont {Talkner}, \citenamefont {Hänggi},\
  and\ \citenamefont {Łuczka}}]{Kostur2008}%
  \BibitemOpen
  \bibfield  {author} {\bibinfo {author} {\bibfnamefont {M.}~\bibnamefont
  {Kostur}}, \bibinfo {author} {\bibfnamefont {L.}~\bibnamefont {Machura}},
  \bibinfo {author} {\bibfnamefont {P.}~\bibnamefont {Talkner}}, \bibinfo
  {author} {\bibfnamefont {P.}~\bibnamefont {Hänggi}},\ and\ \bibinfo {author}
  {\bibfnamefont {J.}~\bibnamefont {Łuczka}},\ }\bibfield  {title} {\bibinfo
  {title} {{Anomalous transport in biased ac-driven Josephson junctions:
  Negative conductances}},\ }\href {https://doi.org/10.1103/PhysRevB.77.104509}
  {\bibfield  {journal} {\bibinfo  {journal} {Phys. Rev. B}\ }\textbf {\bibinfo
  {volume} {77}},\ \bibinfo {pages} {104509} (\bibinfo {year}
  {2008})}\BibitemShut {NoStop}%
\bibitem [{\citenamefont {Doh}\ \emph {et~al.}(2008)\citenamefont {Doh},
  \citenamefont {De~Franceschi}, \citenamefont {Bakkers},\ and\ \citenamefont
  {Kouwenhoven}}]{Doh2008}%
  \BibitemOpen
  \bibfield  {author} {\bibinfo {author} {\bibfnamefont {Y.~J.}\ \bibnamefont
  {Doh}}, \bibinfo {author} {\bibfnamefont {S.}~\bibnamefont {De~Franceschi}},
  \bibinfo {author} {\bibfnamefont {E.~P. A.~M.}\ \bibnamefont {Bakkers}},\
  and\ \bibinfo {author} {\bibfnamefont {L.~P.}\ \bibnamefont {Kouwenhoven}},\
  }\bibfield  {title} {\bibinfo {title} {{Andreev Reflection versus Coulomb
  Blockade in Hybrid Semiconductor Nanowire Devices}},\ }\href
  {https://doi.org/10.1021/nl801454k} {\bibfield  {journal} {\bibinfo
  {journal} {Nano Lett.}\ }\textbf {\bibinfo {volume} {8}},\ \bibinfo {pages}
  {4098} (\bibinfo {year} {2008})}\BibitemShut {NoStop}%
\bibitem [{\citenamefont {{Martín-Rodero}}\ and\ \citenamefont
  {Levy~Yeyati}(2011)}]{martin-rodero_josephson_2011}%
  \BibitemOpen
  \bibfield  {author} {\bibinfo {author} {\bibfnamefont {A.}~\bibnamefont
  {{Martín-Rodero}}}\ and\ \bibinfo {author} {\bibfnamefont {A.}~\bibnamefont
  {Levy~Yeyati}},\ }\bibfield  {title} {\bibinfo {title} {{Josephson and
  Andreev transport through quantum dots}},\ }\href
  {https://doi.org/10.1080/00018732.2011.624266} {\bibfield  {journal}
  {\bibinfo  {journal} {Advances in Physics}\ }\textbf {\bibinfo {volume}
  {60}},\ \bibinfo {pages} {899} (\bibinfo {year} {2011})}\BibitemShut
  {NoStop}%
\bibitem [{\citenamefont {{K\"onig}}\ \emph {et~al.}(1996)\citenamefont
  {{K\"onig}}, \citenamefont {Schmid}, \citenamefont {Schoeller},\ and\
  \citenamefont {{Sch\"on}}}]{Konig1996b}%
  \BibitemOpen
  \bibfield  {author} {\bibinfo {author} {\bibfnamefont {J.}~\bibnamefont
  {{K\"onig}}}, \bibinfo {author} {\bibfnamefont {J.}~\bibnamefont {Schmid}},
  \bibinfo {author} {\bibfnamefont {H.}~\bibnamefont {Schoeller}},\ and\
  \bibinfo {author} {\bibfnamefont {G.}~\bibnamefont {{Sch\"on}}},\ }\bibfield
  {title} {\bibinfo {title} {{Resonant tunneling through ultrasmall quantum
  dots: Zero-bias anomalies, magnetic-field dependence, and boson-assisted
  transport}},\ }\href {https://doi.org/10.1103/PhysRevB.54.16820} {\bibfield
  {journal} {\bibinfo  {journal} {Phys. Rev. B}\ }\textbf {\bibinfo {volume}
  {54}},\ \bibinfo {pages} {16820} (\bibinfo {year} {1996})}\BibitemShut
  {NoStop}%
\bibitem [{\citenamefont {Kern}\ and\ \citenamefont
  {Grifoni}(2013)}]{Kern2013}%
  \BibitemOpen
  \bibfield  {author} {\bibinfo {author} {\bibfnamefont {J.}~\bibnamefont
  {Kern}}\ and\ \bibinfo {author} {\bibfnamefont {M.}~\bibnamefont {Grifoni}},\
  }\bibfield  {title} {\bibinfo {title} {{Transport across an Anderson quantum
  dot in the intermediate coupling regime}},\ }\bibfield  {journal} {\bibinfo
  {journal} {Eur Phys J B}\ }\textbf {\bibinfo {volume} {86}},\ \href
  {https://doi.org/10.1140/epjb/e2013-40618-9} {10.1140/epjb/e2013-40618-9}
  (\bibinfo {year} {2013})\BibitemShut {NoStop}%
\bibitem [{\citenamefont {Glazman}\ and\ \citenamefont
  {Matveev}(1989)}]{Glazman1989}%
  \BibitemOpen
  \bibfield  {author} {\bibinfo {author} {\bibfnamefont {L.~I.}\ \bibnamefont
  {Glazman}}\ and\ \bibinfo {author} {\bibfnamefont {K.~A.}\ \bibnamefont
  {Matveev}},\ }\bibfield  {title} {\bibinfo {title} {{Resonant Josephson
  current through Kondo impurities in a tunnel barrier }},\ }\href
  {http://jetpletters.ru/ps/1121/article_16988.shtml} {\bibfield  {journal}
  {\bibinfo  {journal} {JETP Lett.}\ }\textbf {\bibinfo {volume} {49}},\
  \bibinfo {pages} {570} (\bibinfo {year} {1989})}\BibitemShut {NoStop}%
\bibitem [{\citenamefont {Kautz}(1996)}]{Kautz1996}%
  \BibitemOpen
  \bibfield  {author} {\bibinfo {author} {\bibfnamefont {R.~L.}\ \bibnamefont
  {Kautz}},\ }\bibfield  {title} {\bibinfo {title} {{Noise, chaos, and the
  Josephson voltage standard}},\ }\href
  {https://doi.org/10.1088/0034-4885/59/8/001} {\bibfield  {journal} {\bibinfo
  {journal} {Rep. Prog. Phys.}\ }\textbf {\bibinfo {volume} {59}},\ \bibinfo
  {pages} {935} (\bibinfo {year} {1996})}\BibitemShut {NoStop}%
\bibitem [{\citenamefont {{Dom\'{\i}nguez}}\ \emph {et~al.}(2012)\citenamefont
  {{Dom\'{\i}nguez}}, \citenamefont {Hassler},\ and\ \citenamefont
  {Platero}}]{Dominguez2012}%
  \BibitemOpen
  \bibfield  {author} {\bibinfo {author} {\bibfnamefont {F.}~\bibnamefont
  {{Dom\'{\i}nguez}}}, \bibinfo {author} {\bibfnamefont {F.}~\bibnamefont
  {Hassler}},\ and\ \bibinfo {author} {\bibfnamefont {G.}~\bibnamefont
  {Platero}},\ }\bibfield  {title} {\bibinfo {title} {{Dynamical detection of
  Majorana fermions in current-biased nanowires}},\ }\href
  {https://doi.org/10.1103/PhysRevB.86.140503} {\bibfield  {journal} {\bibinfo
  {journal} {Phys. Rev. B}\ }\textbf {\bibinfo {volume} {86}},\ \bibinfo
  {pages} {140503} (\bibinfo {year} {2012})}\BibitemShut {NoStop}%
\bibitem [{\citenamefont {{Pic\'o-Cort\'es}}\ \emph {et~al.}(2017)\citenamefont
  {{Pic\'o-Cort\'es}}, \citenamefont {{Dom\'{\i}nguez}},\ and\ \citenamefont
  {Platero}}]{PicoCortes2017}%
  \BibitemOpen
  \bibfield  {author} {\bibinfo {author} {\bibfnamefont {J.}~\bibnamefont
  {{Pic\'o-Cort\'es}}}, \bibinfo {author} {\bibfnamefont {F.}~\bibnamefont
  {{Dom\'{\i}nguez}}},\ and\ \bibinfo {author} {\bibfnamefont {G.}~\bibnamefont
  {Platero}},\ }\bibfield  {title} {\bibinfo {title} {{Signatures of a
  $4\ensuremath{\pi}$-periodic supercurrent in the voltage response of
  capacitively shunted topological Josephson junctions}},\ }\href
  {https://doi.org/10.1103/PhysRevB.96.125438} {\bibfield  {journal} {\bibinfo
  {journal} {Phys. Rev. B}\ }\textbf {\bibinfo {volume} {96}},\ \bibinfo
  {pages} {125438} (\bibinfo {year} {2017})}\BibitemShut {NoStop}%
\bibitem [{\citenamefont {Park}\ \emph {et~al.}(2021)\citenamefont {Park},
  \citenamefont {Choi}, \citenamefont {Lee},\ and\ \citenamefont
  {Lee}}]{Park2021}%
  \BibitemOpen
  \bibfield  {author} {\bibinfo {author} {\bibfnamefont {J.}~\bibnamefont
  {Park}}, \bibinfo {author} {\bibfnamefont {Y.}~\bibnamefont {Choi}}, \bibinfo
  {author} {\bibfnamefont {G.}~\bibnamefont {Lee}},\ and\ \bibinfo {author}
  {\bibfnamefont {H.}~\bibnamefont {Lee}},\ }\bibfield  {title} {\bibinfo
  {title} {{Characterization of Shapiro steps in the presence of a
  $4\pi$-periodic Josephson current}},\ }\href
  {https://doi.org/10.1103/PhysRevB.103.235428} {\bibfield  {journal} {\bibinfo
   {journal} {Phys. Rev. B}\ }\textbf {\bibinfo {volume} {103}},\ \bibinfo
  {pages} {235428} (\bibinfo {year} {2021})}\BibitemShut {NoStop}%
\bibitem [{\citenamefont {Kwon}\ \emph {et~al.}(2004)\citenamefont {Kwon},
  \citenamefont {Yakovenko},\ and\ \citenamefont {Sengupta}}]{Kwon2004}%
  \BibitemOpen
  \bibfield  {author} {\bibinfo {author} {\bibfnamefont {H.~J.}\ \bibnamefont
  {Kwon}}, \bibinfo {author} {\bibfnamefont {V.~M.}\ \bibnamefont
  {Yakovenko}},\ and\ \bibinfo {author} {\bibfnamefont {K.}~\bibnamefont
  {Sengupta}},\ }\bibfield  {title} {\bibinfo {title} {{Fractional ac Josephson
  effect in unconventional superconductors}},\ }\href
  {https://doi.org/10.1063/1.1789931} {\bibfield  {journal} {\bibinfo
  {journal} {Low Temperature Physics}\ }\textbf {\bibinfo {volume} {30}},\
  \bibinfo {pages} {613} (\bibinfo {year} {2004})}\BibitemShut {NoStop}%
\bibitem [{\citenamefont {Virtanen}\ and\ \citenamefont
  {Recher}(2013)}]{Virtanen2013}%
  \BibitemOpen
  \bibfield  {author} {\bibinfo {author} {\bibfnamefont {P.}~\bibnamefont
  {Virtanen}}\ and\ \bibinfo {author} {\bibfnamefont {P.}~\bibnamefont
  {Recher}},\ }\bibfield  {title} {\bibinfo {title} {{Microwave spectroscopy of
  Josephson junctions in topological superconductors}},\ }\href
  {https://doi.org/10.1103/PhysRevB.88.144507} {\bibfield  {journal} {\bibinfo
  {journal} {Phys. Rev. B}\ }\textbf {\bibinfo {volume} {88}},\ \bibinfo
  {pages} {144507} (\bibinfo {year} {2013})}\BibitemShut {NoStop}%
\bibitem [{\citenamefont {Li}\ \emph {et~al.}(2018)\citenamefont {Li},
  \citenamefont {Song}, \citenamefont {Liu}, \citenamefont {Jiang},
  \citenamefont {Sun},\ and\ \citenamefont {Xie}}]{Li2018}%
  \BibitemOpen
  \bibfield  {author} {\bibinfo {author} {\bibfnamefont {Y.}~\bibnamefont
  {Li}}, \bibinfo {author} {\bibfnamefont {J.}~\bibnamefont {Song}}, \bibinfo
  {author} {\bibfnamefont {J.}~\bibnamefont {Liu}}, \bibinfo {author}
  {\bibfnamefont {H.}~\bibnamefont {Jiang}}, \bibinfo {author} {\bibfnamefont
  {Q.}~\bibnamefont {Sun}},\ and\ \bibinfo {author} {\bibfnamefont {X.~C.}\
  \bibnamefont {Xie}},\ }\bibfield  {title} {\bibinfo {title} {{Doubled Shapiro
  steps in a topological Josephson junction}},\ }\href
  {https://doi.org/10.1103/PhysRevB.97.045423} {\bibfield  {journal} {\bibinfo
  {journal} {Phys. Rev. B}\ }\textbf {\bibinfo {volume} {97}},\ \bibinfo
  {pages} {045423} (\bibinfo {year} {2018})}\BibitemShut {NoStop}%
\bibitem [{\citenamefont {Galaktionov}\ and\ \citenamefont
  {Zaikin}(2021)}]{Galaktionov2021}%
  \BibitemOpen
  \bibfield  {author} {\bibinfo {author} {\bibfnamefont {A.~V.}\ \bibnamefont
  {Galaktionov}}\ and\ \bibinfo {author} {\bibfnamefont {A.~D.}\ \bibnamefont
  {Zaikin}},\ }\bibfield  {title} {\bibinfo {title} {{Fractional Shapiro steps
  without fractional Josephson effect}},\ }\href
  {https://doi.org/10.1103/PhysRevB.104.054521} {\bibfield  {journal} {\bibinfo
   {journal} {Phys. Rev. B}\ }\textbf {\bibinfo {volume} {104}},\ \bibinfo
  {pages} {054521} (\bibinfo {year} {2021})}\BibitemShut {NoStop}%
\bibitem [{\citenamefont {Saptsov}\ and\ \citenamefont
  {Wegewijs}(2012)}]{Saptsov2012}%
  \BibitemOpen
  \bibfield  {author} {\bibinfo {author} {\bibfnamefont {R.~B.}\ \bibnamefont
  {Saptsov}}\ and\ \bibinfo {author} {\bibfnamefont {M.~R.}\ \bibnamefont
  {Wegewijs}},\ }\bibfield  {title} {\bibinfo {title} {{Fermionic
  superoperators for zero-temperature nonlinear transport: Real-time
  perturbation theory and renormalization group for Anderson quantum dots}},\
  }\href {https://doi.org/10.1103/PhysRevB.86.235432} {\bibfield  {journal}
  {\bibinfo  {journal} {Phys. Rev. B}\ }\textbf {\bibinfo {volume} {86}},\
  \bibinfo {pages} {235432} (\bibinfo {year} {2012})}\BibitemShut {NoStop}%
\bibitem [{\citenamefont {Rohrmeier}\ and\ \citenamefont
  {Donarini}(2021)}]{Rohrmeier2021}%
  \BibitemOpen
  \bibfield  {author} {\bibinfo {author} {\bibfnamefont {C.}~\bibnamefont
  {Rohrmeier}}\ and\ \bibinfo {author} {\bibfnamefont {A.}~\bibnamefont
  {Donarini}},\ }\bibfield  {title} {\bibinfo {title} {{Pseudospin resonances
  reveal synthetic spin-orbit interaction}},\ }\href
  {https://doi.org/10.1103/PhysRevB.103.205420} {\bibfield  {journal} {\bibinfo
   {journal} {Phys. Rev. B}\ }\textbf {\bibinfo {volume} {103}},\ \bibinfo
  {pages} {205420} (\bibinfo {year} {2021})}\BibitemShut {NoStop}%
\bibitem [{\citenamefont {Breuer}\ and\ \citenamefont
  {Petruccione}(2007)}]{Breuer2007}%
  \BibitemOpen
  \bibfield  {author} {\bibinfo {author} {\bibfnamefont {H.-P.}\ \bibnamefont
  {Breuer}}\ and\ \bibinfo {author} {\bibfnamefont {F.}~\bibnamefont
  {Petruccione}},\ }\href
  {https://doi.org/10.1093/acprof:oso/9780199213900.001.0001} {\emph {\bibinfo
  {title} {{The {Theory} of {Open} {Quantum} {Systems}}}}}\ (\bibinfo
  {publisher} {Oxford University Press},\ \bibinfo {year} {2007})\BibitemShut
  {NoStop}%
\bibitem [{\citenamefont {{Žonda}}\ \emph {et~al.}(2016)\citenamefont
  {{Žonda}}, \citenamefont {Pokorný}, \citenamefont {Janiš},\ and\
  \citenamefont {Novotný}}]{zonda_perturbation_2016}%
  \BibitemOpen
  \bibfield  {author} {\bibinfo {author} {\bibfnamefont {M.}~\bibnamefont
  {{Žonda}}}, \bibinfo {author} {\bibfnamefont {V.}~\bibnamefont {Pokorný}},
  \bibinfo {author} {\bibfnamefont {V.}~\bibnamefont {Janiš}},\ and\ \bibinfo
  {author} {\bibfnamefont {T.}~\bibnamefont {Novotný}},\ }\bibfield  {title}
  {\bibinfo {title} {{Perturbation theory for an Anderson quantum dot
  asymmetrically attached to two superconducting leads}},\ }\href
  {https://doi.org/10.1103/PhysRevB.93.024523} {\bibfield  {journal} {\bibinfo
  {journal} {Phys. Rev. B}\ }\textbf {\bibinfo {volume} {93}},\ \bibinfo
  {pages} {024523} (\bibinfo {year} {2016})}\BibitemShut {NoStop}%
\bibitem [{\citenamefont {Pic\'o-Cort\'es}(2021)}]{PicoCortesTh2021}%
  \BibitemOpen
  \bibfield  {author} {\bibinfo {author} {\bibfnamefont {J.}~\bibnamefont
  {Pic\'o-Cort\'es}},\ }\emph {\bibinfo {title} {{AC dynamics of quantum dots
  and Josephson junctions for quantum technologies}}},\ \href
  {http://hdl.handle.net/10486/700881} {Ph.D. thesis},\ \bibinfo  {school}
  {Universidad Aut\'onoma de Madrid} (\bibinfo {year} {2021})\BibitemShut
  {NoStop}%
\end{thebibliography}
\end{document}